\documentclass{aa}  

\usepackage{graphicx}
\usepackage{txfonts}	
\usepackage{amsmath}	
\usepackage{longfigure}
\usepackage{subfig}
\usepackage[colorinlistoftodos]{todonotes}
\usepackage[export]{adjustbox}
\usepackage{makecell}
\usepackage[bookmarks=false]{hyperref}

\definecolor{Gold}{HTML}{F2AB27}
\definecolor{Silver}{HTML}{C0C0C0}
\definecolor{Bronze}{HTML}{B8422D}

\newcommand\encircle[1]{%
  \tikz[baseline=(X.base)] 
    \node (X) [draw, shape=circle, inner sep=0] {\strut #1};}

\newcommand{\gold}{\textcolor{Gold}{\bf \encircle{g}}}
\newcommand{\silver}{\textcolor{Silver}{\bf \encircle{s}}}
\newcommand{\bronze}{\textcolor{Bronze}{\bf \encircle{b}}}

\newcommand{\Halpha}{H$_\alpha$}
\newcommand{\OIIIdoublet}{[OIII] $\lambda \lambda 4959,5007$}
\newcommand{\OIII}{[OIII]}
\newcommand{\OIIlambda}{[OII] $\lambda 3727$}
\newcommand{\OII}{[OII]}
\newcommand{\Deltarms}[1]{$\Delta_{\rm r m s}=0\farcs#1$}

\defcitealias{Bergamini23}{Be23}
\defcitealias{Pearls23}{Di24}

\begin{document}

   \title{CANUCS: Constraining the MACS J0416.1-2403 Strong Lensing Model with JWST NIRISS, NIRSpec and NIRCam}

   \author{G. Rihtaršič
          \inst{1}
          \and
          M. Brada\v{c}
          \inst{1,2}
          \and
         G. Desprez
         \inst{3}
         \and
          A. Harshan
          \inst{1}
          \and
          G. Noirot
          \inst{3,4}
          \and
          V. Estrada-Carpenter
          \inst{3}
          \and
          N. S. Martis
          \inst{1}
          \and
          R. G. Abraham
          \inst{5,6}
          \and
          Y. Asada
          \inst{3,7}
          \and
          G. Brammer
          \inst{8,9}
          \and
          K. G. Iyer
          \inst{10}
          \and
          J. Matharu
          \inst{8,9}
          \and
          L. Mowla
          \inst{11}
          \and
          A. Muzzin
          \inst{12}
          \and
          G. T. E. Sarrouh
          \inst{12}
          \and
          M. Sawicki
          \inst{3}
          \and
          V. Strait
          \inst{8,9}
          \and
          C. J. Willott
          \inst{13}
          \and
          R. Gledhill
          \inst{8,9}
          V. Markov
          \inst{1}
          \and
          R. Tripodi
          \inst{1}
          }

   \institute{Faculty of Mathematics and Physics, Jadranska ulica 19, SI-1000 Ljubljana, Slovenia\\
              \email{gregor.rihtarsic@fmf.uni-lj.si}
         \and
             Department of Physics and Astronomy, University of California Davis, 1 Shields Avenue, Davis, CA 95616, USA
             \and
             Department of Astronomy \& Physics and Institute for Computational Astrophysics, Saint Mary's University, 923 Robie Street, Halifax, NS B3H 3C3, Canada
             \and
             Space Telescope Science Institute, 3700 San Martin Drive, Baltimore, Maryland 21218, USA
            \and 
            David A. Dunlap Department of Astronomy and Astrophysics, University of Toronto, 50 St. George Street, Toronto, Ontario, M5S 3H4, Canada
            \and 
            Dunlap Institute for Astronomy and Astrophysics, 50 St. George Street, Toronto, Ontario, M5S 3H4, Canada
            \and 
            Department of Astronomy, Kyoto University, Sakyo-ku, Kyoto 606-8502, Japan
            \and
            Cosmic Dawn Center (DAWN), Denmark
            \and 
            Niels Bohr Institute, University of Copenhagen, Jagtvej 128, DK-2200 Copenhagen N, Denmark
            \and
            Columbia Astrophysics Laboratory, Columbia University, 550 West 120th Street, New York, NY 10027, USA
            \and
            Whitin Observatory, Department of Physics and Astronomy, Wellesley College, 106 Central Street, Wellesley, MA 02481, USA
            \and
            Department of Physics and Astronomy, York University, 4700 Keele St. Toronto, Ontario, M3J 1P3, Canada
            \and
            National Research Council of Canada, Herzberg Astronomy \& Astrophysics Research Centre, 5071 West Saanich Road, Victoria, BC, V9E 2E7, Canada
             }

   \date{Received 15 June 2024; accepted 23 January 2025}

 
  \abstract
   {Strong gravitational lensing in galaxy clusters has become an essential tool in astrophysics, as it allows one to directly probe the dark matter distribution and study magnified background sources. The precision and reliability of strong lensing models rely heavily on the number and quality of multiple images of background sources with spectroscopic redshifts. 
  }
   {We present an updated strong lensing model of the galaxy cluster MACS J0416.1-2403 with the largest sample of multiple images with spectroscopic redshifts in a galaxy cluster field to date. Furthermore, we aim to demonstrate the effectiveness of JWST, particularly its NIRISS camera, for strong lensing studies. }
   {We used JWST's NIRCam imaging and NIRSpec and NIRISS spectroscopy from the CAnadian NIRISS Unbiased Cluster Survey (CANUCS). The cluster mass model was constrained using \texttt{Lenstool} software.}
   {Our new dataset, which we used for constraining the lens model, comprises 303 secure multiple images with spectroscopic redshifts from 111 background sources and includes 95 systems with previously known MUSE redshift and 16 systems (with 46 multiple images) for which we obtained spectroscopic redshift for the first time using NIRISS and NIRSpec spectroscopy. Three of the spectroscopic systems were not identified by previous JWST studies. The total number of secure spectroscopic systems is more than $ 20 \%$ higher than in the previous strong lensing studies of this cluster. The derived strong lensing model can reproduce multiple images with the root-mean-square distance of $\sim 0\farcs 52$. We also provide a full catalogue with 415 multiple images, including less reliable candidates. In total, we provide 15 new multiple-image system candidates (with 38 multiple images) not reported in previous studies. Furthermore, we demonstrate the effectiveness of JWST, particularly NIRISS, for obtaining spectroscopic redshifts of multiple images. As NIRISS F115W, F150W, and F200W grism spectroscopy captures at least two of the \OIIlambda{}, \OIIIdoublet,  and \Halpha\ lines at $1\lesssim z \lesssim 3$ (a redshift range particularly relevant for strong lensing studies) without target pre-selection, it complements MUSE and NIRSpec observations extremely well.}
   {}

                
\titlerunning{CANUCS: Constraining the MACS J0416.1-2403 strong lensing model with JWST}
\authorrunning{G. Rihtaršič et al.}
   \maketitle
%

\section{Introduction}
\label{sec:intro}
Strong gravitational lensing of cluster-sized lenses has become an important tool in astrophysics. It provides large magnifications that allow one to study faint galaxies and sub-kiloparsec regions at high redshifts that would otherwise be inaccessible with current telescopes. It thus offers an insight into the formation of first galaxies and the sources of reionisation \citep[e.g.][]{2018Natur.557..392H,2023A&A...678A.173V,2023ApJ...943....2W,2023MNRAS.523L..40A,2023ApJ...949L..23S,fireflysparkle,2024Natur.632..513A,2024ApJ...961L..21B,2024arXiv240218543F}. In extreme cases, it even enables studies of individual stars over cosmological distances \citep[e.g.][]{2022Natur.603..815W,2023ApJ...944L...6M,Fudamoto_2025,2024MNRAS.527L...7F}. Since gravitational lensing allows one to estimate mass distribution independently of the hydrostatic equilibrium and dynamical equilibrium of cluster galaxies,
it is also an essential tool for investigating the interplay between baryons and dark matter on different scales and characterising properties of merging and out-of-equilibrium clusters (see \citealt{2024SSRv..220...19N} for a review). Finally, since lensing depends on the geometry of the Universe, it can also be utilised to constrain values of cosmological parameters  \citep[e.g.][]{1964MNRAS.128..307R,2010Sci...329..924J,2018ApJ...865..122M,2022A&A...657A..83C}. 

The various applications of strong lensing rely on an accurate reconstruction of the mass distribution, which in turn relies on the quality and abundance of data, namely, positions and accurate redshifts of multiple images of strongly lensed background sources. This first requires high-resolution imaging in several photometric bands, which enable precise position measurements and multiple-image system identification based on colours and morphologies. It also requires spectroscopic redshift measurements, which are essential for confirming multiple images and reducing systematic errors \citep[e.g.][]{2016ApJ...832...82J}. Until recently, the progress in the field of strong lensing in galaxy clusters was spearheaded by the \textit{Hubble} Space Telescope (HST) through its large field of view and high-resolution imaging in the visible wavelengths and complemented by ground-based spectroscopic observations, for example, with Multi-Unit Spectroscopic Explorer (MUSE; \citealt{2012Msngr.147....4B}) at the Very Large Telescope (VLT). The \textit{James Webb} Space Telescope (JWST), launched in 2021, expanded those capabilities to near-infrared wavelengths, paving the way for a new generation of lens models \citep[e.g.][]{2022A&A...666L...9C,2023ApJ...945...49M,2023MNRAS.523.4568F,Pearls23,Gledhill24}. 

The CAnadian NIRISS Unbiased Cluster Survey (CANUCS) is a $\sim 200$ h survey that aims to use the unprecedented capabilities of JWST in combination with strong gravitational lensing in galaxy clusters to study high redshift galaxies. With three of the four instruments on board JWST, CANUCS observed five massive clusters known to be effective gravitational lenses, including MACS J0416.1-2403 (hereafter MACS0416).


MACS0416 is a massive galaxy cluster at $z=0.397$ discovered by the MAssive Cluster Survey (MACS; \citealt{2010MNRAS.407...83E}). It is a cluster merger featuring a two-peaked X-ray surface brightness distribution \citep{2012MNRAS.420.2120M} coinciding with a bimodal mass distribution and is likely in a pre-collisional phase \citep{2016ApJS..224...33B}. Due to its elongated mass distribution and large Einstein radius, MACS0416 is a very effective gravitational lens with an extraordinary number of multiple images \citep[e.g.][]{2013ApJ...762L..30Z,2016ApJS..224...33B,2016ApJ...831..182H,Caminha17,Pearls23}. It has also sparked interest due to its abundance of high magnification transient events, enabling the studies of stars at $z\sim1$ \citep{2018NatAs...2..324R,2019ApJ...881....8C,2019ApJ...880...58K,2023ApJS..269...43Y,Pearls23}. Since its discovery, the cluster has been targeted by several cluster surveys. It was selected as a part of the Cluster Lensing and Supernova survey with \textit{Hubble} (CLASH; \citealt{2012ApJS..199...25P}), which provided constraints for the first detailed mass model by \citet{2013ApJ...762L..30Z}. Due to its essential lensing properties, MACS0416 was among the six clusters observed as a part of the Hubble Frontier Fields (HFF) program \citep{2017ApJ...837...97L,2014AAS...22325401L}, which provided deep ( $\lesssim28.5$ mag) imaging data in seven HST ACS/WFC3 bands of the central cluster region. The HFF imaging data has since been used in numerous lensing analyses \citep[e.g.][]{2014MNRAS.443.1549J,2015MNRAS.446.4132J,2014ApJ...797...48J,2015MNRAS.447.3130D,2016ApJ...819..114K,2016MNRAS.461.2126S}. The extended cluster region was also observed by The Beyond Ultra-deep Frontier Fields and Legacy Observations (BUFFALO) program \citep{2020ApJS..247...64S}, and it was included in the Flashlights program \citep{2019hst..prop15936K}.

MACS0416 has also been targeted by several spectroscopic follow-up observations. The cluster was observed with VLT's VIsible MultiObject Spectrograph (VIMOS) as a part of the CLASH-VLT spectroscopic campaign (presented in \citealt{2016ApJS..224...33B}), resulting in a lens model constrained with 30 spectroscopically confirmed multiple images belonging to ten systems (\citealt{2015ApJ...800...38G}; see our Table \ref{tab:previous_catalogs}). This spectroscopic dataset contributed to the development of \citet{2014MNRAS.444..268R,2014ApJ...797...48J} and \citet{2014MNRAS.443.1549J} lens models. The HST/WFC3 infrared grism spectroscopy obtained by the Grism Lens-Amplified Survey from Space (GLASS; \citealt{2015ApJ...812..114T,2014ApJ...782L..36S}) led to a non-parametric lens model and a dataset of 30 spectroscopic multiple images of 15 background sources \citep{2016ApJ...831..182H}. A substantial improvement of the multiple-image catalogue was facilitated by several VLT/MUSE observations. The southwest region of the cluster was observed with MUSE as a part of the programme 094.A-0525 (PI: Bauer) with 11h of exposure time. The northeast region has been targeted by the programs GTO 094.A-0115B (PI: Richard) and 0100.A-0764 (PI: Vanzella, \citealt{Vanzella21}), reaching a total integration time of 17.1h. This makes MACS0416 a cluster with some of the deepest MUSE observations to date. 

Leveraging the first MUSE observations, \citet{Caminha17} presented a catalogue of 102 spectroscopically confirmed multiple images from 37 sources. Their model was improved by including a hot X-ray emitting gas component \citep{2017ApJ...842..132B,2018ApJ...864...98B} and information on galaxy kinematics \citep{Bergamini19}. The dataset of multiple images was further expanded by \citet{bergamini21} by leveraging the data presented in \citet{Vanzella21}. \cite{Richard21} published a MUSE catalogue of 198 multiple images from 71 systems. This number was further increased to 237 spectroscopically confirmed multiple images by \citet[hereafter \citetalias{Bergamini23}]{Bergamini23}. Another catalogue of 214 multiple images was compiled by \cite{Diego23} by re-evaluating systems from existing literature. Before the onset of JWST, these catalogues presented the most extensive dataset of spectroscopically confirmed multiple images for this cluster, and they served as a starting point for constructing the initial catalogue used in our search for additional multiple-image systems. Apart from the aforementioned works, MUSE datasets have been used for several other lens reconstructions \citep[][]{2019MNRAS.486.5414V,2018ApJ...855....4K,2022A&A...664A..90L,2023ApJ...951..140C,2025MNRAS.536.2690P}.

\begin{table}
\caption{\label{tab:previous_catalogs}Number of systems $N_{\rm sys}$ and multiple images $N_{\rm im}$ with known spectroscopic redshift from several strong lensing studies.}
    \centering
    \renewcommand{\arraystretch}{1.2}
    \begin{tabular}{l c c c}
        Catalogue & $N_{\rm sys}$ & $N_{\rm im}$ & Spec. measurement \\
        \hline
        \hline
        \citet{2015ApJ...800...38G} & 10& 30 & VLT/VIMOS\\
        \hline
        \citet{2016ApJ...831..182H} & 15 & 30 & HST/GLASS \\
        \hline
        \citet{Caminha17} & 37 & 102 & VLT/MUSE \\
        \citet{bergamini21} & 66 & 182 & \\
        \citet{Richard21}& 71 & 198 \\
        \citetalias{Bergamini23}& 88 & 237 \\
        \citetalias{Pearls23} (PEARLS)& 77 & 226 \\
        \hline
        This work - gold &111 &303 & JWST/CANUCS \\
        This work - all &124 & 349 &  \\
    \end{tabular}
    \tablefoot{The last column represents the name of the survey or the instrument used to improve the catalogue of spectroscopic systems. Some studies use the same data but differ in the number of systems and images, either due to different quality assessments or due to the inclusion of multiply imaged clumps within a single background galaxy. The number does not include systems without spectroscopic redshifts, which were also used to constrain the lens model in some works (e.g. \citetalias{Pearls23}). Note that the last line (this work - all) also includes images from the quartz catalogue with MUSE redshift, which are not used for lens modelling. }
    
\end{table}

MACS0416 was recently observed with the JWST Near Infrared Camera (NIRCam) by the CANUCS survey and the Prime Extragalactic Areas for Reionization and Lensing Science (PEARLS) project (\citealt{2023AJ....165...13W}), covering the wavelength range from 0.8 to 5 $\mu\mathrm{m}$. Three epochs of observations of the PEARLS program with an additional CANUCS epoch have enabled studies of transient events in highly magnified regions of lensed galaxies \citep[e.g.][]{2023ApJS..269...43Y,2023A&A...679A..31D}. During the preparation of this paper, the PEARLS collaboration published a new lens model constrained by leveraging the new JWST imaging data (\citealt{Pearls23}, henceforth \citetalias{Pearls23}). With 343 multiple-image candidates, belonging to 119 multiple-image systems, the \citetalias{Pearls23} catalogue represents the largest dataset of its kind to date. It contains all previously known systems with MUSE spectroscopic redshift as well as 41 additional candidates without spectroscopic confirmation. We used \citetalias{Pearls23} catalogue to supplement our dataset, which we further expanded with our imaging and spectroscopic data. For easier comparison of our work with theirs, we cross-matched the catalogues and adopted the names in \citetalias{Pearls23} of the multiple-image systems. 

In addition to NIRCam imaging, the CANUCS program also includes Near Infrared Imager and Slitless Spectrograph (NIRISS; \citealt{2012SPIE.8442E..2RD}) wide-field slitless spectroscopy in F115W, F150W, and F200W bands as well as Near Infrared Spectrograph (NIRSpec) multi-object prism spectroscopy. The new spectroscopic data obtained with JWST are the basis of this work. 

In Table \ref{tab:previous_catalogs}, we list catalogues from several strong lensing studies with the number of multiple images and multiple-image systems. The combination of having a number of strong lensing studies and the largest catalogue of reliable spectroscopic multiple-image systems available (303 in this work) also make MACS0416 interesting for investigating the evolution of the lens models due to an increasing abundance of data. A recent study by \cite{2024arXiv241105083P} has provided a comparison of the mass reconstructions using different datasets (including the one presented in this work).

Throughout this work we assume a flat $\Lambda$CMD cosmology with $\Omega_\Lambda=0.7$, $\Omega_{\rm m}=0.3$, and $H_0=70~{\rm km\,s^{-1}\,Mpc^{-1}}$. At the cluster redshift $z=0.396$, a projected distance of $1''$ corresponds to a physical scale of $5.340$~kpc. Magnitudes are given in the AB system \citep{1983ApJ...266..713O}.

\section{Data}

\subsection{Imaging and photometry}
\label{sec:photometry}
In this work we use CANUCS NIRCam observations in filters F090W, F115W, F150W, F200W, F277W, F356W, F410M, and F444W with $6.4~\mbox{ks}$ exposure time per filter. In addition, we use the archival HST/ACS imaging data in F435W, F606W, and F814W filters and HST/WFC3 data in F105W, F110W, F125W, F140W, F160W filters from the HFF \citep{2017ApJ...837...97L,2014AAS...22325401L} and CLASH \citep{2012ApJS..199...25P} programs. The data reduction and production of the photometric catalogues follows the procedure outlined in \citet{Noirot23} and \citet{2024MNRAS.52711372A}. The NIRCam and WFC3 images were first processed with the official STScI JWST pipeline and \texttt{Grizli} \citep{grizli} and drizzled on the same pixel scale (40 mas/pixel) and aligned with Gaia DR3 astrometry \citep{2016A&A...595A...2G,2023A&A...674A...1G}. The diffuse cluster light and the bright cluster galaxies were modelled and removed following the procedure described in \citet{Martis24}. The bright cluster galaxy subtracted images were then PSF-homogenised to match the PSF in the F444W filter. Source detection was performed on the $\chi_{\rm mean}$ detection image \citep{2018ApJS..235...33D}. Photometric catalogues were produced with the \texttt{Photutils} package \citep{photutils} using Kron \citep{1980ApJS...43..305K} and fixed apertures of different sizes.  Photometric redshifts were computed with \texttt{EAZY-py} \citep{2008ApJ...686.1503B} using  the latest standard templates (\texttt{tweak\_fsps\_QSF\_12\_v3}) and templates from \citet{2023ApJ...958..141L}. \texttt{EAZY-py} also returns the "risk" value \texttt{z\_phot\_risk}, as defined in \citet{2018PASJ...70S...9T}, which we use as an indicator of the photometric redshift reliability. In this work, we only show photometric redshifts with \texttt{z\_phot\_risk}$<0.2$ (Sect. \ref{sec:new_sys}).

\subsection{NIRISS grism spectroscopy}
\label{sec:grism}
The cluster was observed with NIRISS using wide-field slitless spectroscopy mode \citep{Willott2022} in F115W, F150W and F200W filters and two orthogonal grisms (GR150C and GR150R, $R\sim150$) with 9.7 ks exposure time per configuration. Data reduction follows the procedure described in \citet{Noirot23} and includes the initial processing and source modelling with \texttt{Grizli}. The contamination of the cluster galaxies is modelled using the isophotal models of the bright cluster galaxies (Sect. \ref{sec:photometry}) and subtracted from the spectra (the procedure is described in \citealt{Estrada-Carpenter24}). We use our \texttt{Photutils} catalogues derived from NIRCam images to obtain the positions of sources and their spectra. The continuum of each source is then modelled using an iterative polynomial fit. The contamination from other sources is subtracted from the spectrum of each source in the catalogue. The redshift is fitted using the procedure described in \cite{Noirot23}. Each source is fitted with three different methods. The first method fits the spectra using a narrow spectral range
(1.03 - 1.26 $\mu$m in F115W, 1.35 - 1.65 $\mu$m in F150W, 1.79 - 2.20 $\mu$m in F200W) to avoid the less sensitive regions at the edges of each filter response curve. The second method uses the full spectral range to reliably fit sources with emission lines with wavelengths close to the edges of the filter response curves. The third method is a joint fit of the spectro-photometric data with \texttt{Grizli} using all available JWST NIRISS, NIRCam and HST filters. The obtained grism redshifts $z_{\rm niriss}$ are prioritised in that order and are selected upon visual inspection of each spectral fit and assessment of its quality. For some sources, none of the three fits returned a reliable redshift. Those sources were refitted with custom modification of the data range (e.g. by removing configurations with visible unsubtracted contamination) or by applying narrower redshift priors and then visually reinspected. The grism spectra used in this work are provided in Appendix \ref{Appendix:grism spectra} together with details of the individual fits. 

While \texttt{Grizli} provides a redshift probability density function, its width likely underestimates the true redshift uncertainty (see \citealt{Noirot23}). To account for the possible systematic errors we compare grism redshifts of known multiple images with their MUSE redshifts $z_{\rm muse}$ from \citet{Richard21} catalogue. We find that all grism redshifts apart from the two outliers K8 and K5 (see Sect. \ref{sec:known_images}) match MUSE redshift within $\delta z_{\rm niriss}=\pm 0.005 (1+z_{\mathrm{spec}})$ which we use as an error estimate. This error encompasses a small systematic offset - we found that our NIRISS redshifts $z_{\rm niriss}$ may be, on average, underestimated by $\sim 0.002 (1+z_{\mathrm{spec}})$ when compared with $z_{\rm muse}$. While the issue will be further investigated before the release of the CANUCS NIRISS redshift catalogues, we note that this difference is of negligible consequence for the strong lensing analysis.

In Sect. \ref{sec:known_images} (see Fig. \ref{fig:d43}), we also utilise spatially resolved emission line maps to match sub-components of multiple images.  Maps are obtained as described in \citet{Estrada-Carpenter24}. To this end, the extended source was first segmented based on the pixel signal-to-noise ratio. All regions were then fitted simultaneously with priors obtained with broadband photometry and by using a linear combination of spectral templates. The emission line maps were obtained by forward modelling the galaxy with the best-fit model, excluding the line of interest, subtracting it, and dithering the residuals.

\subsection{NIRSpec prism spectroscopy}
\label{sec:nirspec}
Following the NIRCam and NIRISS observations, their reduction and subsequent target selection, MACS0416 was observed with NIRSpec using the PRISM/CLEAR disperser with nominal resolving power $\sim 100$       and the approximate wavelength coverage between 0.6  and 5.3 $\mu\mathrm{m}$. Targets were selected for a variety of science goals, including multiple-image system candidates for this project (see Sect. \ref{sec:new_sys}), and were observed through the Micro-Shutter Assembly (MSA) with 2.9 ks exposures for each MSA configuration. The cluster was observed with three MSA configurations. The data is then processed using the STScI JWST pipeline and the \texttt{msaexp} package \citep{msaexp}. This includes mitigating the snowball residuals, masking out pixels approaching saturation, standard wavelength calibration, flat-field and path-loss corrections, photometric calibration and the extraction of 1-D spectra (e.g. see \citealt{2023ApJ...958L..14W}, and \citealt{2024MNRAS.530.2935D} for a more detailed outline of the procedure).  The redshifts $z_{\rm nirspec}$ were obtained using \texttt{msaexp}. In Appendix \ref{Appendix:NIRSpec spectra}, we show the NIRSpec spectra used in this work and describe cases that required distinct treatment.

The NIRSpec redshift uncertainty was estimated by comparing NIRSpec and MUSE redshift from \citet{Richard21}, similarly to Sect. \ref{sec:grism}. We found that the difference between MUSE and NIRSpec redshift is below $0.002 (1+z_{\mathrm{spec}})$ for more than 68$\%$ of multiple images. We use this as an error estimate, which takes into account potential small systematic deviations \citep[e.g.][]{2024A&A...690A.288B}. As a test, we also estimated the uncertainty by fitting a Gaussian profile to the strongest spectral line in spectra of several objects and obtained the line position uncertainty. We find it to be consistent with our error estimate. 

\section{Catalogue of multiple images}
\label{sectioncatalogue}

\begin{figure*}
    \centering
    \includegraphics[width=\linewidth]{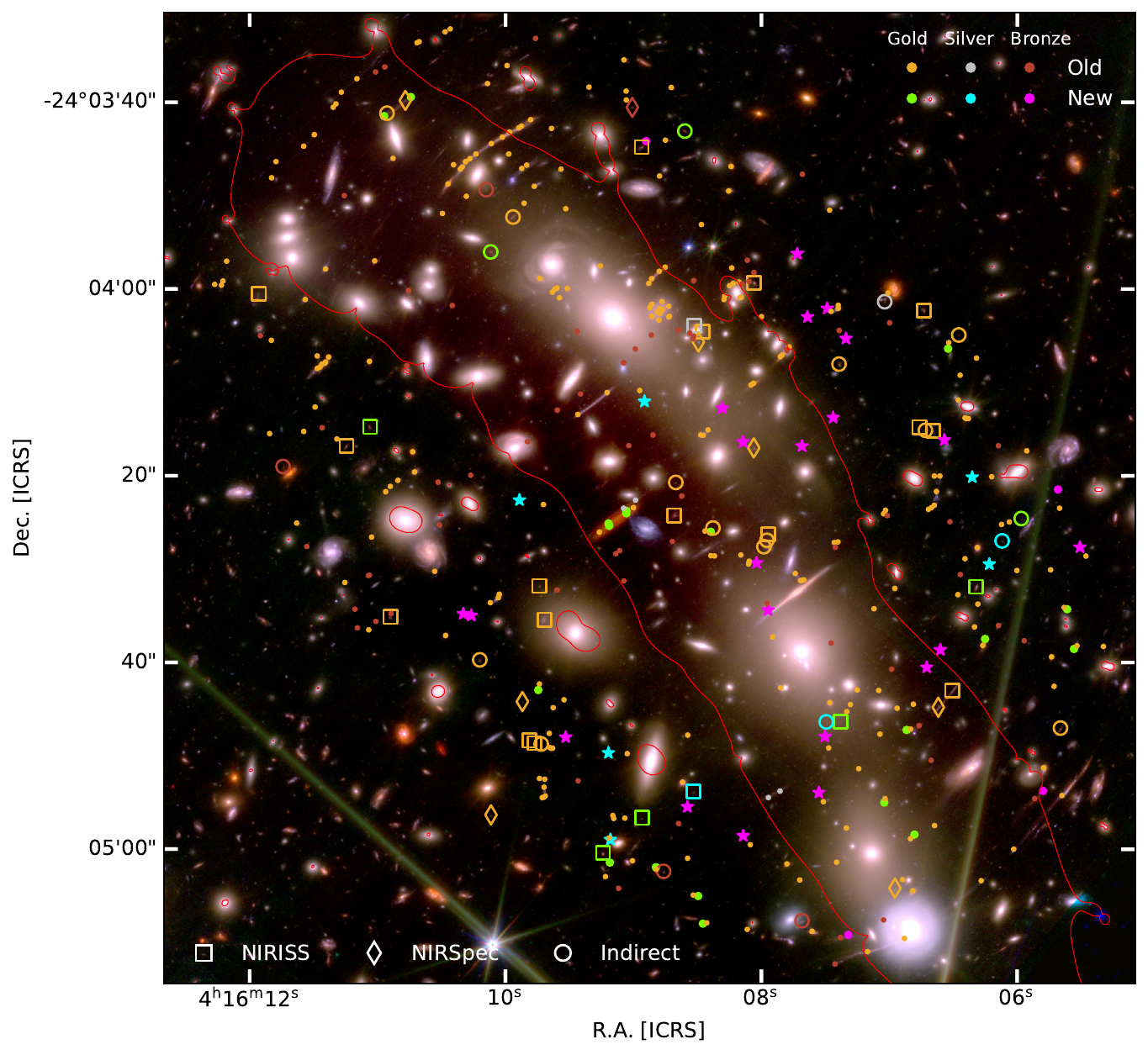}
    \caption{RGB image of the central region of MACS0416 with multiple images. For the RGB image, we used several JWST/NIRCam and HST/WCS filters. F277W, F356W, F410M, and F444W are shown in red; F115W, F150W, and F200W are in green; and F814W, F606W, F435W, and F090W are in blue. Newly identified systems or multiply imaged clumps belonging to the gold, silver, or bronze category are shown in green, cyan, or magenta, respectively. Known systems and clumps from \citetalias{Pearls23} and \citetalias{Bergamini23} are depicted in gold, silver, or bronze, with a colour corresponding to their category in this work. The bronze category also includes quartz images. Squares and diamonds represent images with new spectroscopic redshift, obtained with NIRISS and NIRSpec, respectively. Empty circles represent images for which we could not measure spectroscopic redshift directly but have new spectroscopic redshifts inferred from other images of the system. Stars represent newly identified candidate systems for which we could not obtain spectroscopic redshift. The red solid line represents the tangential (outermost) critical curve from our lensing model for redshift $z=9$.} 
    \label{fig:field}
\end{figure*}

In this section we describe our catalogue of strong lensing constraints and the methods used to obtain them. Based on reliability, we divide our catalogue in several categories. The gold category, which is used for lens model optimisation, is the most secure and encompasses reliable images from existing catalogues with MUSE redshift $z_{\rm muse}$ (see Sect. \ref{sec:known_images})  and new spectroscopic systems, which we trust due to multiple spectroscopic redshift measurements with NIRISS or NIRSpec, morphology, photometric redshift, or/and lensing configuration. It contains 303 images from 111 multiple-image systems.  We also provide a catalogue of less reliable sources which may be of interest for future studies. These sources are divided into silver or bronze categories with intermediate and low degrees of confidence, respectively. We also add a fourth category of quartz images. Those images appear in previous works with measured MUSE redshifts but are excluded from our gold category because we cannot reliably identify or confirm them with our NIRCam imaging (see Sect. \ref{sec:known_images}). The total number of multiple images in all catalogues is 415, belonging to 150 systems (124 with spectroscopic redshifts). This makes our catalogue the largest catalogue of multiple images in a galaxy cluster field to date. The number of secure spectroscopic multiple images in our gold category transcends the number obtained with MUSE data (\citetalias{Bergamini23}) by 28\%. The catalogues of spectroscopic lensing constraints from previous studies are summarised in Table~\ref{tab:previous_catalogs}. In Fig.~\ref{fig:field}, we show the HST \& JWST image of the central region of MACS0416, overlaid with our lensing catalogue. 

Multiple images in our catalogue are labelled as follows. New CANUCS systems are denoted with the prefix C. Previously known systems are denoted with the prefix K and the identifier from \citetalias{Pearls23}, allowing for a straightforward cross-matching of the catalogues. The lowercase letter following the main ID is the label of the multiply imaged clump. We note that each clump is considered a separate system when reporting the total system count. The decimal digit represents the image ID. For instance, image K14b.3 is the third image of the second (b) clump of a galaxy with ID 14 in \citetalias{Pearls23} catalogue.

We assign systems with at least one spectroscopic redshift measurement, a system redshift $z_{\rm sys}$, which is used for lens modelling. If MUSE redshift is available, we set $z_{\rm sys}=z_{\rm muse}$ due to its higher spectral resolution compared to NIRSpec and NIRISS (except in systems K5 and K8 where we update the spectroscopic redshift; see Sect. \ref{sec:known_images}). If $z_{\rm muse}$ is unavailable, we use $z_{\rm niriss}$, $z_{\rm nirspec}$ or the mean of the two if both are available, due to their comparable spectral resolutions. We note that the redshift differences between different measurements are small and have a negligible impact on the lensing analysis. 

Gold catalogue in \texttt{Lenstool} format, as well as the full catalogue with indicated gold, silver and bronze categories are available on CANUCS website\footnote{\url{https://niriss.github.io/}} and will be available on Mikulski Archive for Space Telescopes (MAST)\footnote{\url{https://archive.stsci.edu/}} with DOI \texttt{10.17909/ph4n-6n76} together with other CANUCS data products and lens models. Any relevant changes to the previous lensing catalogues are listed and described in the following sub-sections.

\subsection{New multiple images}
The search for new multiple images and multiple-image systems was performed by leveraging the strong lensing model from \citetalias{Bergamini23}, which was constrained with over a hundred spectroscopically confirmed multiple images, uniformly distributed across the cluster field and covering a wide redshift range $1 \lesssim z \lesssim 6$. The model reproduced positions of their multiple images with root-mean-square distance  \Deltarms{43}. We used \citetalias{Bergamini23} best model parameters to predict the positions of potential counter images of all sources in our photometric catalogues brighter than the 28.5 magnitude in F200W filter, using \texttt{Lenstool} and \texttt{Eazy} redshifts. The magnitude limit was applied to reduce computational and manual inspection time. It is motivated by the fact that identifying multiple-image candidates based on colour and morphology similarity becomes increasingly difficult at fainter magnitudes. Each source with predicted multiple images was then visually inspected. We examined a region around the predicted counter images in several RGB combinations of the HST and JWST filters, trying to identify sources with similar colours, morphology or photometric redshift. The visually inspected region around each source was broad, spanning over $\sim 15 ''$ - much broader than $\Delta_{rms}$ of the model. This strategy was adopted to prevent any potential lensing model inaccuracies from biasing our dataset, as well as to account for errors in photometric redshifts. The regions around the source and image predictions were also checked for any neighbouring sources with similar configurations of multiple images, not predicted by the lens model either due to catastrophic redshift errors (e.g. with redshift below cluster redshift) or due to them not being included in our photometric catalogues. We note that a vast majority of multiple-image systems in this cluster follow a similar configuration, forming a sequence of three multiple images oriented perpendicularly to the elongated cluster morphology.

After compiling a list of candidates for multiple-image systems, we inspected their grism spectra and their fitted redshifts (see Sect. \ref{sec:grism}). The strategy for obtaining grism redshifts differed from the strategy used to obtain CANUCS grism redshift catalogues (Noirot et al., in prep.), where each spectrum is evaluated independently. We jointly evaluated the spectra of each multiple-image system and used all of them to assess the reliability of individual redshifts or spectral features. For some images, we could identify a potential emission line in a single orientation only (for instance, if the other orientation was too contaminated). If the redshift obtained from such emission lines matched the spectroscopic redshift of other images within redshift uncertainty, we considered the candidate spectroscopically confirmed. This approach allowed us to extract useful information from several spectra that would, on their own, prove inconclusive (see Appendix \ref{Appendix:grism spectra}). The newly found CANUCS systems with NIRISS spectroscopic redshifts are listed in  Table \ref{tab:gold_canucs}.

After the initial search for multiple images we compiled the list of targets for spectroscopic follow-up observations with NIRSpec. The candidates were assigned priority values based on their reliability and unavailable grism redshifts. We note that our degree of confidence in some systems was high enough (e.g. due to colour and morphology similarity) that they did not necessarily require spectroscopic confirmation of all images. In such cases, a single spectroscopic redshift was enough to include the system in the gold category of multiple images with $z_{\rm spec}$. Due to the limited possible configurations of the three MSA masks and the large number of follow-up candidates, not all of our lensing candidates were observed. On the other hand,  we obtained NIRSpec spectra of some images that were initially not identified as multiple-image candidates (e.g. K100.1 at $z_{\rm nirspec}=7.24 \pm 0.02$).  

After its release, we used  the multiple-image catalogue from the PEARLS collaboration (\citetalias{Pearls23}), which leveraged JWST  imaging data,  to assess the effectiveness of our multiple-image identification method. As mentioned above, we cross-matched the catalogues and renamed our candidates to match the \citetalias{Pearls23} catalogues for a straightforward comparison. We independently found 20 multiple-image system candidates from \citetalias{Pearls23}. For 11 of them, we also obtained spectroscopic redshifts, which are listed in Table \ref{tab:multsys_canucs} with their class. Then, we visually inspected the remaining 22 system candidates from \citetalias{Pearls23}, which we previously missed, and evaluated them, following the same procedure as before. We included systems K100, K99 and image K115a.3 in the gold catalogue. For systems K100 and K99 we obtained spectroscopic redshifts (see Table \ref{tab:multsys_canucs}). We also added 1 new silver system and 4 new bronze systems. For consistency, we did not include other system candidates from \citetalias{Pearls23} in our catalogue as they were too faint or otherwise could not be confirmed with the same degree of confidence as used for our candidates. Regardless, those candidates may still be interesting for future studies, so we direct the reader to \citetalias{Pearls23} for a full list. We also note that our spectroscopic redshifts of the \citetalias{Pearls23} systems are in good agreement with \citetalias{Pearls23} geometric redshift, predicted by their lens model - with relative differences below a few percent for most systems (see Fig. B.1 in \citetalias{Pearls23}).

\begin{table} 
    \centering
    \renewcommand{\arraystretch}{1.2}
    \caption{\label{tab:gold_canucs}New CANUCS systems of multiple images with spectroscopic redshifts. }
    \begin{tabular}{c c c c c}
    
        id & ra & dec &$z_{\mathrm{niriss}}$&class \\
      
\hline
\hline
C1.1 & 64.026343 & -24.075524 & 2.91 & \gold \\
C1.2 & 64.038488 & -24.083446 & 2.91 &   \\
\hline
C2.1 & 64.037222 & -24.082402 & 3.07 & \gold \\
C2.2 & 64.024868 & -24.073494 &   &   \\
C2.3 & 64.030749 & -24.079544 & 3.07 &   \\
\hline
C3.1 & 64.035827 & -24.061968 &   & \gold \\
C3.2 & 64.042150 & -24.065565 &   &   \\
C3.3 & 64.046081 & -24.070770 & 2.56 &   \\
\hline
C4.1 & 64.035547 & -24.081619 & 1.86 & \silver \\
C4.2 & 64.031217 & -24.079546 &   &   \\
C4.3 & 64.025491 & -24.074159 &   &   \\
    \end{tabular}
\tablefoot{Columns represent multiple-image IDs, their ICRS coordinates, NIRISS redshifts and their class (gold or silver, depending on their reliability). Grism redshift uncertainty is estimated as $0.005 (1+z)$ (see Sect. \ref{sec:grism}).}
\end{table}

\begin{table*}[h]
    \centering
    \renewcommand{\arraystretch}{1.2}
    \caption{\label{tab:multsys_canucs}Multiple image candidates from \citetalias{Pearls23} with new CANUCS spectroscopic redshifts.  }
    \begin{tabular}{c c c c c c c}
        id & ra & dec &$z_{\mathrm{niriss}}$&$z_{\mathrm{nirspec}}$&$z_{\mathrm{sys}}$&class \\

\hline
\hline
K82.1 & 64.027117 & -24.078617 & 2.03 &   & 2.04 & \gold \\
K82.2 & 64.027568 & -24.079111 &   & 2.043 &   &   \\
K82.3 & 64.036516 & -24.083995 &   &   &   & \bronze \\
\hline
K87.1 & 64.028044 & -24.067304 & 2.91 &   & 2.91 & \gold \\
K87.2 & 64.036172 & -24.073404 & 2.91 &   &   &   \\
K87.3 & 64.040400 & -24.076506 & 2.91 &   &   &   \\
\hline
K89.1 & 64.027990 & -24.070869 &   &   & 3.08 & \gold \\
K89.2 & 64.033151 & -24.074143 &   &   &   &   \\
K89.3 & 64.040716 & -24.080179 & 3.07 & 3.082 &   &   \\
\hline
K90.1 & 64.028186 & -24.070777 & 3.06 &   & 3.07 & \gold \\
K90.2 & 64.033103 & -24.073967 & 3.07 &   &   &   \\
K90.3 & 64.040882 & -24.080101 & 3.07 &   &   &   \\
\hline
K91.1 & 64.027739 & -24.070888 & 3.07 &   & 3.07 & \gold \\
K91.2 & 64.033250 & -24.074333 &   &   &   &   \\
K91.3 & 64.040505 & -24.080204 &   &   &   &   \\
\hline
K95.1 & 64.037234 & -24.062446 & 2.33 &   & 2.33 & \gold \\
K95.2 & 64.041423 & -24.064525 &   &  &   &   \\
K95.3 & 64.046847 & -24.071344 & 2.33 & 2.330 &   &   \\
\hline
K98.1 & 64.033581 & -24.071393 &   & 2.303 & 2.30 & \gold \\
K98.2 & 64.030804 & -24.068900 &   &   &   &   \\
K98.3 & 64.042504 & -24.077700 &   &   &   &   \\
\hline
K99.1 & 64.040567 & -24.075502 & 2.51 &   & 2.51 & \gold \\
K99.2 & 64.036119 & -24.072422 &   &   &   &   \\
K99.3 & 64.029316 & -24.067045 &   &   &   & \silver \\
\hline
K100.1 & 64.041116 & -24.078943 &   & 7.24 & 7.24 & \gold \\
K100.2 & 64.034910 & -24.073778 &   &   &   &   \\
K100.3 & 64.026904 & -24.068034 &   &   &   &   \\
\hline
K106.1 & 64.033572 & -24.066483 & 2.70 &   & 2.71 & \gold \\
K106.2 & 64.035385 & -24.068243 &   & 2.712 &   &   \\
K106.3 & 64.045416 & -24.076423 & 2.70 &   &   &   \\
\hline
K109.1 & 64.032006 & -24.085463 &   &   & 2.16 & \bronze \\
K109.2 & 64.028979 & -24.084492 &   & 2.160 &   & \gold \\
K109.3 & 64.023589 & -24.079729 &   &   &   & \gold \\
\hline
K115a.1 & 64.044930 & -24.061056 &   & 2.298 & 2.30 & \gold \\
K115a.2 & 64.045528 & -24.061436 &   &   &   &   \\
K115a.3 & 64.049699 & -24.066808 & 2.30 &   &   &   \\
K115b.1 & 64.044751 & -24.060955 &   &   &   &   \\
K115b.2 & 64.045622 & -24.061524 &   &   &   &   \\
\hline
K97.1 & 64.037538 & -24.061262 &   & 5.99 & 5.99 & \bronze \\
K97.2 & 64.042299 & -24.063712 &   &   &   &   \\
K97.3 & 64.048913 & -24.071950 &   &   &   &   \\
    \end{tabular}

\tablefoot{System redshift $z_\mathrm{sys}$, used for lens modelling is the average of NIRISS redshift $z_\mathrm{niriss}$ and NIRSpec redshift $z_\mathrm{nirspec}$. All images of the system are in the same class (gold, silver or bronze, depending on their reliability) as the first image, unless explicitly indicated otherwise. The redshift uncertainty is estimated as $0.005 (1+z)$ and $0.002 (1+z)$ for  $z_{\mathrm{niriss}}$ and $z_{\mathrm{nirspec}}$, respectively (see Sects. \ref{sec:grism} and \ref{sec:nirspec}).}
\end{table*}

\begin{table}
    \centering
    \renewcommand{\arraystretch}{1.2}
    \caption{New CANUCS system candidates without spectroscopic redshifts.}
    \begin{tabular}{c c c c c c}

        id & ra & dec &$z_{\mathrm{phot}}$\\
      
\hline
\hline
C5.1 & 64.037140 & -24.070012 & $1.41^{+0.09}_{-0.12}$ & \silver  \\
C5.2 & 64.041213 & -24.072947 & $1.35^{+0.18}_{-0.11}$&    \\
C5.3 & 64.032164 & -24.065615 &   & \bronze  \\
\hline
C6.1 & 64.025908 & -24.074851 & $3.02^{+0.06}_{-0.05}$ & \silver  \\
C6.2 & 64.038250 & -24.083050 &   &    \\
\hline
C7.1 & 64.026463 & -24.072264 & & \silver  \\
C7.2 & 64.038313 & -24.080469 & $2.08^{+0.20}_{-0.35}$ &    \\
C7.3 & 64.033115 & -24.076220 &   & \bronze  \\
\hline
C8.1 & 64.042781 & -24.076381 & $2.01^{+0.12}_{-0.11}$ & \bronze  \\
C8.2 & 64.031192 & -24.067244 &  $1.99^{+0.18}_{-0.10}$ &    \\
\hline
C9.1 & 64.033488 & -24.074815 & $2.78^{+0.19}_{-0.25}$ & \bronze  \\
C9.2 & 64.039707 & -24.080002 & $2.91^{+0.14}_{-0.26}$ &    \\
C9.3 & 64.027375 & -24.071157 &  &    \\
\hline
C10.1 & 64.031831 & -24.067497 &  $2.05^{+0.11}_{-0.12}$ & \bronze  \\
C10.2 & 64.034607 & -24.070214 &  &    \\
C10.3 & 64.043043 & -24.076331 & $1.95^{+0.17}_{-0.15}$ &    \\
\hline
C11.1 & 64.027502 & -24.077402 &  & \bronze  \\
C11.2 & 64.027948 & -24.077922 &  &    \\
\hline
C12.1 & 64.035741 & -24.082074 & $2.65^{+0.17}_{-0.33}$ & \bronze  \\
C12.2 & 64.031260 & -24.079980 & $2.60^{+0.08}_{-2.40}$  &    \\
\hline
C13.1 & 64.030579 & -24.068140 &  *$3.40^{+0.03}_{-0.09}$  & \bronze  \\
C13.2 & 64.033930 & -24.071216 &  *$2.60^{+0.04}_{-0.03}$  &    \\
\hline
C14.1 & 64.033924 & -24.082935 & $3.32^{+0.10}_{-0.09}$ & \bronze  \\
C14.2 & 64.031458 & -24.081651 & $3.41^{+0.05}_{-0.07}$ &    \\
C14.3 & 64.022955 & -24.074347 & $3.40^{+0.04}_{-0.08}$ &    \\
\hline
C15.1 & 64.030995 & -24.070493 & *$3.48^{+0.03}_{-0.04}$ & \bronze  \\
C15.2 & 64.032010 & -24.071340 & *$3.53^{+0.08}_{-0.08}$ &    \\
    \end{tabular}
    \tablefoot{Columns represent multiple image IDs, their ICRS coordinates, \texttt{Eazy} photometric redshifts and category (silver or bronze, depending on their reliability). We only show multiple images from our photometric catalogues with \texttt{z\_phot\_risk}$<0.2$ and with redshifts higher than the cluster redshift. Redshifts marked with * have \texttt{z\_phot\_risk} below the threshold but have SNR>10 in less than 4 wide NIRCam bands.}
    \label{tab:multsys_canucs_bronze2}
\end{table}

In addition, we also found 15 systems not reported in \citetalias{Pearls23}. For systems C1, C2, C3 and C4, we obtained grism redshifts and included C1, C2 and C3 in the gold category and C4 in the silver category. We provide a list of new multiple-image systems with spectroscopic redshifts in Table \ref{tab:gold_canucs}. For candidates from C5 to C15, we did not obtain spectroscopic redshifts. In Table \ref{tab:multsys_canucs_bronze2}, we provide their photometric redshifts and class. The class of the new multiple images was assigned upon qualitative assessment of the similarity of their colours in several HST and JWST RGB combinations (in Appendix \ref{Appendix:grism spectra}, we show cutouts of the new images), morphological features, their lensing configuration and photometric and spectroscopic redshifts.

In total, we identified eight gold, nine silver, and 21 bronze new multiple images. We also applied our method of finding multiple images to the other cluster of the CANUCS survey (e.g. \citealt[][]{Gledhill24}, Desprez et al. in prep).

\label{sec:new_sys}

\subsection{Known systems with MUSE redshift}
\label{sec:known_images}
Our initial catalogue of secure multiple images was compiled from \citetalias{Bergamini23}, \cite{Diego23}, and \cite{Richard21} catalogues, which were all based on HST imaging and VLT/MUSE observations, described in Sect. \ref{sec:intro}. A small correction ($\sim 0\farcs 2$) was applied to positions from the previous catalogues to match the positions in our JWST images. We adopted the system names from \citetalias{Pearls23}. We also include additional multiply lensed clumps from \citetalias{Bergamini23}, which provide additional information on image deformation and are particularly useful for constraining critical curves \citep[e.g.][]{bergamini21,2016ApJ...822...78G}. We keep the same order of clumps as in \citetalias{Bergamini23} while indexing them with letters instead of numerals. For example, system 12.6 from \citetalias{Bergamini23} corresponds to the clump K1f where the name K1 indicates the galaxy with ID 1 in \citetalias{Pearls23}. We visually inspected JWST data of all systems and assigned them to the gold or quartz category. The quartz systems, which were considered reliable in \citetalias{Pearls23}, but are excluded from the gold category in this work, include systems K27, K48, K51, K56, K58, K59, K60, K61, K75 and multiple images K1a.3, K17.3, K25.2, K45.3, K57.3. and K77.4. Some of those sources were excluded from the gold category because they are not detected or are extremely faint (e.g. K51, K61, K75) or obscured by a foreground galaxy (K27.1). We also downgraded clump c of system K62 (corresponding to images 21.3b and 21.3c in \citetalias{Bergamini23}) to the quartz category due to faintness.  We note that the candidates in the quartz system were not rejected by our study; we excluded them from the gold category since our data alone did not provide a sufficient degree of confidence. In some cases \citetalias{Bergamini23} and \citetalias{Pearls23} select different candidates for the same image or report significantly different positions ($\gtrsim 0\farcs 3$). In such cases, we picked a more reliable candidate based on similar morphology or clear detection and added it to the quartz catalogue (K45.3, K58.1, K60.1, K61.3). We excluded the alternative candidates from all our catalogues (see \citetalias{Bergamini23} and \citetalias{Pearls23} for a full list of candidates). The two images of system K59 at redshift $z=4.530$ have different positions in the \citetalias{Bergamini23} and \citetalias{Pearls23} catalogues, with K59.2 candidates separated by more than $3''$. Despite clear detection of the \citetalias{Pearls23} candidate and absence of the \citetalias{Bergamini23} candidate in the NIRCam data, we opted to include the latter in the quartz catalogue, as it more closely matches the MUSE detection (see \citealt{Richard21}). We also excluded the K59.3 candidate from all catalogues, as it was detected as a possible counter image of \citetalias{Pearls23} candidates.

In a few spectroscopically confirmed extended systems we were not able to unambiguously identify multiply imaged features with our imaging data. An example is system K1 at redshift $z=0.94$ \citep{2016ApJ...831..182H,Caminha17}, which contains a large extended arc. The arc is close to the northern brightest cluster galaxy (BCG), which increases the probability of microlensing events. Hence, it has been studied extensively due to its abundance of caustic-crossing transients
\citep{2019ApJ...881....8C,2019ApJ...880...58K,2019ApJ...881....8C,2023ApJS..269...43Y}. The counterimage of the arc (K1a.3) is spectroscopically confirmed \citep{Caminha17} and is lensed by the cluster member with ID 8971 (ID from \citetalias{Bergamini23}; see Sect. \ref{sec:lensmodel}). However, as it is difficult to securely identify exact counterparts of individual clumps in the arc with HST and JWST images, we decided to include the counterimage in the bronze category using the position from \citetalias{Pearls23}  (the image was excluded on the same grounds from \citealt{bergamini21} and \citetalias{Bergamini23} lens models).

\begin{figure*}
    \centering
    \includegraphics[width=\linewidth]{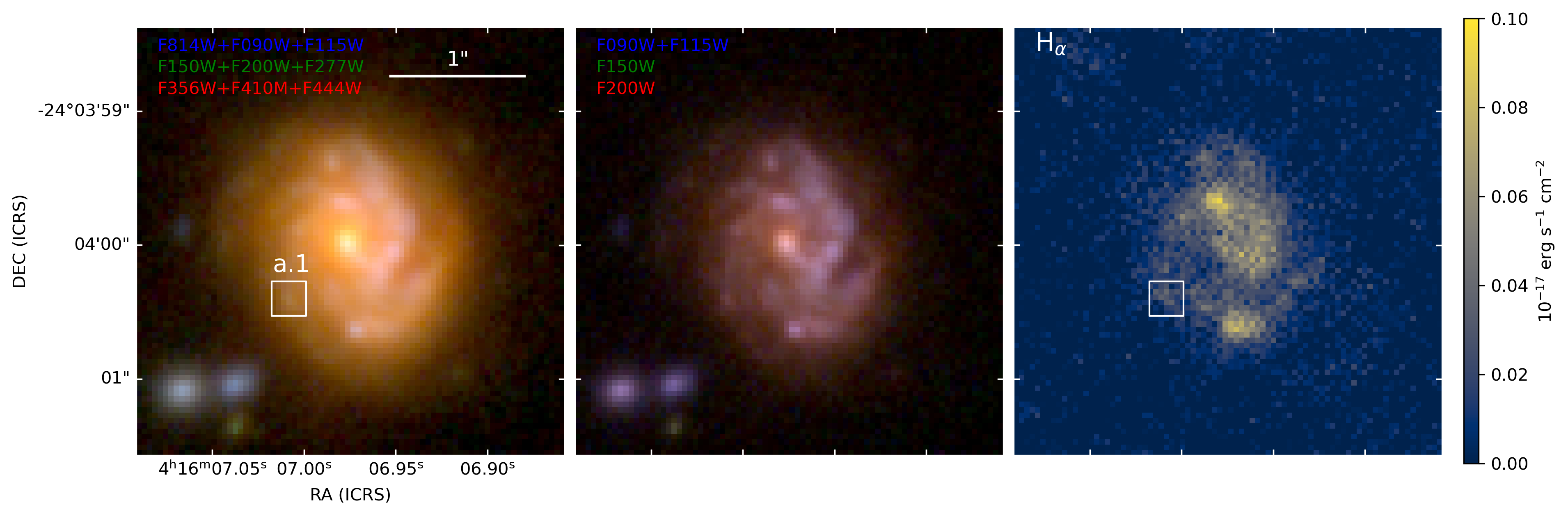}\\
    \includegraphics[width=\linewidth]{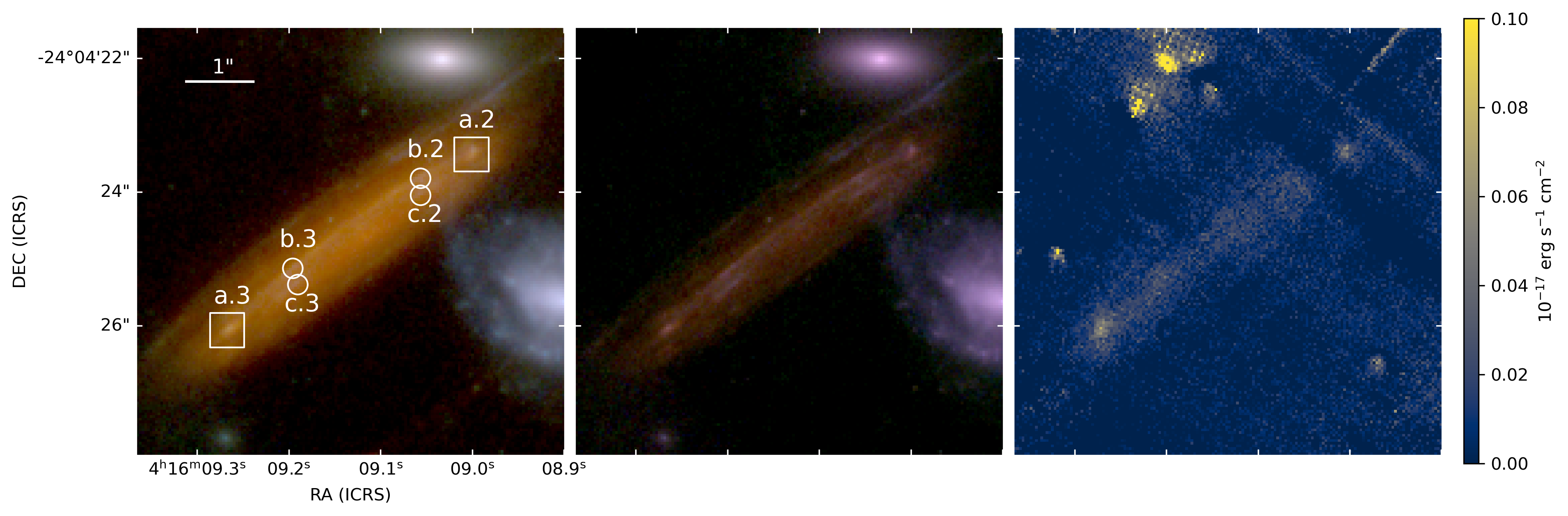}
  
    \caption{Multiply imaged clumps a, b, and c of system K8. The lower panels show the extended arc, with two counter images of the spiral galaxy shown in the upper panels. The two panels on the left show two RGB combinations of NIRCam images. In the right panel, we show \Halpha\ emission line flux obtained from the F200W grism spectra (see Appendix \ref{Appendix:grism spectra}), which was used to identify image K8a.1. }
    \label{fig:d43}
\end{figure*}

We find a similar situation in system K8, which comprises a very red arc and a spiral galaxy at redshift $z=2.04$. The arc has also been investigated by \citetalias{Pearls23} as it hosted two extremely magnified star candidates. The system has a strong \Halpha\ emission line in the F200W NIRISS grism channel, enabling a reliable spectroscopic redshift measurement with NIRISS in both the arc and counterimage. A part of the arc has also been observed with NIRSpec, confirming the spectroscopic redshift of $z_{\rm nirspec}=2.046 \pm 0.001$ (see Table \ref{tab:previouscats} and Fig. \ref{fig:nirspec_canucs_old} and \ref{fig:nirspec_canucs4} in the Appendix). The updated redshift differs from the previously measured MUSE redshift ($z_{\rm muse} = 1.953$; see Table B.4 in \citealt{Richard21}) by 0.09, which is above our redshift uncertainty. Furthermore, we identified the counterpart of the brightest clump of the arc in the counterimage. To this end, we examined several RGB combinations of JWST and HST filters, two of which are shown in Fig. \ref{fig:d43}. Additionally, we produced the map of \Halpha\ emission line flux from F200W grism spectrum (spectrum is shown in Fig. \ref{fig:nirspec_canucs4} and the emission line map in the right panel of Fig. \ref{fig:d43}). The brightest clumps a.2 and a.3 in the arc show clear \Halpha\ emission. The emission is expected to be higher than its counterpart a.1 in the spiral image due to higher magnifications in the arc. This rules out two of the brightest \Halpha\ emitting clumps above and below the galactic centre. Based on colour similarity and faint \Halpha\ emission, we identify the clump on the bottom left relative to the galactic centre to be the most likely candidate for clump a.1. We verified that given the shape of the caustic and the positions of bright clumps in the source plane, the selected a.1 clump is the farthest below the caustic line; in other words, it is one of the most likely to be multiply imaged. Unlike K8a.1, we could not securely identify fainter counter images K8b.1 and K8c.1 to clumps K8b and K8c in the arc. 

We did a minor correction to the spectroscopic redshift of system K5. Each of the three images of system K5 shows clear \OIIIdoublet\ (blended at the grism resolution), \OIIlambda\ and \Halpha\ emission lines in the grism spectrum (see Fig. \ref{fig:nirspec_canucs4}), yielding spectroscopic redshift of $z_{\rm niriss}=1.84 \pm 0.01$, which differs from $z_{\rm muse}=1.8178$ \citep{Richard21} by more than our redshift uncertainty estimate.

We also spectroscopically confirmed image K62b.3, which was reported in \citetalias{Pearls23} (62.c) as a candidate. We measured $z_{\rm nirspec}=5.14 \pm 0.01$ and included it in the gold category with system redshift set to $z_{\rm muse}=5.106$. In system K26 we identified an alternative candidate to K26.3 from \citetalias{Pearls23}. The grism spectrum (see Fig. \ref{fig:nirspec_canucs4}) of our candidate shows the \OIII\ emission, yielding $z_{\rm niriss}=2.92 \pm 0.02$. This agrees with $z_{\rm muse}=2.9259$; hence, we included our candidate in the gold catalogue. All relevant updates to known systems with MUSE redshifts are summarised in Table \ref{tab:previouscats}. We also modified some of the known systems without spectroscopic redshift from \citetalias{Pearls23}. We provide our alternative candidate K81.3 to image 81.c from \citetalias{Pearls23} as well as a new counter image K96.3 of system K96. Both image candidates were found using the method described in Sect. \ref{sec:new_sys} and are listed in Table \ref{tab:previouscats}.  

\begin{table*}
    \centering
    \renewcommand{\arraystretch}{1.3}
    \caption{\label{tab:previouscats}Images belonging to the old systems from \citetalias{Pearls23}, added, updated or confirmed with our data. }
    \begin{tabular}{c c c c c c c}

        id & ra & dec &$z_{\mathrm{niriss}}$&$z_{\mathrm{nirspec}}$&$z_{\mathrm{sys}}$&class \\
\hline
\hline
K5.1 & 64.02388 & -24.07761 & 1.84$^*$ &   & 1.84$^*$ & \gold \\
K5.2 & 64.03056 & -24.0827 & 1.84$^*$ &   &  &  \\
K5.3 & 64.0325 & -24.08377 & 1.84$^*$ &   &  &  \\
\hline
K8a.1 & 64.029208$^*$ & -24.066782$^*$ & 2.03$^*$ &   & 2.04$^*$ & \gold \\
K8a.3 & 64.038610 & -24.073900 & 2.04$^*$ & 2.046$^*$ &  &  \\
\hline
K26.3$^*$ & 64.038267 & -24.083734 & 2.92$^*$ &   & 2.93 & \gold \\

\hline
K62b.3 & 64.042139 & -24.082314 &   & 5.14$^*$ & 5.11 & \gold \\
\hline
K81.3$^*$ & 64.023671 & -24.072630 &   &   &   & \bronze \\
\hline
K96.3$^*$ & 64.037090 & -24.062266 &   &   &   & \bronze \\
    \end{tabular}

\tablefoot{The asterisk ($*$) indicates relevant changes. Asterisk in the id means newly added counterimages, including the spectroscopically confirmed alternative to K26.3 image from \citetalias{Pearls23}. We have updated the system redshift of systems K5 and K8 while also changing the exact position of K8a.1 (see Sect. \ref{sec:known_images}). We spectroscopically confirmed K62b.3.  The redshift uncertainty is estimated as $0.005 (1+z)$ and $0.002 (1+z)$ for  $z_{\mathrm{niriss}}$ and $z_{\mathrm{nirspec}}$, respectively (see Sects. \ref{sec:grism} and \ref{sec:nirspec}). }
\end{table*}

In some lensed sources with resolved structure in our imaging data, we included different features as separate multiple-image systems. This includes several features in highly magnified arcs, included in \citetalias{Bergamini23} as well as additional new families K8b, K8c, K22b, K49b, K55b, K64b, K85b and K115b. Out of those, K8b, K8c and K115b are situated in strongly lensed arcs and are useful for locally constraining the critical curves. Others belong to well-separated images and provide additional information on their orientation and local deformations. 

\begin{figure}
    \centering
    \includegraphics[width=\linewidth]{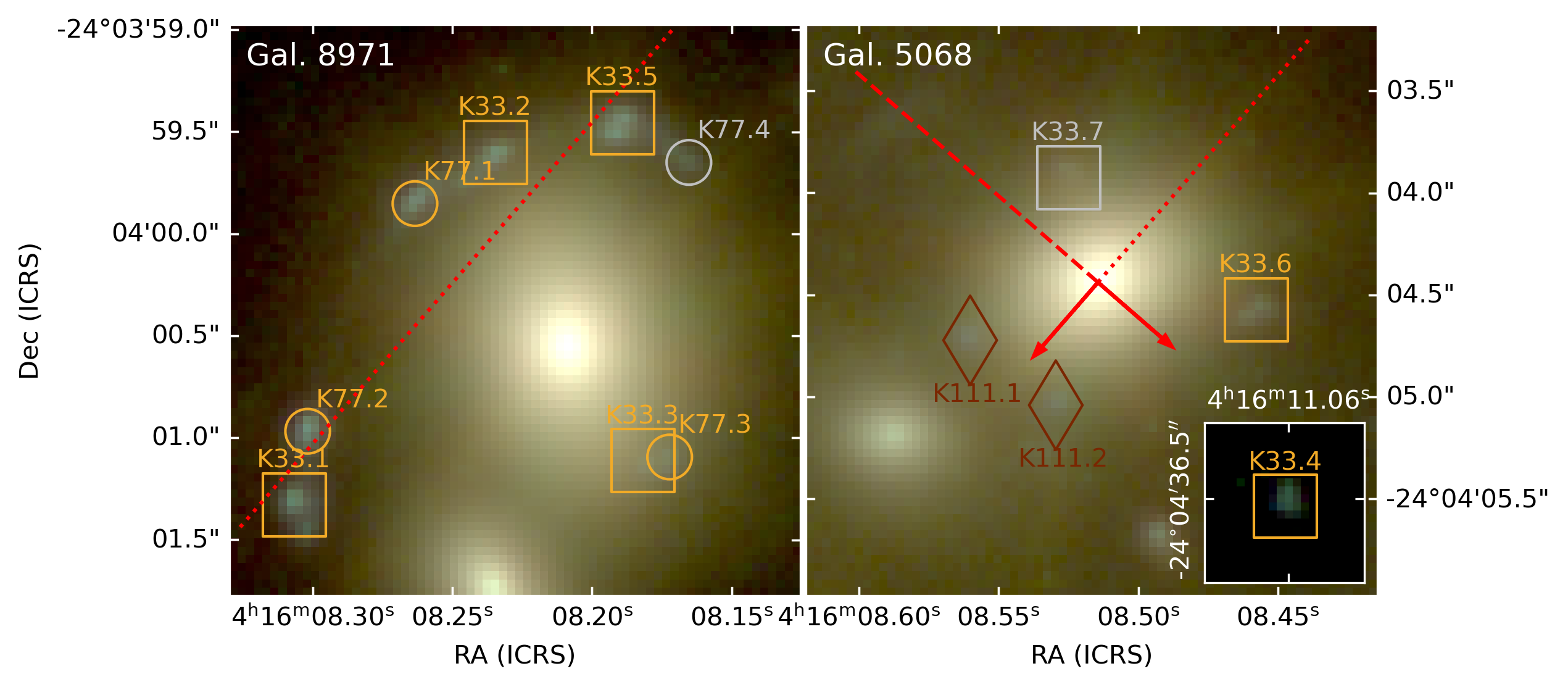}
    \includegraphics[width=\linewidth]{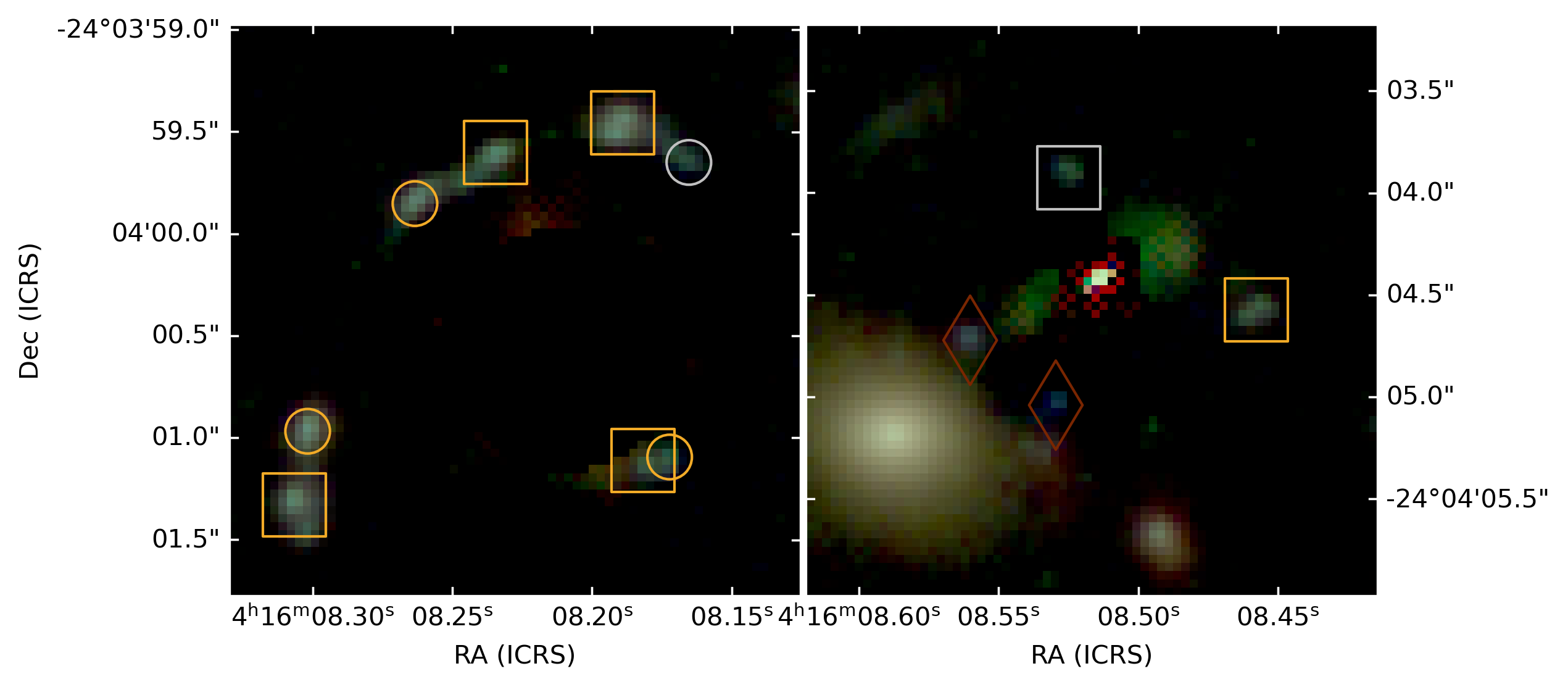}
    \includegraphics[width=0.9\linewidth]{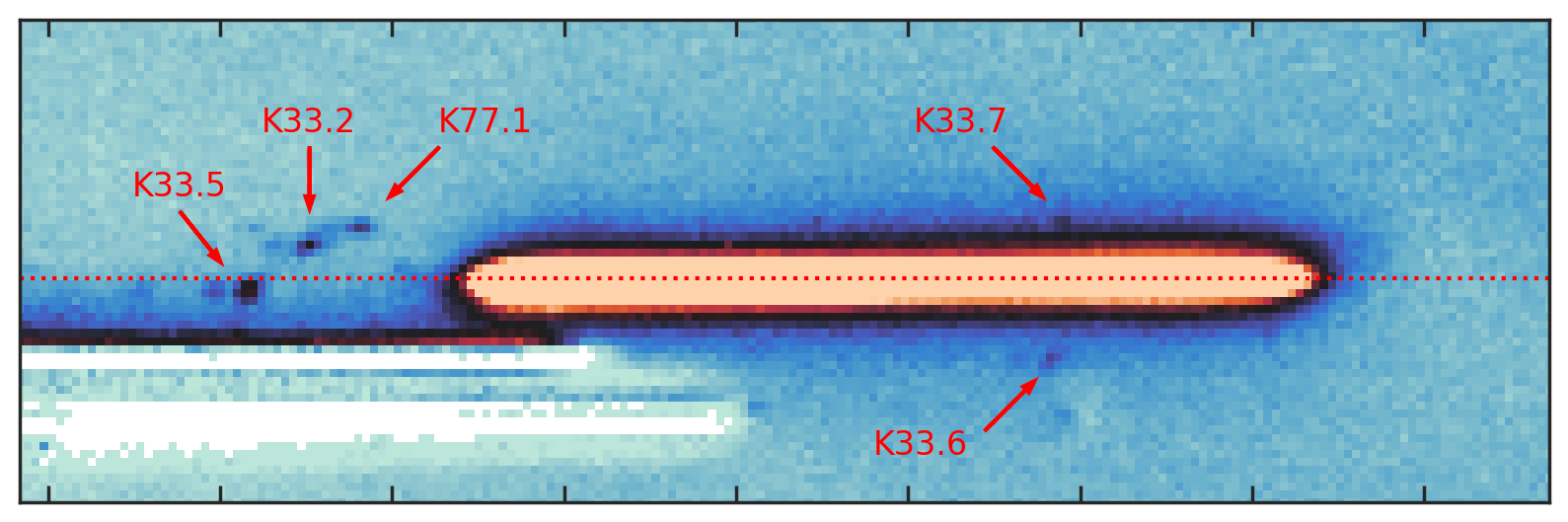}
    \includegraphics[width=0.9\linewidth]{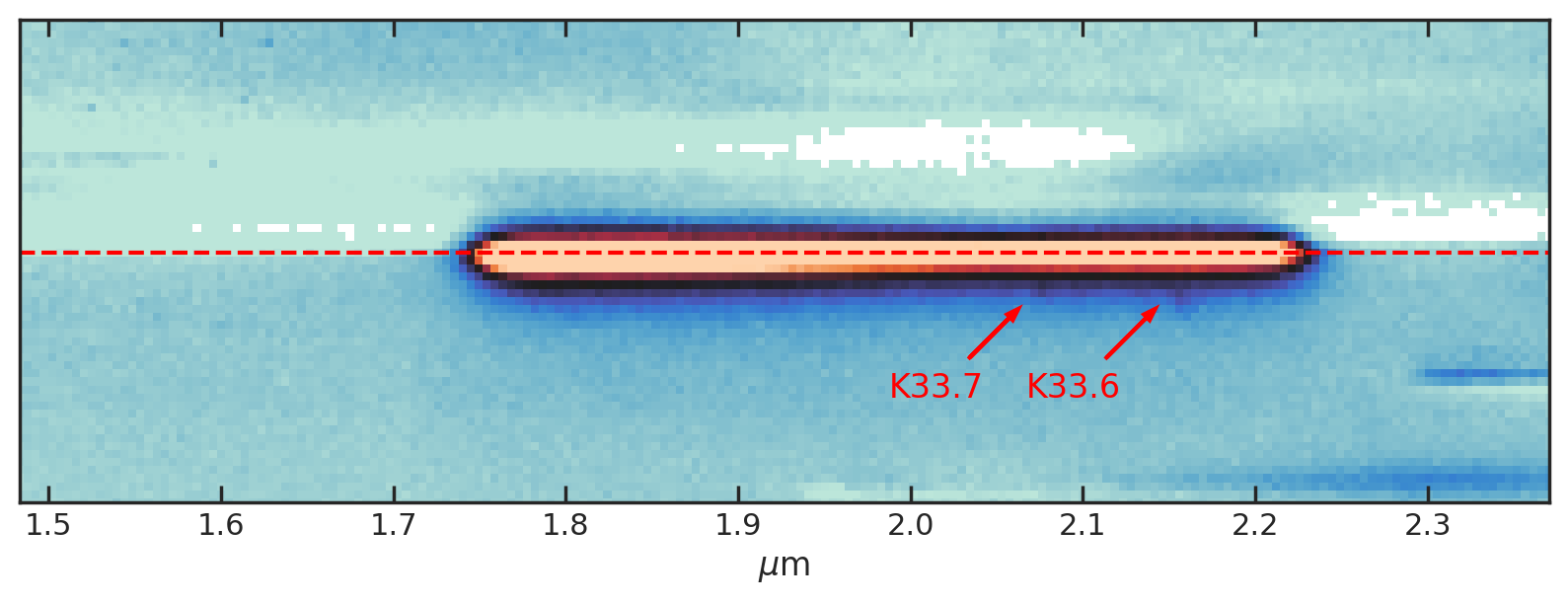}
    \caption{Upper panel: RGB cutouts (with the same filter combinations as used in Fig. \ref{fig:field}) of cluster members 8971 and 5068 with surrounding multiple-image systems K33, K77, and K111. The inset cutout shows the counterimage K33.4.  Gold, silver or bronze colours indicate the category (reliability) of multiple images. Dotted and dashed lines with arrows represent the orientation of 150C and 150R grism spectrum, respectively. They are also indicated on the bottom two panels. The dotted line indicating the 150C spectrum orientation extends to galaxy 8971, positioned $6''$ to the northeast of galaxy 5068. Second panel: RGB cutouts with removed cluster galaxies, enabling a better colour comparison. Third panel: F200W 150C grism spectrum of galaxy 5068 showing potential  \OIII\ emission lines observed at $2.1\mu\mathrm{m}$ from K33.6 and K33.7, and emission from K33.2, K77.1, and K33.5 on the left. The wavelength calibration matches the spectrum of galaxy 5068. Fourth panel: F200W 150R grism spectrum of galaxy 5068, indicating the \OIII\ emission of K33.6 and K33.7.}
    \label{fig:B14}
\end{figure}

K33 and K77 (14.1 and 14.2 in \citetalias{Bergamini23}) are a unique lensing system consisting of two compact low luminosity star-forming stellar complexes at redshift 3.221 \citep{2017ApJ...842...47V}. The sources are first multiply lensed by the global cluster profile and then further split into several images by two cluster members, resulting in a system of seven multiple-image candidates (shown in Fig. \ref{fig:B14}). The systems are included as in \citetalias{Pearls23} with a few modifications. We excluded K77.4 (as given in \citetalias{Pearls23}) since it is indistinguishable from K33.4. We added our K77.4 candidate, which belongs to the same multiple image as K33.5. However, due to its uncertain position, we include it in the quartz category. We included system candidate 110 from \citetalias{Pearls23} to system K33 as images K33.6 and K33.7. Of the two, we include K33.6 to our gold catalogue, following \citetalias{Bergamini23}. For image K33.7, which has not been included as a reliable candidate in previous studies, we provide a tentative spectroscopic confirmation based on the \OIII\ emission line at around 2.1 $\mu$m in the grism spectra, shown in Fig.~\ref{fig:B14} (see also the NIRSpec spectrum of K33.1 in Fig. \ref{fig:nirspec_canucs_old}). The emission can also be seen in other images of the system. This makes system K33 a very likely system with 7 multiple images. In this work, we add K33.7 to the silver category and do not use it to constrain the model. 

In Table \ref{tab:catalogdiffs} we summarise the differences between our and \citetalias{Pearls23} catalogues. We indicate sources that are missing in either catalogue and sources that were upgraded to or downgraded from the most reliable categories.

\begin{table}
    \renewcommand{\arraystretch}{1.6}
    \caption{\label{tab:catalogdiffs_summary}Differences between the \citetalias{Pearls23} multiple-image catalogue and this work.}
    \begin{tabular}{c | c | c} 
    This work & \citetalias{Pearls23} & Systems \\
    \hline
    \hline
    \gold  & {\textcolor{Bronze}{BCD}}  & \makecell{K26.3, K62.3, K71.3$^*$, K82, K87, K89,\\ K90, K91, K95, K98, K99, K100,\\ K106, K109, K115} \\
    \hline
    \gold  & /  & \makecell{C1, C2, C3} \\

    \hline
    \silver\bronze  &/  & \makecell{K14.5$^*$, K14.6$^*$, K96.3, C4, C5,\\ C6, C7, C8, C9, C10, C11, C12,\\ C13, C14, C15} \\

    \hline
    \hline
    \makecell{\silver\bronze\\(quartz)}  & {\textcolor{Gold}A}  & \makecell{K1.3, K17.3, K25.2, K27, K45.3,\\ K48, K51, K56, K57.3, K58, K59\\ K60, K61, K75, K77.4} \\
 \hline
    /  & {\textcolor{Bronze}{BCD}}  & \makecell{K3.3, K27.3, K47.3, K56.3, K59.3,\\ K60.3, K73.3, K78.3, K79.3, K83.3,\\ K86, K92, K93, K94, K104,\\ K108, K112, K113, K114, K115.4,\\ K116, K117, K118, K119 } 
\label{tab:catalogdiffs}
  \end{tabular}

    \tablefoot{The first column represents the category of the system in this work (gold, silver, bronze or not included in our catalogues). The second column represents the category in \citetalias{Pearls23} (A for spectroscopically confirmed images, B and C for good candidates without spectroscopic confirmation and D for least reliable candidates). The first part of the table represents newly discovered systems (or individual images) and systems promoted into our gold category based on spectroscopic redshifts. The second part of the table represents systems we excluded or downgraded to silver or bronze categories upon our inspection. The \citetalias{Pearls23} IDs are renamed as described at the beginning of Sect. \ref{sectioncatalogue}. Multiple images marked with $^*$ are added/upgraded based on \citetalias{Bergamini23} catalogue. Multiple-image systems belonging to the same background object (e.g. K115a and K115b) are written only once for conciseness.}

\end{table}

\section{Lens modelling}
\label{sec:lensmodel}
The strong lensing model was optimised using the parametric strong lens modelling software \texttt{Lenstool}\footnote{\url{https://projets.lam.fr/projects/lenstool/wiki/}} \citep{Kneib96,Jullo07,Jullo09}, which uses Markov chain Monte Carlo sampler \texttt{BayeSys} \citep{skilling2004} to sample the posterior probability of the model parameters ${\pmb \xi}$. The posterior probability is computed as 
\begin{equation}
    P(\pmb{\xi}| \pmb{r}^o)= P( \pmb{r}^o | \pmb{\xi}) P(\pmb{\xi}),  
\end{equation}
where the prior probability $P(\pmb{\xi})$ is the product of either uniform or Gaussian probabilities for each parameter. The likelihood $P( \pmb{r}^o | \pmb{\xi})$ given the positions of $N$ multiple images $\pmb{r}^o$, each with uncertainty $\sigma_p$ is defined as 

\begin{equation}
P( \pmb{r}^o | \pmb{\xi})= \left(2 \pi \sigma_p^2   \right)^{-\frac{N}{2}} \exp \left({-\frac{\chi^2(\pmb{r}^o, \pmb{\xi})}{2} } \right).
\end{equation}
 The $\chi^2$ is calculated from the distance between the observed positions $\pmb{r}^o$ and model predicted positions $\pmb{r}^m$: 
\begin{equation}
  \chi^2(\pmb{r}^o, \pmb{\xi})= \sum_{j=1}^{n_{\rm sys}} \sum_{i=1}^{n_{\rm im}^j} \frac{\lVert \pmb{r}_{i,j}^o - \pmb{r}_{i,j}^m (\pmb{\xi}) \rVert^2}{\sigma_p^2},
\end{equation}
where the sum goes over $n_{\rm sys}$ multiple-image systems, each containing $n_{\rm im}^j$ multiple images. 
For optimisation, \texttt{Lenstool} uses 10 interlinked chains to avoid local minima and varies the likelihood influence during the burn-in phase to control the convergence speed (see \citealt{Jullo07}).

To evaluate the accuracy of the lens model, we used the root-mean-square distance from the observed positions of multiple images and positions predicted by the best model with parameters $\pmb{\xi}_{\rm best}$:
\begin{equation}
    \Delta_{\rm rms}=\sqrt{\frac{1}{N} \sum_{i=1}^{N} \lVert \pmb{r}_{i}^o - \pmb{r}_{i}^m (\pmb{\xi}_{\rm best}) \rVert^2}.
\end{equation}

In this work, we use the mass distribution parameterisation from \citetalias{Bergamini23}, which was introduced as the best-performing lens model (\texttt{LM-4HALOS}) in \citet{bergamini21}. It describes the total (baryonic and dark) matter distribution as a superposition of several elliptical halos with dual pseudo isothermal elliptical (dPIE) density profile \citep{2005MNRAS.356..309L,2007arXiv0710.5636E}. Each dPIE halo is parametrised with 8 parameters: coordinates $x$ and $y$ of the centre, redshift $z$ (set to 0.397 for all halos),  ellipticity $e$, orientation $\theta$ , core radius $r_{\rm core}$, cut radius $r_{\rm cut}$ and the \texttt{Lenstool} fiducial velocity dispersion $\sigma_{\rm lt}$, which is related to central velocity dispersion $\sigma_0$ of the dPIE profile as $\sigma_{\rm lt}=\sqrt{2/3}\sigma_0$. Positions $x$ and $y$ in this work are given in arcsec relative to the reference position $\alpha=64.038142^\circ$, $\delta=-24.067472^\circ$, which coincides with 
northern BCG. The total mass of each dPIE halo is computed as 
\begin{equation}
M_{\text {tot }}=\frac{\pi \sigma_0^2 r_{\text {cut }}}{G} .
\label{eq:piemdmass}
\end{equation}

The global mass component in our lensing model features 4 large-scale cluster halos. Two are centred on the northeastern and southwestern BCGs with $30''$ and $15''$ wide uniform position priors, respectively (H1 and H2 in Fig. \ref{fig:kappadif}). A second halo is added to the southern region with a wider position prior as a second-order correction to the southern mass component (H4). Following previous studies (\citealt{Caminha17,Bergamini19,bergamini21}; \citetalias{Bergamini23}), another halo (H3) was added to the northeast of the northern BCG with fewer free parameters (the halo is kept circular without free ellipticity and orientation), which serves as a correction to the northeastern cluster component and accounts for a small overdensity of cluster galaxies. All four cluster components have free $\sigma_{\rm lt}$ and $r_{\rm core}$, while we fix their $r_{\rm cut}$ to a large value ($2000''$), as it cannot be constrained by the multiple images in the inner cluster regions.  

All prior ranges of the free parameters are listed in Table \ref{tab:modelparamspriors} and match the values from \citetalias{Bergamini23}. Only one modification was required in the position prior of the H3 cluster halo, which had to be narrowed down to  $-39''<x<-25''$ and  $2''<y<16''$. This modification proved necessary as the model with extended prior range ($-55''<x<-25''$,  $0''< y<30''$; see \citetalias{Bergamini23}) had problems with convergence after the addition of new spectroscopic systems - the model placed the $3^{\rm rd}$ cluster halo in a local minimum of $\chi^2$ outside of the region covered with multiple images. After narrowing down the prior range, we found that the new solution, consistent with the best-fit position from \citetalias{Bergamini23}, yielded significantly better $\chi^2$ and $\Delta_{rms}$. We also note that the chosen prior range is nonetheless much larger than the $68\%$ posterior confidence interval (see Table \ref{tab:modelparams}). This also further demonstrated the need for the additional mass component H3 in the overdensity of cluster galaxies to the northeast of the main cluster component (H1). 

The model also includes four dPIE clumps describing the cluster gas distribution, derived from X-ray \textit{Chandra} observations \citep{2018ApJ...864...98B,2017ApJ...842..132B}. They are included as fixed clumps without free parameters, which is justified by a smaller set of assumptions when deriving the gas component mass distribution and smaller statistical errors compared to other components \citep{2017ApJ...842..132B}. The parameters of the four clumps describing the cluster gas can be found in Table 1 of \citetalias{Bergamini23}.

Smaller sub-halos, which are associated with cluster members, were modelled as circular dPIE profiles with a vanishing core radius $r_{\rm core}$. The cut radius $r_{\rm cut \it, i}$ and velocity dispersion $\sigma_{\rm lt, \it i}$ of each cluster member $i$ scale with luminosity $L_i$ of each galaxy were calculated according to the following relations \citep{1996MNRAS.280..167J, 1997MNRAS.287..833N}:
\begin{equation}
\sigma_{\mathrm{lt}, i}=\sigma_{\mathrm{lt}}^{\mathrm{ref}}\left(\frac{L_i}{L_{\mathrm{ref}}}\right)^\alpha,
\label{eq:scaling1}
\end{equation}

\begin{equation}
r_{\rm cut, \it i}=r_{\mathrm{cut}}^{\mathrm{ref}}\left(\frac{L_i}{L_{\mathrm{ref}}}\right)^{\beta_{\rm cut}}.
\label{eq:scaling2}
\end{equation}
In this work, we use the cluster member catalogue with magnitudes from \citet{bergamini21}, comprising 213 galaxies, 171 of which have spectroscopic confirmation. The positions of galaxies were first shifted to match the astrometry of CANUCS images (the correction was small, $\sim 0\farcs 2$, and was also applied to the other mass components).  The slopes of the scaling relations were derived in \citet{bergamini21}, following the procedure described in \citet{Bergamini19}. They leverage measured velocity dispersions of cluster members with MUSE spectroscopy to obtain $\alpha=0.3$, and derive $\beta_{\rm cut}=0.6$ by assuming that the sub-halo mass $M_{\rm tot, \it i}$ scales with luminosity as $M_{\rm tot, \it i} \propto L_i^{0.2}$. As in \citetalias{Bergamini23} and \citet{bergamini21}, we assume a uniform prior between $1''$ and $50''$ for  $r_{\mathrm{cut}}^{\mathrm{ref}}$ , and  a Gaussian prior for $\sigma_{\mathrm{lt}}^{\mathrm{ref}}$ centred at 248 km s$^{-1}$ with standard deviation of 28 km s$^{-1}$.  The Gaussian prior was determined with MUSE spectroscopy as described in \citet{Bergamini19} and helps break the degeneracy between $\sigma_{\mathrm{lt}}^{\mathrm{ref}}$ and $r_{\mathrm{cut}}^{\mathrm{ref}}$, allowing for more accurate characterisation of the sizes of the cluster members. Luminosity $L_i$ of the cluster members was estimated with HST F160W magnitudes from \citet{bergamini21}, and for $L_{\rm ref}$, the reference magnitude of 17.02 was chosen. As a test, we optimised a lens model of the upper half of the cluster (to reduce the dataset size and computational time). We used measured ellipticities instead of circular sub-halos, as well as using NIRCam F090W magnitudes in place of F160W magnitudes. Neither case resulted in a $\chi^2$ improvement compared to an identical run with the old cluster member catalogue; hence, we did not implement these changes in the final model. We leave the refinement of the cluster member catalogue with new JWST data for a future study, for instance, with new NIRISS spectroscopic catalogues of the full cluster field (Noirot et al., in prep.).

Two galaxy-scale halos were not modelled with scaling relations (Eqs. \ref{eq:scaling1} and \ref{eq:scaling2}). A foreground galaxy ($z$= 0.112), which perturbs the lensing potential south of the H4 cluster halo, was modelled as a circular dPIE halo at the cluster redshift with free $\sigma_{\rm lt}$ and $r_{\rm cut}$ and vanishing $r_{\rm core}$. Cluster member 8971, which produces the galaxy-galaxy lensing of systems K77 and K33 (left panel of Fig. \ref{fig:B14}, discussed in Sect. \ref{sec:known_images}), was included as an elliptical dPIE halo with free ellipticity $e$, orientation $\theta$, $\sigma_{\rm lt}$ and $r_{\rm cut}$ with vanishing $r_{\rm core}$. See Table \ref{tab:modelparamspriors} for the prior values.

The parameter optimisation was done in two steps. For the first run of the model, we used the position uncertainty $\sigma_p=0\farcs 3$. We note that this is larger than the actual position uncertainty of the multiple images from the HST and JWST data. However, the actual accuracy ($\sim 40$ mas) is beyond the current lens modelling capabilities, which do not consider contributions such as from the line-of-sight mass distribution or the scatter around the mass-to-light relation of the cluster galaxies \citep[e.g.][]{2010Sci...329..924J}. The limitations of the lens model were further reflected in high $\chi^2$ values after the first run. After obtaining the minimum of the $\chi^2$, we rescaled the position uncertainties $\sigma_p$ by a constant factor so that the minimal $\chi^2$ was approximately equal to the number of degrees of freedom (DoF): 
\begin{equation}
    \mathrm{DoF}=2 N - 2n_{\rm sys} - N_{\rm par}.
\label{eq:dof}
\end{equation}
The term $N_{\rm par}$ in Eq. \ref{eq:dof} represents the number of free parameters of the mass distribution (30 in this work), and  $2n_{\rm sys}$ is equal to the number of source positions of multiple-image systems, which represent an additional 222 free parameters. With $2N=606$ measured coordinate positions of multiple images in our gold catalogue, we obtained $\mathrm{DoF}=354$. After rescaling $\sigma_p$ so that $\chi^2/\mathrm{DoF}\approx 1$, we reran the model. The position uncertainty used in the final optimisation was 0\farcs 49. This rescaling is standard practice in strong lensing studies (e.g. \citetalias{Bergamini23}, \citealt{2016ApJ...822...78G}), and it serves as an attempt to account for systematic inaccuracies of the lens model. We emphasise that rescaling the uncertainty does not influence the best-fit model; rather, it provides better-approximated uncertainties of the model parameters.

After the 83000 burn-in steps, during which the $\chi^2$ value reached a stable value showing no significant average variation over large number of steps, we sampled the posterior with 40000 samples. The optimisation required over two weeks of computational time on 130 CPU cores. The best-fit model resulted in \Deltarms{52}. We do not use other fit quality criteria such as the BIC or AIC (e.g. see \citealt{bergamini21}) as the comparison of different model parameterisations is beyond the scope of this work. We also note the dependence of those criteria on the adopted $\sigma_p$, which does not reflect the uncertainty of the measurements in our case. In Table \ref{tab:modelparams}, we provide the median values and the $68\%$ confidence intervals of the lensing model parameters. We note that the median parameter values may differ from the best model values, which we provide separately as a \texttt{Lenstool} parameter file on MAST. However, it enables a more straightforward comparison with the analogous table in \citetalias{Bergamini23}. 

\begin{table*}
\centering
\renewcommand{\arraystretch}{1.5}
\caption{Input \texttt{Lenstool} parameter values and prior ranges of the four cluster halos, four gas halos, scaling relations, and two galaxy-scale halos (galaxy 8971 and the foreground galaxy), which are modelled separately.}
\label{tab:modelparamspriors}
\begin{tabular}{c|c c c c c c c }

   & $x$ [$''$] & $y$ [$''$] & $e$ & $\theta\left[^{\circ}\right]$ &$r_{\text {core }}['']$ & $\sigma_{\rm lt}\left[\mathrm{~km} \mathrm{~s}^{-1}\right]$ &  $r_{\text {cut }}['']$ \\
\hline
\hline

$1^{\text {st}}$ Cluster Halo &$-15 \div 15$&	$-15 \div 15$&	$0.2 \div 0.9$ &	$100 \div  180$&	$0 \div 20$ &	$350 \div 1000$& 2000.0 \\
  $2^{\text {nd }}$ Cluster Halo & $15 \div 30$&	$-45 \div -30$&	$0.2 \div 0.9$ &	$90 \div  170$&	$0 \div 25$&	$350 \div 1200$& 2000.0 \\
 $3^{\text{rd}}$ Cluster Halo & $-39^* \div -25$&	$2^*\div 16^*$&0&0&	$0 \div 35$&	$50 \div 750$& 2000.0 \\

 $4^{\text {th }}$ Cluster Halo & $-10 \div 50$&	$-75\div -15$&	$0.2 \div 0.9$ &	$0 \div  180$&	$0 \div 20$ &	$100 \div 1000$& 2000.0 \\
\hline

 $1^{\text {st}}$ Gas Halo & $-18.1$ & $-12.1$ & 0.12 & $-156.8$ & 433 & 149.2 & 149.8 \\
 $2^{\text {nd }}$ Gas Halo & 30.8 & -48.7 & 0.42 & $-71.5$ & 249 & 34.8 & 165.8 \\
 $3^{r d}$ Gas Halo & $-2.4$ & $-1.3$ & 0.42 & -54.7 & 102 & 8.3 & 37.6 \\
 $4^{\text {th }}$ Gas Halo & $-20.1$ & 14.7 & 0.40 & $-49.3$ & 282 & 51.7 & 52.3 \\

\hline  Foreground. gal. & 32.0 & -65.6 & 0 & 0 & 0.0001& $50 \div 350$&	$5 \div 100$\\

 Gal-8971 & 13.3 & 2.6 &$0 \div 60$&	$-90 \div  90$&0.0001&	$60 \div 200$&	$0 \div 50$  \\

\hline
Scaling relations & $N_{\text {gal }}=212$ & $m_{\mathrm{F} 160 \mathrm{~W}}^{\text {ref }}=17.02$ & $\alpha=0.30$ & $\sigma_{\rm lt}^{\rm r e f}=[248,28]$& $\beta_{\text {cut }}=0.60$ & $r_{\text {cut }} = 1 \div 50 $& $\gamma=0.20$ \\

\end{tabular}
\tablefoot{A single number is given where the parameter values are fixed, the $\div$ sign represents uniform prior ranges, and square brackets contain the centre and the width of the Gaussian prior for $\sigma_{\rm lt}^{\rm r e f}$. The terms $N_{\text{gal}}$ and $m_{\mathrm{F} 160 \mathrm{~W}}^{\text {ref }}$ indicate the number of sub-halos modelled with the scaling relations and the reference magnitude for which the reference parameters are given, respectively. Positions $x$ and $y$ are given relative to the northern BCG. Apart from the prior range of the third cluster halo (marked with $^*$), all values are identical to values used by \citetalias{Bergamini23} (see Table 1 in \citetalias{Bergamini23}).}
\end{table*}

\begin{table*}
\centering
\renewcommand{\arraystretch}{2.5}

\caption{Optimised \texttt{Lenstool} values of parameters given in Table \ref{tab:modelparamspriors}.} 
\label{tab:modelparams}

\begin{tabular}{c|c c c c c c c }

   & $x$ [$''$] & $y$ [$''$] & $e$ & $\theta\left[^{\circ}\right]$ &$r_{\text {core }}['']$ & $\sigma_{\rm lt}\left[\mathrm{~km} \mathrm{~s}^{-1}\right]$ &  $r_{\text {cut }}['']$ \\
\hline
\hline
$1^{\text {st}}$ Cluster Halo  & \makecell{ $0.6_{-0.8}^{+0.7}$ \\$ [1.2]$} &	\makecell{$-0.4_{-0.4}^{+0.5}$ \\ $ [-0.6]$}  &	\makecell{$0.81_{-0.01}^{+0.01} $\\$[0.80]$} &	\makecell{$142.4_{-0.5}^{+0.6}$\\$ [142.2]$} &	\makecell{$7.5_{-0.3}^{+0.3}$\\$ [7.4]$}&	\makecell{$586_{-10}^{+10}$\\$ [574]$}& 2000.0 \\
  $2^{\text {nd}}$ Cluster Halo  & \makecell{$22.6_{-0.5}^{+0.4}$\\$ [22.0]$} &	\makecell{$-34.3_{-0.8}^{+0.7}$\\$[-33.5]$} &	\makecell{$0.83_{-0.04}^{+0.03}$\\$[0.86]$} &	\makecell{$130.8_{-3.3}^{+1.7}$\\$[131.6]$} &\makecell{$7.1_{-0.5}^{+1.1}$\\$ [7.4]$}&\makecell{$559_{-33}^{+100}$\\$[560]$} & 2000.0 \\
 $3^{\text {rd}}$ Cluster Halo  & \makecell{$-30.7_{-0.6}^{+0.5}$\\$ [-30.4]$} &	\makecell{$9.3_{-0.6}^{+0.4}$\\$ [9.6]$} &0&0&	\makecell{$7.2_{-0.8}^{+0.9}$\\$ [8.0]$}&	\makecell{$351_{-20}^{+17}$\\$ [371]$} & 2000.0 \\

 $4^{\text {th}}$ Cluster Halo & \makecell{$22.0_{-1.2}^{+1.1}$\\$[21.7]$} &	\makecell{$-49.5_{-2.5}^{+1.3}$\\$[-48.4]$} &	\makecell{$0.76_{-0.03}^{+0.04}$\\$[0.71]$} &	\makecell{$126.2_{-1.8}^{+3.7}$\\$[125.4]$} &	\makecell{$11.4_{-0.8}^{+0.8}$\\$[10.5]$} &	\makecell{$595_{-98}^{+38}$\\$[587]$}  & 2000.0 \\

\hline  Foreground gal. & 32.0 & -65.6 & 0 & 0 & 0.0001& \makecell{$99_{-32}^{+27}$\\$[136.0]$} &	\makecell{$54.2_{-26.9}^{+29.6}$\\$[41.1]$}  \\

 Gal-8971 & 13.3 & 2.6 &\makecell{$0.31_{-0.15}^{+0.17}$\\$[0.55]$} &	\makecell{$-52.1_{-16.1}^{+17.7}$\\$[-56.1]$} &0.0001&	\makecell{$112_{-3}^{+3}$\\$[114]$} &	\makecell{$15.7_{-4.9}^{+7.1}$\\$[26.3]$}  \\

\hline  Scaling relations &  \makecell{$\sigma_{\rm lt}^{\rm r e f}=	203_{-5}^{+2}$\\$[206]$} &  \makecell{$r_{\text {cut }}^{\text {ref }}=15.4_{-1.3}^{+1.8}$\\$[16.0]$} & &&& \\
\end{tabular}
\tablefoot{Median parameter values are given with uncertainties (indicating 68\% confidence interval). Best-fit parameter values are shown in square brackets. The table is analogous to the second part of Table 1 in \citetalias{Bergamini23}.   }
\end{table*}

We also note that similar to the \citet{bergamini21} model, our model also predicts the multiple-image K33.7, for which we provided a tentative spectroscopic confirmation in Sect. \ref{sec:known_images} and included it in the silver category. K33.7 is lensed by a cluster member with ID 5068, which is modelled with scaling relations (Eqs. \ref{eq:scaling1} and \ref{eq:scaling2}).

\section{Discussion}
\subsection{Comparisson with Bergamini et al. (2023)}
\begin{figure}
    \centering
    \includegraphics[width=\linewidth]{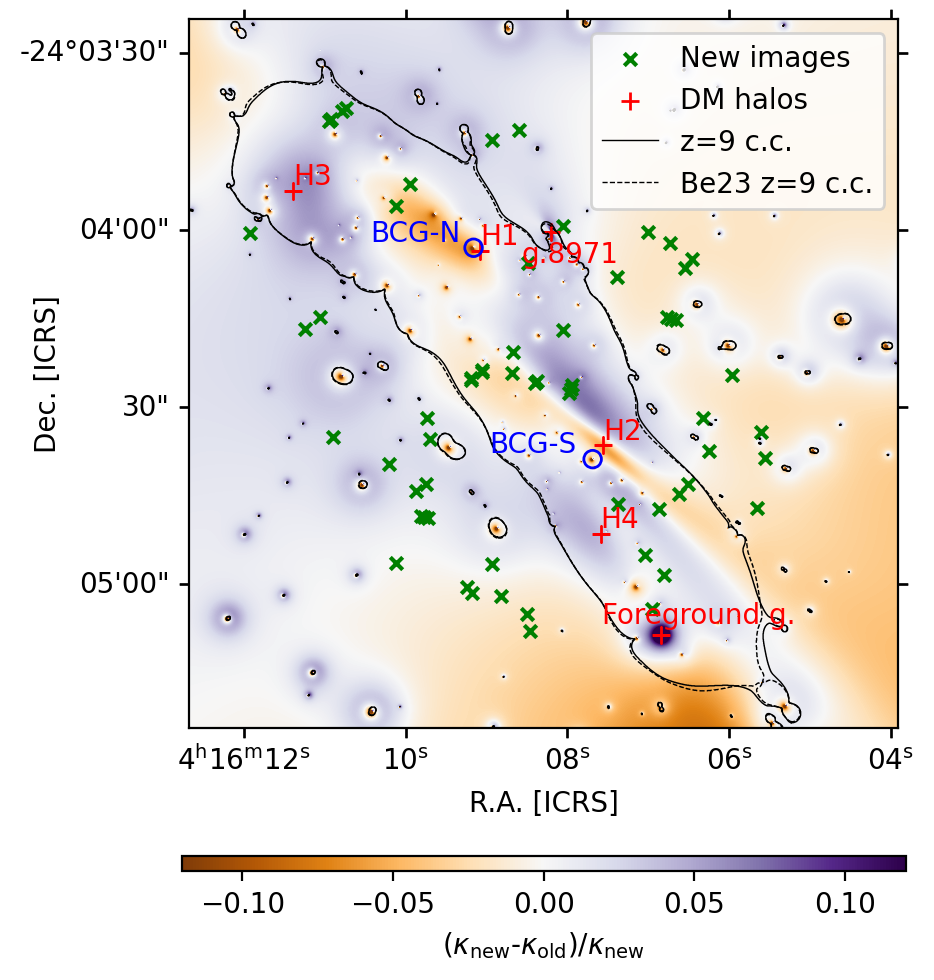}
    \caption{Relative difference between the convergence ($\kappa$) from the updated strong lensing model with JWST data and the lens model from \citetalias{Bergamini23}, which uses the same parameterisation. The positions of four main cluster halos (H1 to H4) and the two galaxies not included in the scaling relations (galaxy 8971 and the foreground galaxy) are marked with red pluses. Positions of the northern and southern BCGs are indicated with blue circles. Green crosses represent new multiple images and old multiple images with new spectroscopic redshifts, not included in the \citetalias{Bergamini23} nor the \citetalias{Pearls23} catalogues. Black solid and dashed lines represent the tangential (outermost) critical curve for sources at redshift 9, derived from the new and \citetalias{Bergamini23} lensing models, respectively.   }
    \label{fig:kappadif}
\end{figure}

In Fig. \ref{fig:kappadif} we show the positions of the main mass components as well the difference of convergence ($\kappa$) maps and the critical curves at redshift $z=9$ between our and \citetalias{Bergamini23} best-fit models. Comparing our mass distribution with \citetalias{Bergamini23}, we can see that the relative $\kappa$ difference between the models is less than a few percent over most of the inner cluster regions. The median difference $|\kappa_{\rm new}-\kappa_{\rm old}|/\kappa_{\rm new}$ in the regions displayed in Fig. \ref{fig:kappadif} is only $1.6 \%$. We note that the close agreement between our density maps is, to a large degree, the consequence of using the same lens modelling method and parameterisation as \citetalias{Bergamini23}. A recent comparison of different MACS0416 lens models (including our, \citetalias{Bergamini23} and \citetalias{Pearls23} models) by \citet{2024arXiv241105083P} shows that the median relative density differences can reach $\sim20\%$ if different modelling methods are used.  
The overall agreement between our and  \citetalias{Bergamini23} mass distributions is also reflected in the positions of the critical curves. As the critical curves trace the regions of extreme magnification, their accurate description is vital for studying high redshift objects. In Fig. \ref{fig:kappadif}, we show the tangential critical curve for sources at $z=9$ to illustrate that the high redshift critical curves derived from our model align well with those from \citetalias{Bergamini23}.

The differences between our and \citetalias{Bergamini23} density maps and critical curves are more drastic in the southern part of the cluster. The foreground galaxy in our best-fit model is significantly more massive than in \citetalias{Bergamini23} (Fig. \ref{fig:kappadif}), which follows from our higher best-fit velocity dispersion $\sigma_{\rm lt}=136 \text{ km s}^{-1}$ (Table \ref{tab:modelparams}) compared to only $77 \text{ km s}^{-1}$ in \citetalias{Bergamini23} best-fit model (see the  $\sigma_{\rm lt}$ dependence of the dPIE mass  in Eq. \ref{eq:piemdmass}). However, we note that our median $\sigma_{\rm lt}$ of the foreground galaxy is significantly lower with large relative uncertainties  ($\sigma_{\rm lt}=99^{+27}_{-32} \text{ km s}^{-1}$). The poorly constrained mass distribution around the foreground galaxy is unsurprising as there are very few constraints in the southern-most cluster regions (see Fig. \ref{fig:field}).

\begin{figure}
    \centering
    \includegraphics[width=\linewidth]{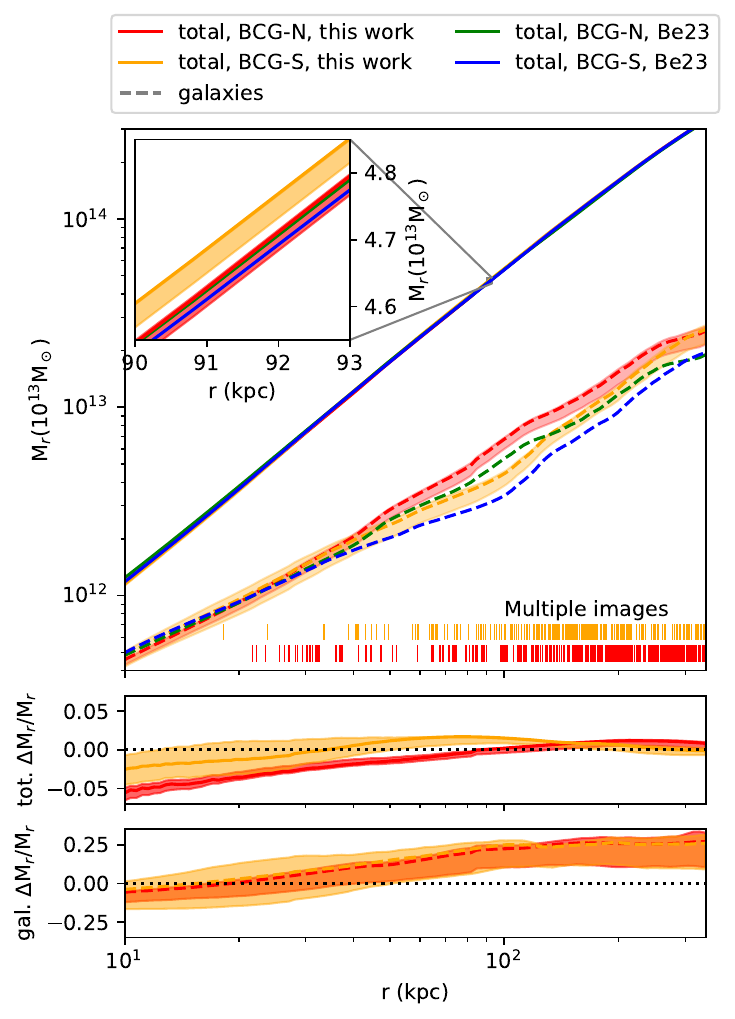}
    \caption{Top: Projected mass $M_r$ in the circular aperture with radius $r$ centred at the northern (red) or the southern BCG (yellow) computed from the best-fit model (solid line) with indicated $68\%$ confidence intervals. The profiles from the best-fit \citetalias{Bergamini23} models are shown in green and blue. The dashed lines represent the cumulative mass $M_r$ contained in the cluster member halos (modelled with the scaling relations). Small vertical bars represent the distances of multiple images from the nothern or southern BCG. Middle: Difference between the total cumulative mass $M_r$ from our model and the \citetalias{Bergamini23} best-fit model relative to our $M_r$. The confidence intervals are computed from our $M_r$ uncertainty. Bottom: Relative difference between the cumulative mass contained in the cluster member halos from our model and that of \citetalias{Bergamini23}.}
    \label{fig:macs0416_cmassprofilebest}
\end{figure}

In Fig. \ref{fig:macs0416_cmassprofilebest} we show the projected total mass $M_r$ enclosed in the circular aperture with radius r (in kpc) from the northern or southern BCG as well as the mass contained in the cluster member halos. The total mass profiles from our model show an excellent agreement  with the \citetalias{Bergamini23} best-fit model, with a relative difference of less than a few percent in the radial range containing multiple images. This is consistent with the literature showing that the circularly averaged total mass profiles are well constrained by lens models - even if different modelling techniques are employed  \citep{2020MNRAS.494.4771R,2017MNRAS.472.3177M}. However, we find a larger difference when comparing the mass contained in the cluster member halos. Our best-fit scaling relation normalisation values $\sigma_{\rm lt}^{\rm r e f}= 206 \text{ km s}^{-1}$ and $r_{\rm cut }^{\rm r e f}=16.0''$ (Table \ref{tab:modelparams}) result in cluster members with $36 \%$ higher masses but more extended profiles with lower central densities than in \citetalias{Bergamini23}  ($\sigma_{\rm lt}^{\rm r e f}= 217 \text{ km s}^{-1}$ , $r_{\rm cut }^{\rm r e f}=10.5''$ ). This can be seen in Fig. \ref{fig:kappadif} and in the last panel of Fig. \ref {fig:macs0416_cmassprofilebest} where the higher total masses of our galaxies and lower mass enclosed in the inner 30 kpc around the BCGs result in a rise of the cumulative mass difference with $r$. Significant cluster member mass differences despite the matching total mass profiles indicate a known degeneracy between the cluster galaxies and the smooth mass component in parametric lens modelling \citep[e.g. ][]{2016A&A...588A..99L}.  Whereas the total mass profiles already seem to be well constrained with 88 multiple-image systems in \citetalias{Bergamini23}, the  cluster member component shows larger variations upon the addition of new systems. We find this despite the Gaussian prior on  $\sigma_{\rm lt}^{\rm r e f}$, which is added to reduce the sub-halo mass uncertainty \citep{Bergamini19}.

We also compared the magnifications of multiple images computed with \citetalias{Pearls23}, \citetalias{Bergamini23} and our lens model. For a meaningful comparison of the lens models, we used the positions and redshifts of spectroscopic multiple images given in Table~B from \citetalias{Pearls23} (while accounting for a slight shift in astrometry between different works). We found that the rms relative magnification difference $2(\mu_1-\mu_2)/(\mu_1+\mu_2)$ between our and \citetalias{Bergamini23} model is 10 \% for images with $\mu < 10$ (in our model) and 35 \% for $\mu>10$.  The larger relative differences at higher $\mu$ are a consequence of rapid magnification variations near the critical curves, which makes those images susceptible to small differences between lens models. Despite different dataset sizes, the magnification differences between our two lens models (both constrained with \texttt{Lenstool)} are relatively small. When comparing our magnifications with those provided by \citetalias{Pearls23}, which used a different modelling code (\texttt{WSLAP+}; \citealt{2005MNRAS.360..477D}), the relative differences are larger, reaching $41\%$ for images with  $\mu < 10$ and $85\%$ for $\mu>10$. For $80\%$ of the multiple images, \citetalias{Pearls23} magnifications are higher than our best model $\mu$.  Large $\mu$ differences when using different modelling methods (which transcend the statistical uncertainties given by individual models) are a known issue that has been investigated with the Hubble Frontier Lens models \citep{2017MNRAS.465.1030P,2020MNRAS.494.4771R}.  As our models differ from  \citetalias{Pearls23} not only in the methods but also in the set of constraints, we cannot meaningfully investigate the impact of the modelling techniques on magnifications in this work. 

When comparing $\Delta_{\rm rms}$ we find that our \Deltarms{52} is larger than \Deltarms{43} in \citetalias{Bergamini23} by $\sim 20 \% $ . The increase is not surprising, considering that the newly derived mass distribution accommodating the extended dataset (with $> 20 \%$ more multiple-image systems) differs from the one that minimised the $\chi^2$ (and $\Delta_{\rm rms}$) of the old dataset using the same model parametrisation.

\subsection{The effectiveness of NIRISS for strong lensing studies }

\begin{figure}
    \centering
    \includegraphics[width=\linewidth]{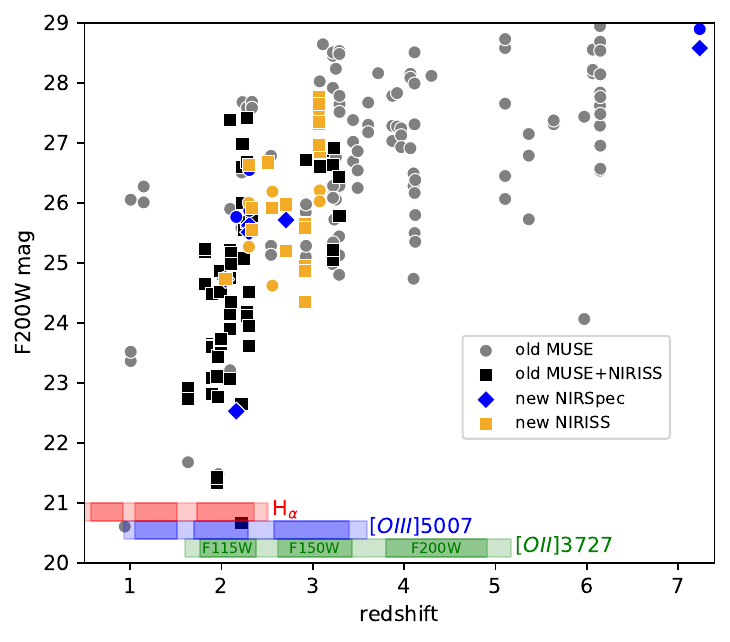}
    \caption{Redshift and F200W magnitude of multiple images in our gold category with known spectroscopic redshifts. Grey and black points represent known images with MUSE redshifts. For black squares, we also obtained NIRISS redshifts. Gold squares represent newly identified images for which we could obtain NIRISS redshift. Blue diamonds represent new images without NIRISS but with NIRSpec redshift. Images for which the redshift could not be reliably measured but with measured system redshift from other images with NIRISS or NIRSpec are marked with gold and blue circles, respectively. Each multiple image is considered once (i.e. different clumps are not considered separately). The coloured bands at the bottom of the figure represent redshift ranges where some of the most prominent emission lines fall in the wavelength range, covered by each of the three NIRISS grism filters, used in this work. The darker shade of the grism range represents a more sensitive region (see Sect. \ref{sec:grism}).}
    \label{fig:z_mag}
\end{figure}

\begin{figure}
    \centering
    \includegraphics[width=\linewidth]{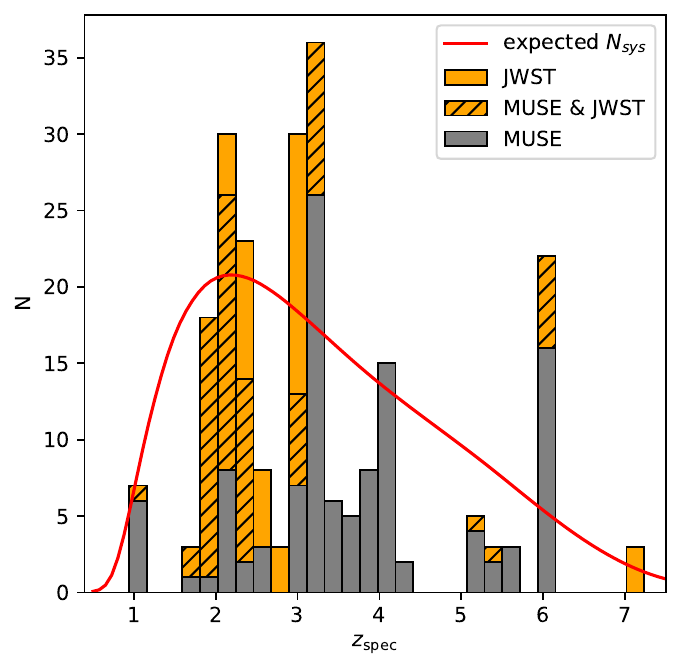}
    \caption{Redshift distribution of multiple images included in our gold catalogue with known spectroscopic redshift. Images for which the spectroscopic redshift was obtained only by MUSE are shown in grey. Images for which we provided spectroscopic redshifts with JWST are shown in orange. Striped regions represent images with both JWST and MUSE redshift measurements. In the histogram, each multiple image is counted once (i.e. different multiply lensed clumps are not considered separately). The red solid line represents the expected redshift distribution of multiple-image systems, considering the redshift distribution of sources in the field and multiply imaged area in the source plane, which increases with redshift. }
    \label{fig:hist_area}
\end{figure}

In Fig. \ref{fig:z_mag}, we show the redshift and F200W magnitude distribution of multiple images with spectroscopic system redshifts. Different markers and colours indicate whether system redshifts of multiple images could be measured with JWST (NIRISS or NIRSpec), MUSE or both. We also indicate the redshift ranges where the most significant emission lines fall in F115W, F150W or F200W grism spectrum. The redshift distribution of multiple images is shown in Fig. \ref{fig:hist_area}. Most new multiple images with NIRISS redshifts are found around redshifts 2 and 3 where at least two of the \OII, \OIII\ and \Halpha\ emission lines fall in the most sensitive range of the grism spectrum. In that range, NIRISS also successfully recovers most redshifts measured by MUSE.

Furthermore, we roughly estimate the expected redshift distribution of multiple-image systems (red line in Fig. \ref{fig:hist_area}). This was done by first calculating the median distribution of photometric redshifts in the NIRCam flanking fields of all 5 CANUCS clusters. We considered all galaxies with reliable photometric redshifts regardless of their magnitudes. The redshift distribution of the sources peaks between redshift 1 and 2 due to the maximum angular diameter distance (and thus the largest volume covered by the same field-of-view). This unlensed distribution was then multiplied by the area of the multiply imaged region in the source plane at each redshift. The area is equal to the area bounded by the tangential caustics, which could be estimated with magnification and deflection maps produced by \texttt{Lenstool}. The area increases rapidly with redshift. The interplay between increasing lensed area, changing angular diameter distance and the decline in the number of detected sources with redshift results in a distribution that peaks at the redshift at which we expect to find the largest number of strongly lensed systems.
 
In Fig. \ref{fig:hist_area}, we show that in MACS0416, we expect to find most multiply lensed sources around redshift 2. At this redshift, \OII, \OIII\ and \Halpha\ emission lines all fall in the range of our F115W, F150W and F200W grism spectra, making NIRISS spectroscopy ideal for finding multiple images up to faint magnitudes (even fainter than magnitude 27 in F200W filter; see Fig. \ref{fig:z_mag}). MUSE spectral range is, on the other hand, less ideal for this redshift range, as it cannot detect any of the above-mentioned emission lines above $z\sim1.5$ and Lyman $\alpha$ emission below $z\sim 3$. Given sufficiently deep observation, such as in this cluster, MUSE is still effective at finding sources based on fainter emission lines (e.g. CIII] at 1907 \r{A}; see the emission line catalogues from \citealt{Richard21}). However, it is evident that NIRISS grism spectroscopy, such as that used in this work, complements even the deepest MUSE observations. 

NIRISS grism spectroscopy also provides emission line maps of bright emission lines (e.g. \Halpha) with high spatial resolution, which can, in combination with photometry, help identify multiply imaged clumps in extended multiple images and giant arcs. High magnifications in the giant arcs make them especially important for probing faint small-scale systems at high redshift. On the other hand, high magnification gradients near the critical curves make magnifications highly uncertain and any additional constraints invaluable. We demonstrated the use of emission line maps with system K8 (Fig. \ref{fig:d43}), where \Halpha\ flux complemented colour information, obtained with broadband NIRCam imaging. It helped us rule out some of the clumps with high \Halpha\ flux in the spiral galaxy as potential counter images of the brightest clump in the giant arc.  We also point out that a significant fraction of systems, including some of the giant arcs (see \citetalias{Pearls23}), are found in the redshift range where at least two of the \OII, \OIII\ and \Halpha\ emission lines fall in our grism range (Fig. \ref{fig:z_mag}). Thus, this method could be extended from using individual emission line maps to using quantities such as emission line ratios, which are independent of magnification (provided they originate from the same source area) and could become a useful tool in future lensing studies.

In addition to NIRISS spectroscopy, this work also demonstrated the value of NIRSpec MSA prism spectroscopy for strong lens modelling. Its wide wavelength coverage enables redshift measurements from multiple emission lines in a wide redshift range (e.g. see the spectra in Fig. \ref{fig:nirspec_canucs_new}). For instance, it allowed us to measure the redshift of system K100, $z_{\rm nirspec}=7.24 \pm 0.02$, which became the highest spectroscopic redshift of a multiple-image system in this cluster. However, unlike NIRISS and MUSE observations, NIRSpec MSA observations target preselected objects, which requires a reliable multiple-image candidate identification based on prior imaging data and photometric redshifts. Furthermore, if other follow-up targets are prioritised based on various science goals, not all multiple images can be observed due to the limited possible configurations of MSA masks. In this work, we obtained only one NIRSpec redshift per multiple-image system (see Table \ref{tab:multsys_canucs}).  Although this is enough for determining the system redshift $z_{\rm sys}$, it is, on its own, insufficient to confirm each multiply imaged candidate. To assess the reliability of our gold multiple-image systems, it was thus crucial to use the NIRSpec data in combination with our imaging data and NIRISS spectroscopy.

\section{Conclusion}
In this work, we have presented an updated strong lensing model of galaxy cluster MACS J0416.1-2403 (MACS0416) that leverages the new JWST imaging and spectroscopic data from CANUCS. We expanded the existing catalogues of multiple images by searching for new multiple images in our NIRCam imaging data. To this end, we first predicted multiple images from sources in CANUCS photometric catalogues and then visually examined the region around each prediction in several HST and JWST/NIRCam RGB combinations. Our search independently discovered most of the already known candidates while also finding 15 previously unknown systems. The catalogue was further augmented by our spectroscopic redshift measurements. We examined the spectra obtained with NIRISS wide-field slitless spectroscopy of all our candidates and observed some of them with the NIRSpec multi-object prism spectroscopy.  We measured spectroscopic redshift for 17 multiple-image systems with previously unknown spectroscopic redshifts. We furthermore corrected the redshift of two known systems and confirmed two new counter images. We also re-evaluated known sources with MUSE redshift (e.g. \citetalias{Bergamini23}, \citealt{Richard21}, \citealt{bergamini21},\citealt{Caminha17}) and candidates found by previous studies that used NIRCam imaging data (\citetalias{Pearls23}). Our complete catalogue of all candidates comprises 415 multiple images from 150 systems, 124 of which are known to have spectroscopic redshift. Of these, we selected 303 reliable multiple images from 111 systems with spectroscopic redshift that we used for constraining the lens model (i.e. the gold catalogue). This makes our catalogue the largest catalogue of spectroscopic multiple images in a galaxy cluster field to date. 

Our lens model was constrained using the parametric lens modelling code \texttt{Lenstool} with the parameterisation presented in \citet{bergamini21} and \citetalias{Bergamini23}. The new model is able to reproduce the multiple-image position with \Deltarms{52} and is in good agreement with the \citetalias{Bergamini23} lens model, which uses the same \texttt{Lenstool} parameterisation and pre-JWST data. The differences between our and the \citetalias{Bergamini23} lens model are larger in poorly constrained southern regions of the cluster. The median relative $\kappa$ difference between our models is $1.6 \%$, and the radial profiles of the total mass in a circular aperture agree within a few percent. We find larger disagreements between our and the \citetalias{Bergamini23} lens model in the mass profiles of the galaxy sub-halo mass component, resulting from different scaling relation normalisation parameter values. This is indicative of a known degeneracy between cluster members and the smooth mass components despite well-constrained total mass distribution \citep[e.g. ][]{2016A&A...588A..99L}. We also compared magnifications of multiple images and found that the relative $\mu$ differences given by our model differ from those obtained with the \citetalias{Bergamini23} model by 10\% for images with our $\mu<10$ (and 30 \% if $\mu>10$). The differences are larger when compared with \citetalias{Pearls23}, which uses a different lens modelling method (as well as a different dataset), amounting to $41\%$ for $\mu<10$ ($85\%$ for $\mu>10$). Large $\mu$ differences between different modelling techniques are consistent with the findings from pre-JWST data \citep{2017MNRAS.465.1030P,2020MNRAS.494.4771R}.

In this work, we also demonstrated the utility of NIRISS wide-field slitless spectroscopy in F115W, F150W, and F200W filters for strong lensing studies. We show that such spectroscopy is the most effective at measuring redshifts between redshifts 2 and 3, where at least two of the three significant emission lines (\Halpha, \OIIIdoublet\ and \OIIlambda) can be detected. With a simple estimate, considering the redshift distribution of galaxies in the field and the multiply lensed area as a function of redshift, we showed that this redshift range is particularly useful for strong lensing studies, as it contains the most multiple images. With system K8, we further demonstrated how spatially resolved NIRISS spectroscopy and emission line maps can aid in understanding of the extended multiple images and giant arcs. Since MACS0416 has been observed by some of the deepest HST imaging and MUSE spectroscopy programs, our study thus serves as a comparison between the results attainable with JWST and results currently attainable with the best ground- or space-based facilities using the largest known dataset in a single cluster.

\section{Data availability}

We provide our catalogue of multiple images used to constrain the model with their magnifications and the full catalogue of multiple image candidates on CANUCS website (\url{https://niriss.github.io/}) together with the lens model presented in this work and a tool for computing magnifications of arbitrary sources. Following the first CANUCS data release, the lens model will also be available on STScI Mikulski Archive for Space Telescopes (MAST) with DOI \texttt{10.17909/ph4n-6n76}.

\begin{acknowledgements}
We thank the anonymous referee whose comments greatly enhanced the quality of this paper. GR, MB, AH, VM and RT acknowledge support from the ERC Grant FIRSTLIGHT and from the Slovenian national research agency ARRS through grants N1-0238, P1-0188. MB acknowledges support
from the program HST-GO-16667, provided through a grant from the STScI under NASA contract NAS5-26555. This research was enabled by grant 18JWST-GTO1 from the Canadian Space Agency and funding from the Natural Sciences and Engineering Research Council of Canada. This research used the Canadian Advanced Network For Astronomy Research (CANFAR) operated in partnership by the Canadian Astronomy Data Centre and The Digital Research Alliance of Canada with support from the National Research Council of Canada the Canadian Space Agency, CANARIE and the Canadian Foundation for Innovation. 
\end{acknowledgements}

\bibliographystyle{aa} 
\bibliography{references}{}

\appendix

\onecolumn

\section{Grism spectra}
\label{Appendix:grism spectra}

\begin{figure*}
\centering
\includegraphics[width=\linewidth]{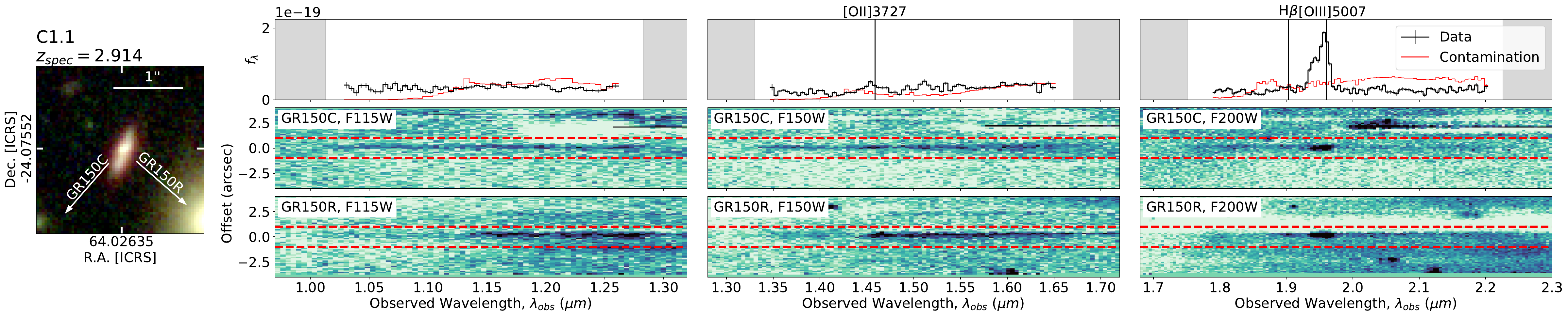}\\ 
\includegraphics[width=\linewidth]{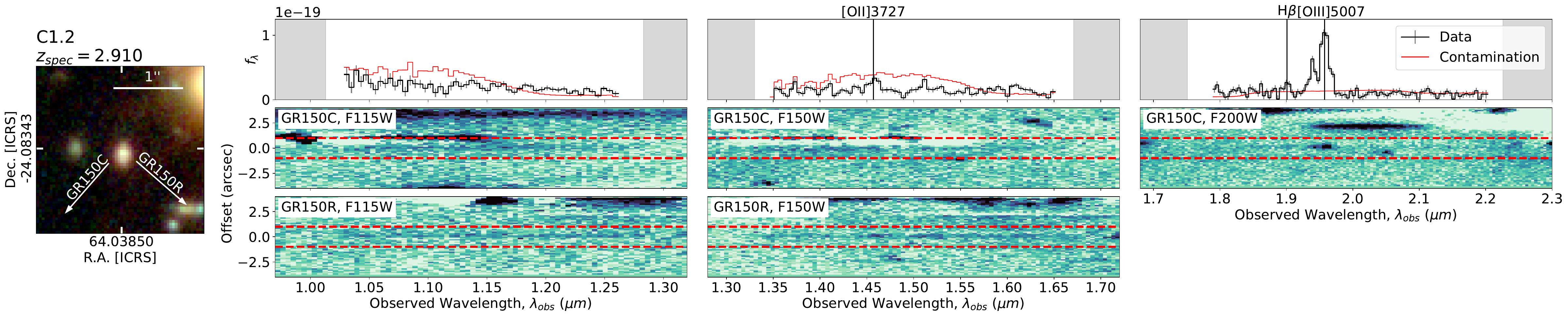}\\ 
\includegraphics[width=\linewidth]{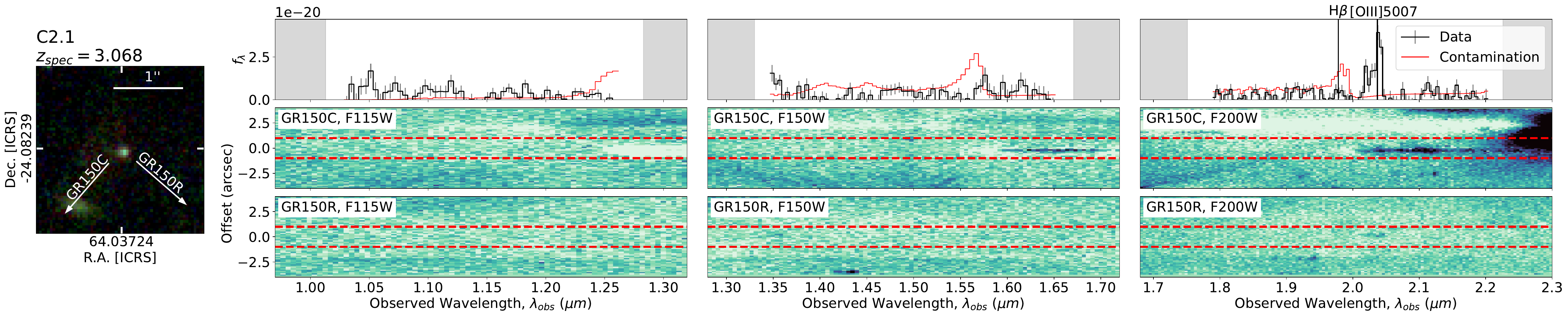}\\ 
\includegraphics[width=\linewidth]{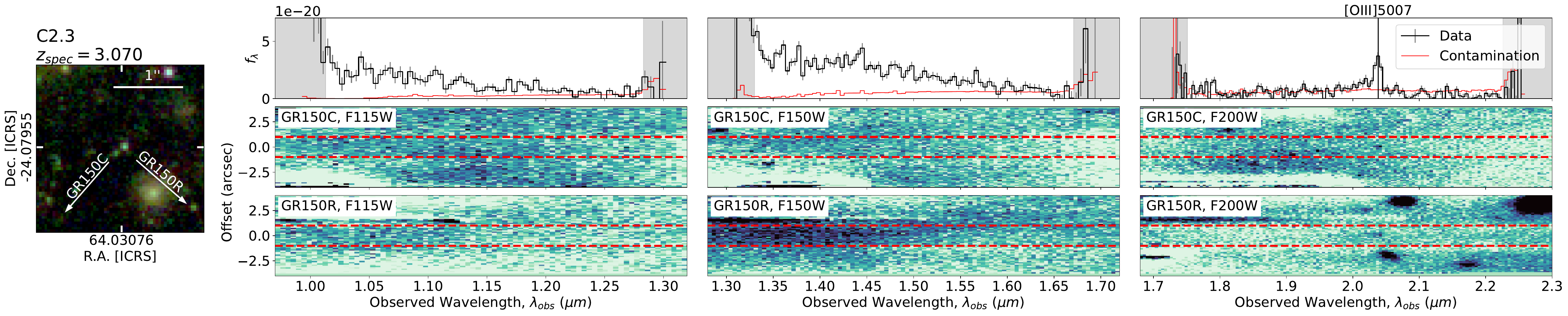}\\ 

\includegraphics[width=\linewidth]{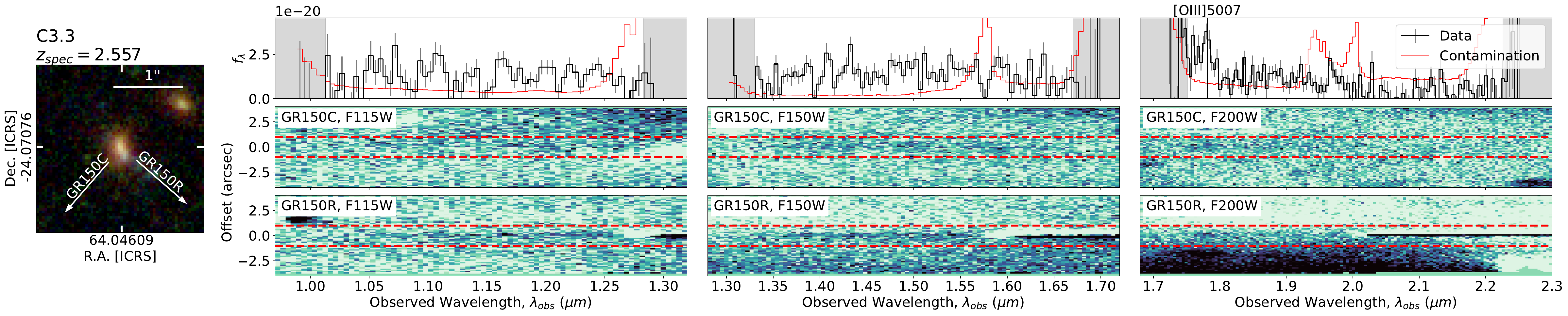}\\ 
\includegraphics[width=0.7\linewidth,left]{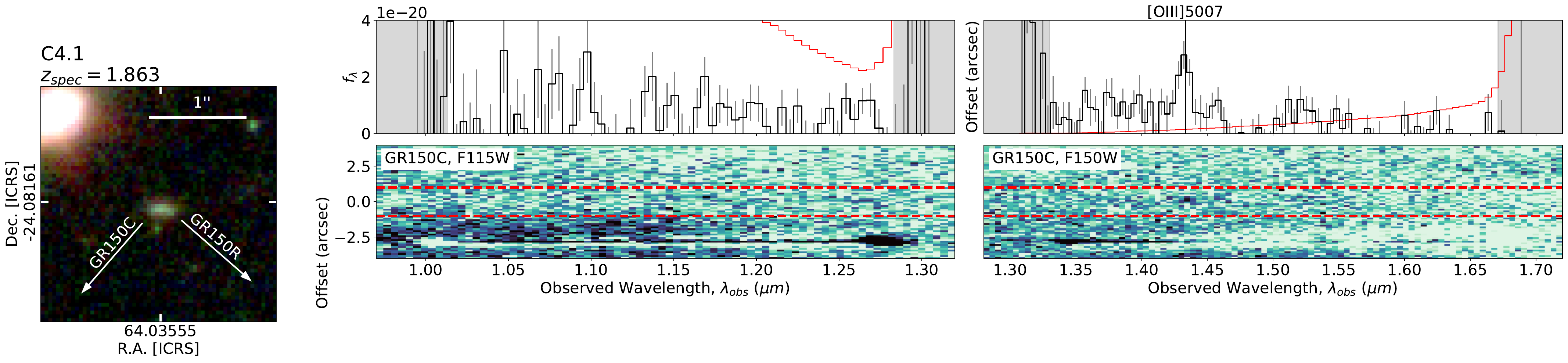}\\ 
\caption{NIRISS F115W, F150W and F200W grism spectra of the newly discovered (CANUCS) multiple images. The RGB cutouts are composed of JWST/NIRCam and HST/WFC3 images (F277W, F356W, F410M, F444W in red, F115W, F150W, F200W in green and F814W, F606W, F435W and F090W in blue) with indicated angular scale and the GR150R and GR150C grism dispersion directions. The upper row represents 1D spectra in different filters, showing only the spectral range used for redshift fitting (see Sect. \ref{sec:grism}). In the plots, we indicate contamination and wavelengths of visible emission lines. The grey-shaded region represents the spectral range where the filter transmission falls below one-half of the maximal value. The second and the third rows include 2D spectra in each filter, and for each orientation that was used to obtain redshift - highly contaminated spectra are not displayed.}
\label{fig:nirspec_canucs0}
\end{figure*}

In Figs. \ref{fig:nirspec_canucs0}, \ref{fig:nirspec_canucs1}, \ref{fig:nirspec_canucs2}, \ref{fig:nirspec_canucs3} and \ref{fig:nirspec_canucs4} we show NIRISS grism spectra used to obtain spectroscopic redshifts of multiple images in this work. All fits, apart from K87.2, used only the grism spectrum. For K87.2, only the combined fit of the grism spectrum and photometry returned the correct redshift in agreement with K87.1 and K87.3. We note that its spectrum contains clear \OIII\ and \OII\ emission in agreement with other images of the system. Images K95.1, K95.3, K90.1, K90.3 and C2.3 were re-fitted with a narrower redshift prior ($\pm 1$ or $\pm 0.5$) where the redshift range was estimated based on other images of the system with more reliable fits (grism or NIRSpec prism) and corroborated by other arguments (several visually discernible emission lines, photometric redshifts etc.). We note that the grism redshifts of those images agree with other multiple images of the system within the spectroscopic redshift uncertainty, which is much narrower than the chosen prior range. We also note that the spectrum of K82.1 in GR150C grism orientation is aligned with the direction of the arc, resulting in a collapsed 1D spectrum without visible emission lines. However, the spectrum in GR150R orientation shows several emission lines, and the combined fit of a narrow spectral range returns a reliable redshift of $2.03\pm0.02$.

\begin{figure*}
\centering
\includegraphics[width=\linewidth]{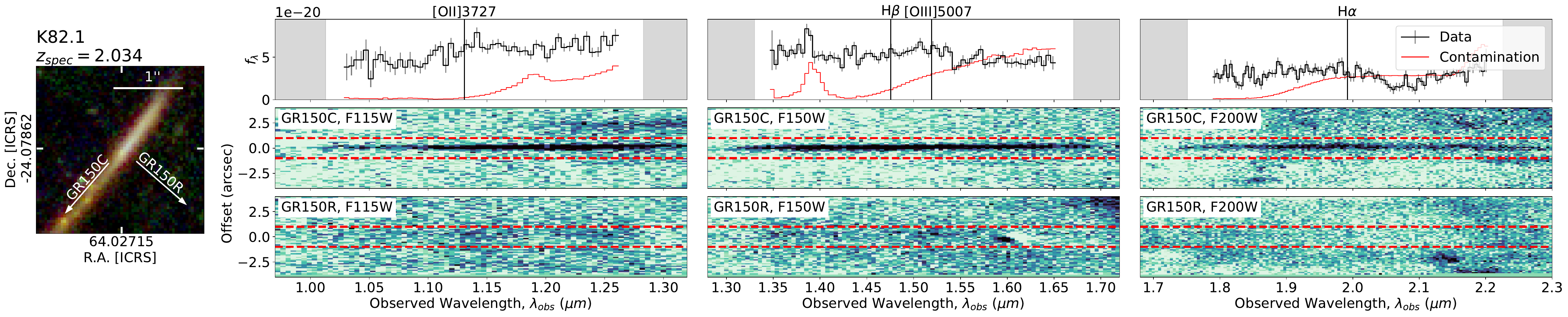}\\ 
\includegraphics[width=\linewidth]{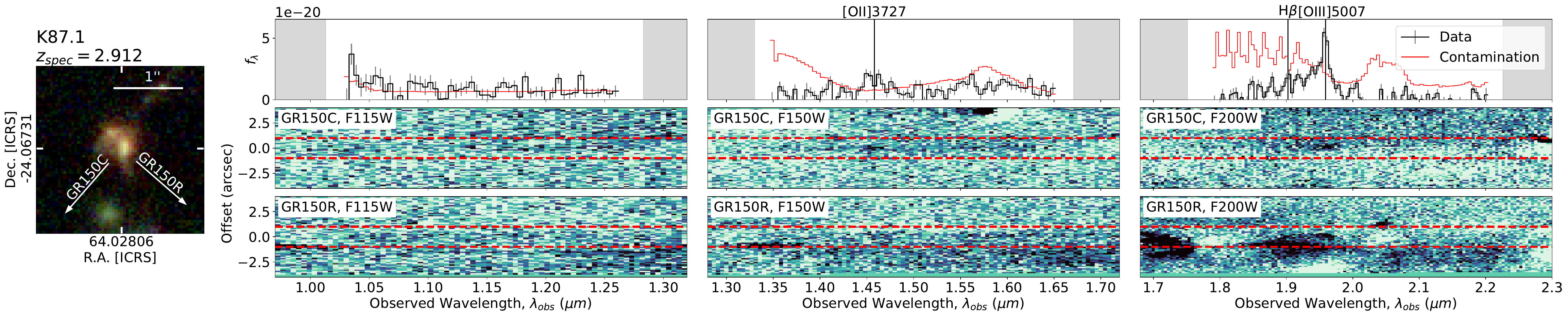}\\ 
\includegraphics[width=\linewidth]{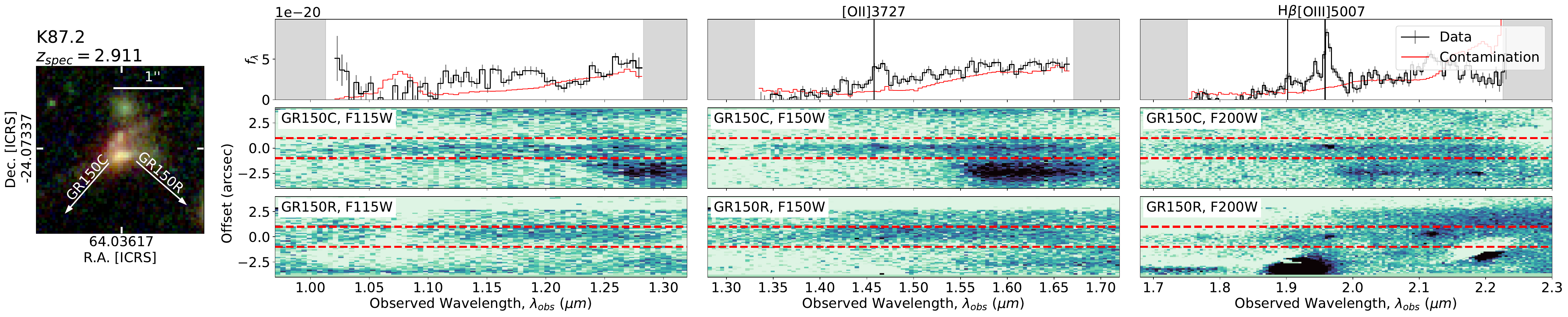}\\ 
\includegraphics[width=\linewidth]{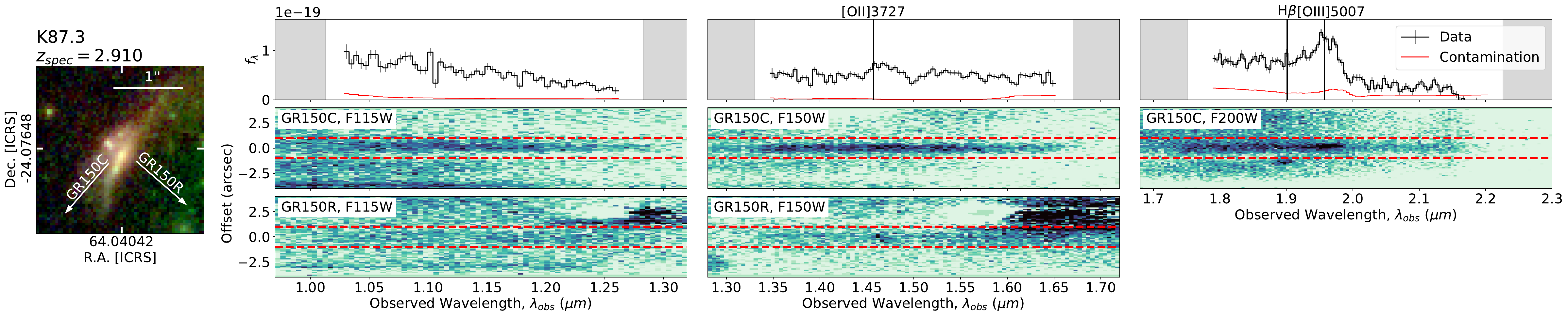}\\ 
\includegraphics[width=\linewidth]{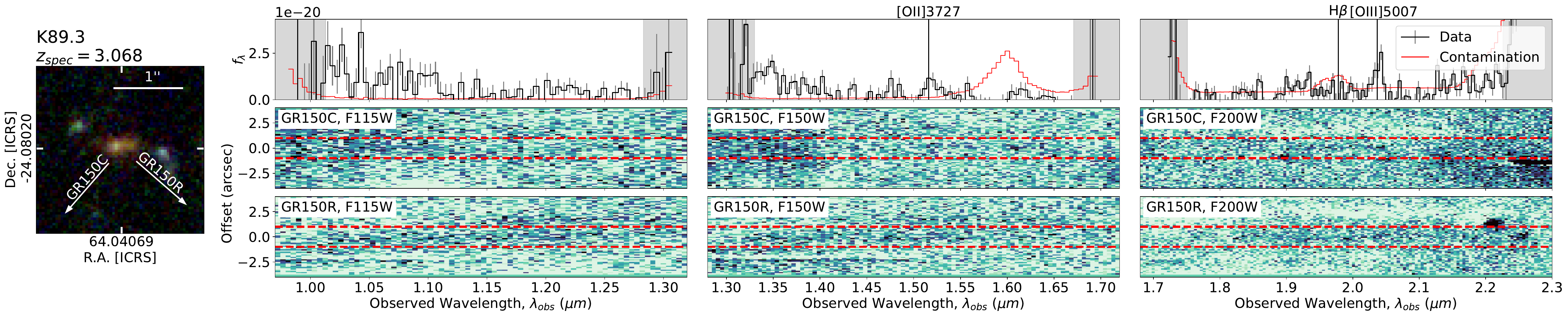}\\ 
\caption{NIRISS spectra of known multiple image candidates with previously unknown system redshift. The figure is analogous to Fig. \ref{fig:nirspec_canucs0}. }
\label{fig:nirspec_canucs1}
\end{figure*}

\begin{figure*}
\ContinuedFloat
\centering

\includegraphics[width=\linewidth]{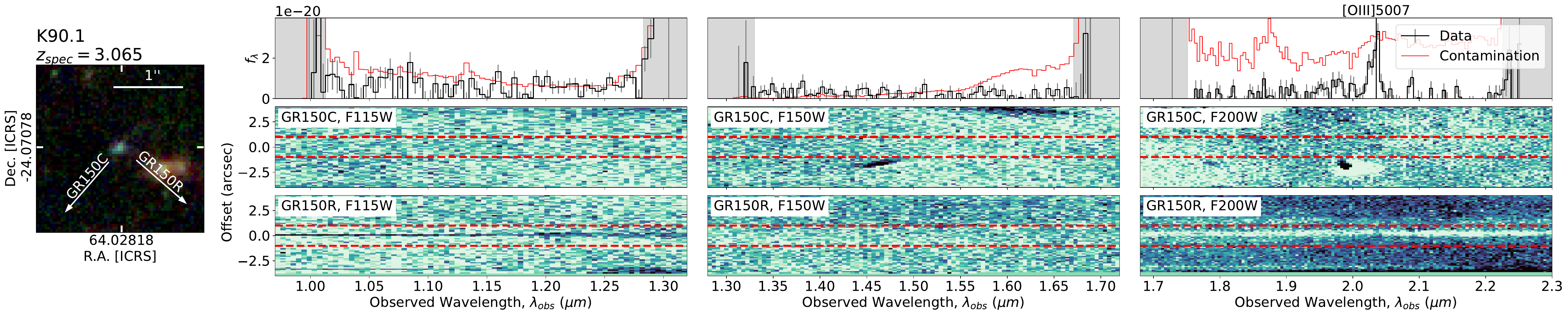}\\ 
\includegraphics[width=\linewidth]{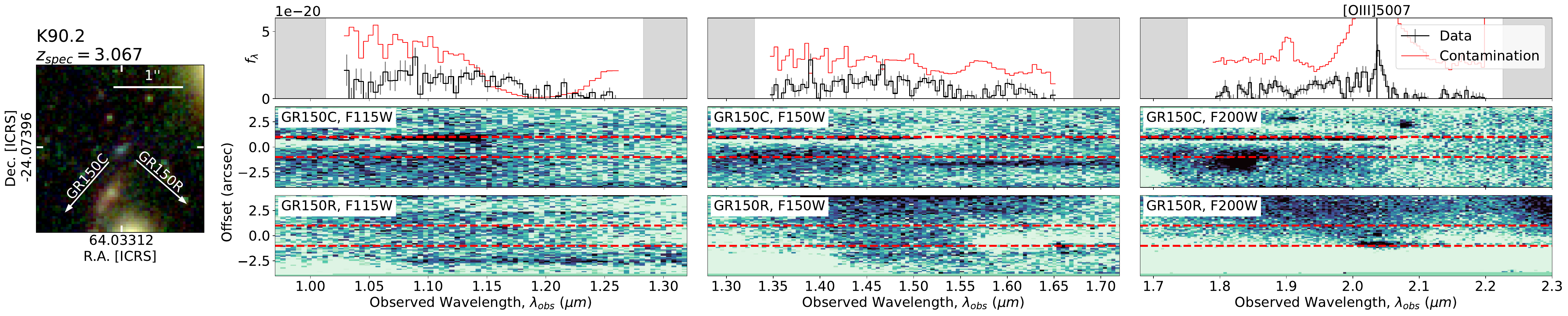}\\ 
\includegraphics[width=\linewidth]{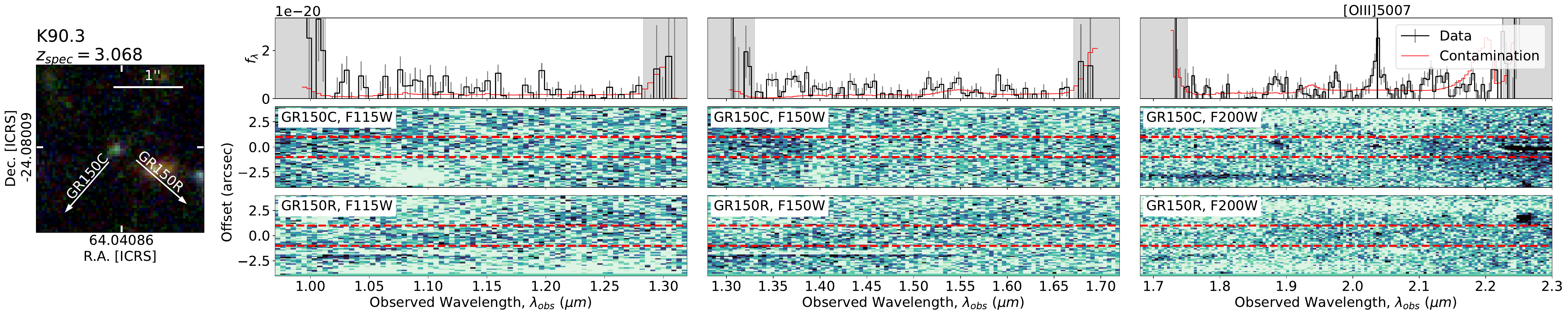}\\ 
\includegraphics[width=\linewidth]{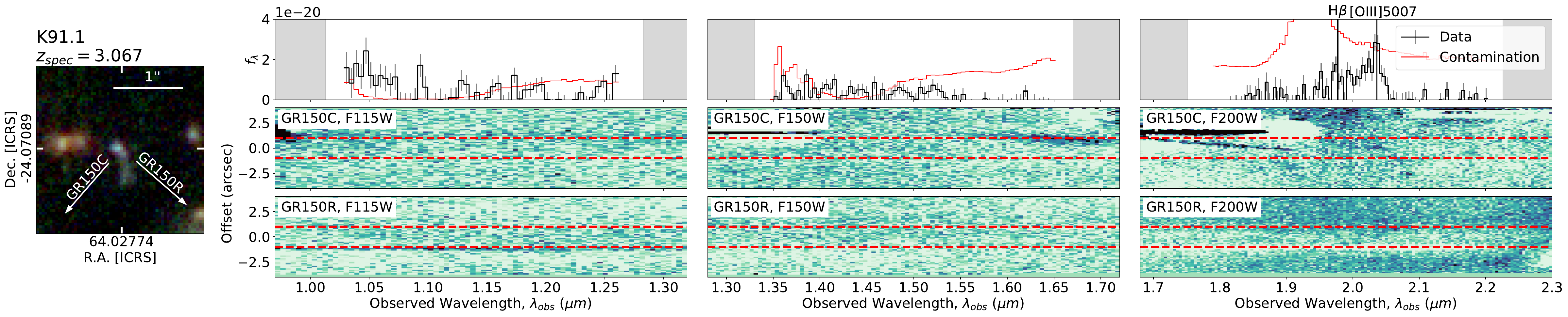}\\ 
\includegraphics[width=\linewidth]{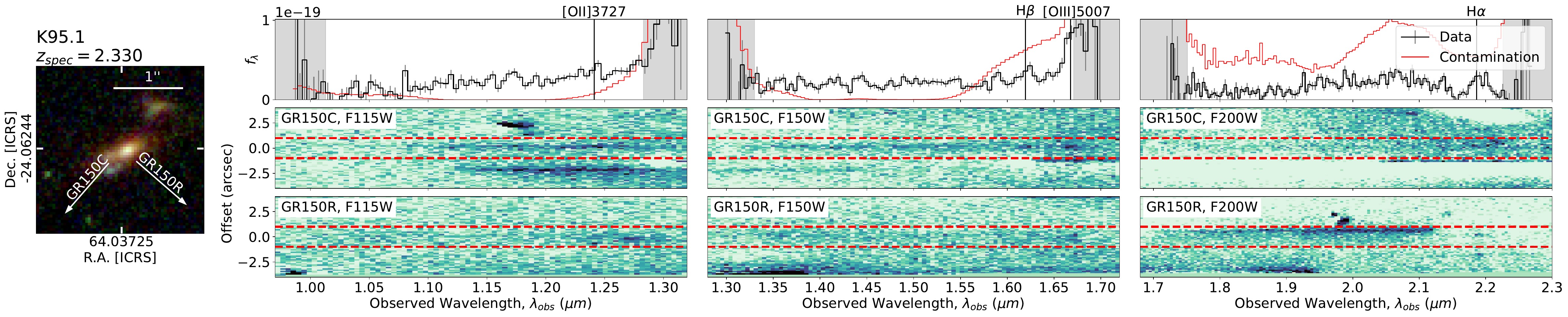}\\ 
\caption{contiuned}
\label{fig:nirspec_canucs2}
\end{figure*}

\begin{figure*}
\ContinuedFloat
\centering
\includegraphics[width=0.53\linewidth,left]{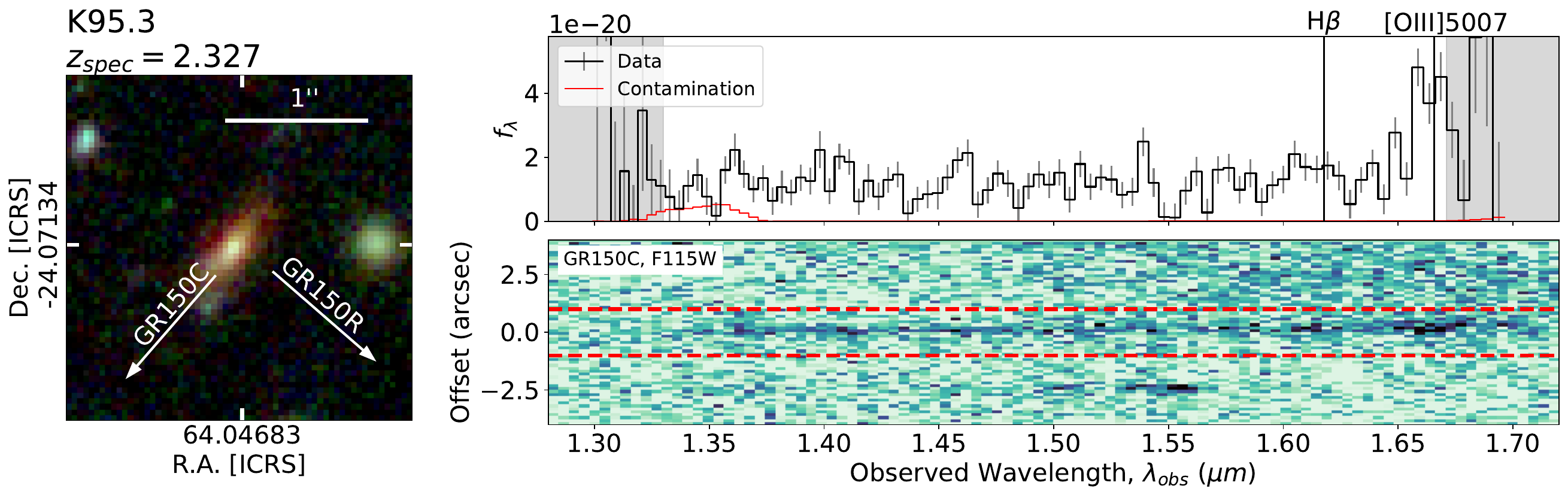}\\ 
\includegraphics[width=\linewidth]{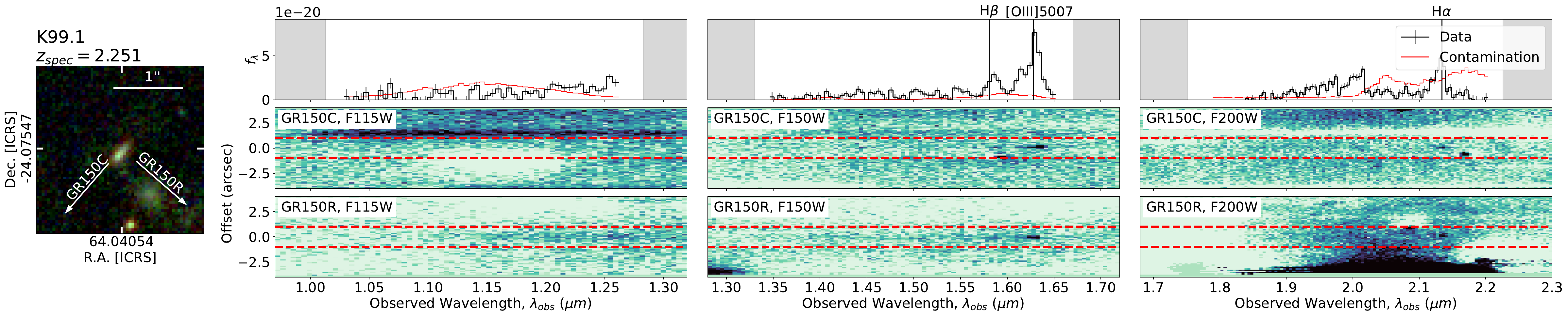}\\ 
\includegraphics[width=\linewidth]{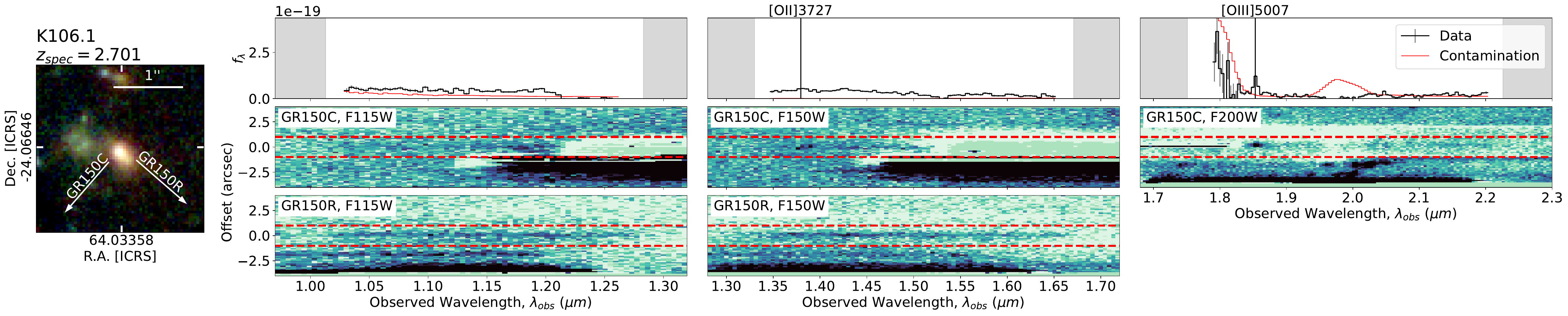}\\ 
\includegraphics[width=\linewidth]{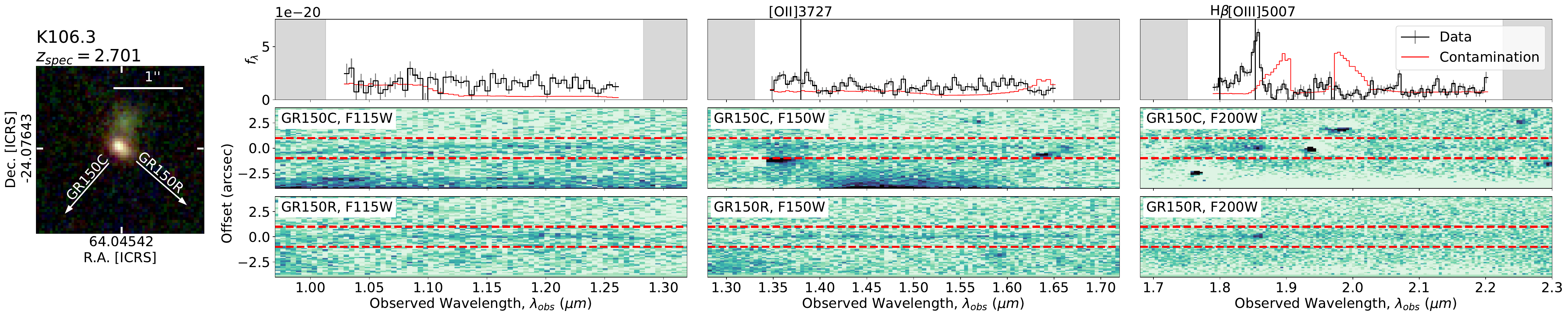}\\ 
\includegraphics[width=\linewidth]{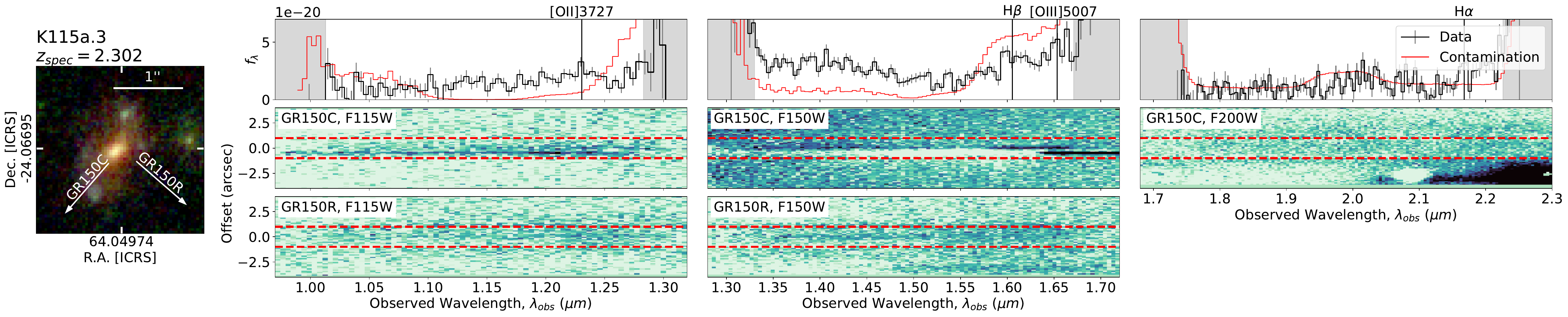}\\ 
\caption{contiuned}
\label{fig:nirspec_canucs3}
\end{figure*}

\begin{figure*}
\centering
\includegraphics[width=\linewidth]{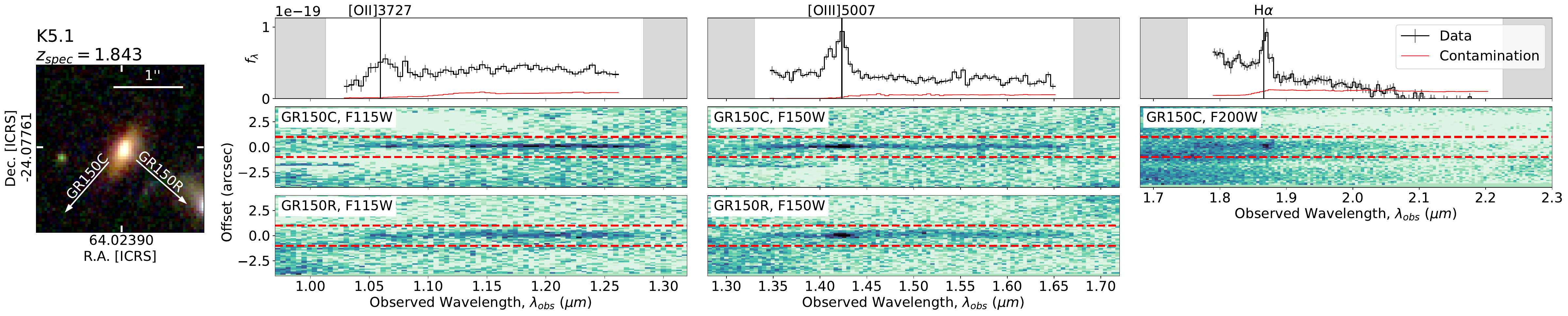}\\ 
\includegraphics[width=\linewidth]{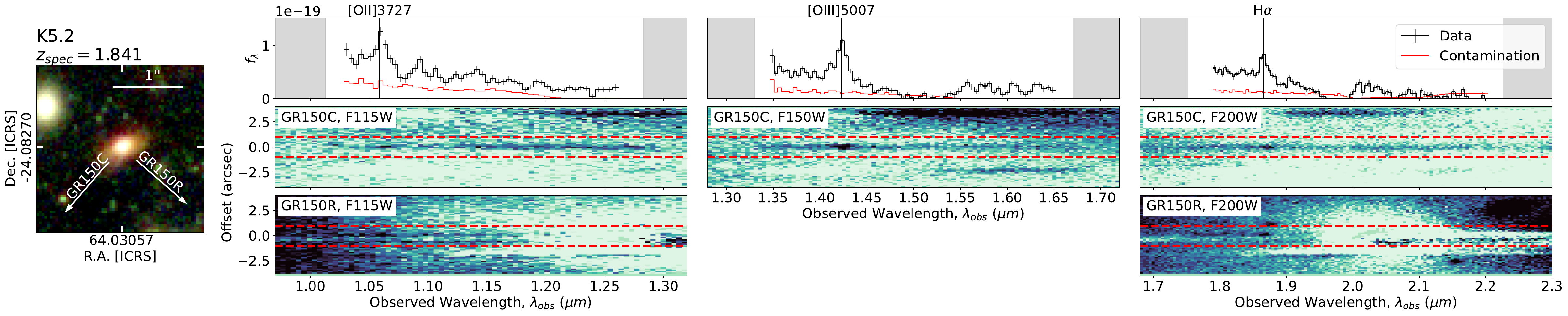}\\ 
\includegraphics[width=\linewidth]{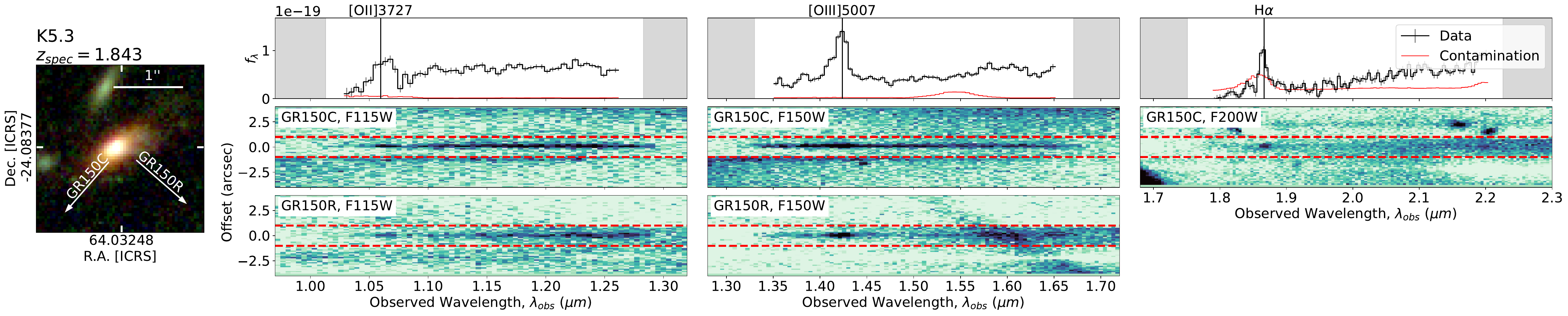}\\
\includegraphics[width=\linewidth]{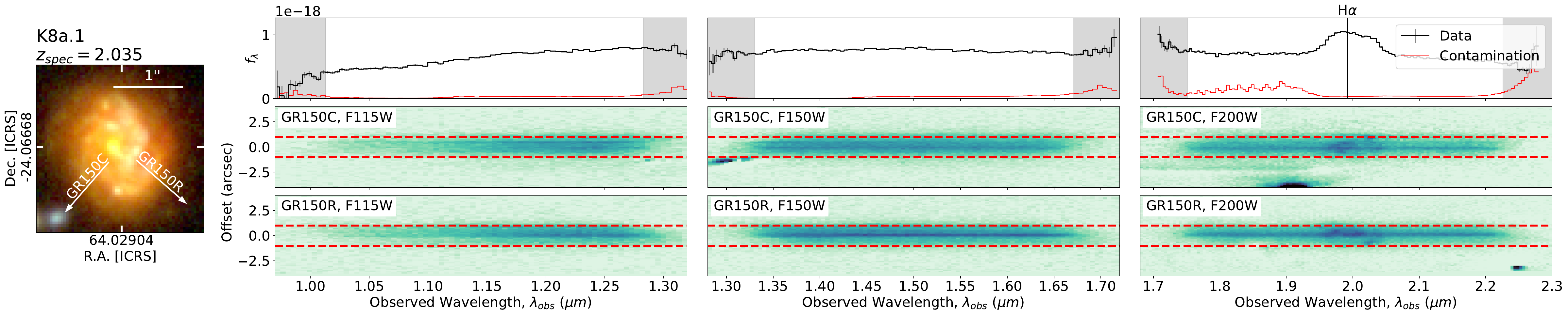}\\ 
\includegraphics[width=\linewidth]{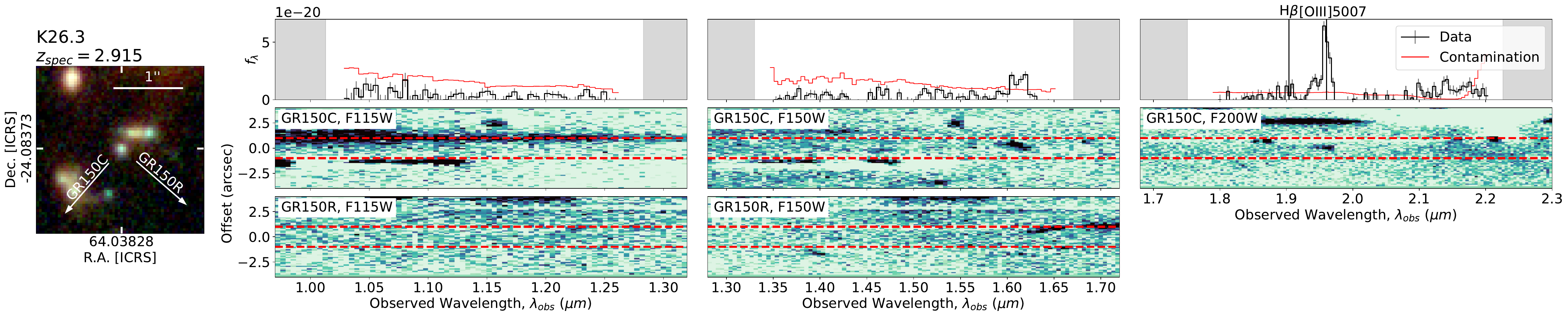}\\ 

\includegraphics[width=\linewidth]{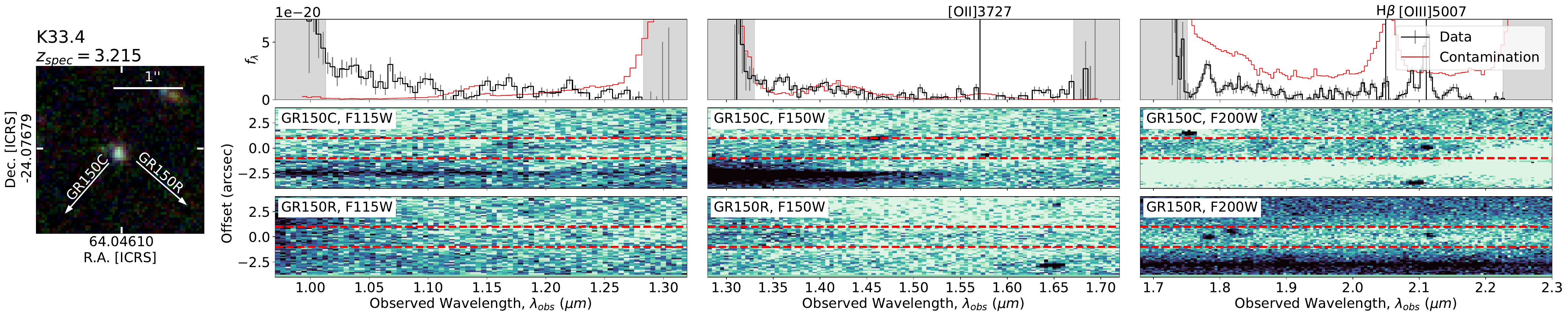}\\ 
\caption{NIRISS spectra, not included in previous figures but also used in this work, either for updating the system redshift (K5 and K8a.1), confirming a new multiple image candidate (K26.3) or showing the spectral line found in the spectra of counter images (K33.4). The figure is analogous to Figs. \ref{fig:nirspec_canucs0} and \ref{fig:nirspec_canucs1}.}
\label{fig:nirspec_canucs4}
\end{figure*}



\twocolumn

\begin{figure}
\includegraphics[width=0.26\linewidth]{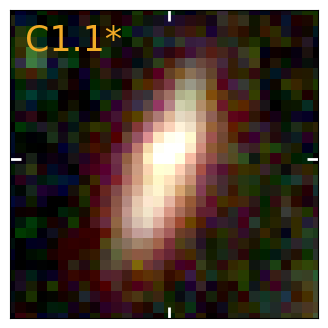}
 \includegraphics[width=0.26\linewidth]{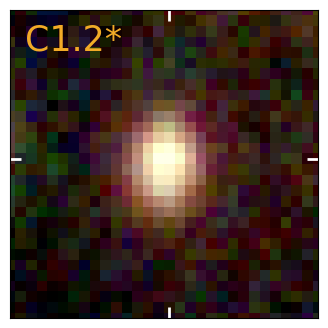}\\
 \includegraphics[width=0.26\linewidth]{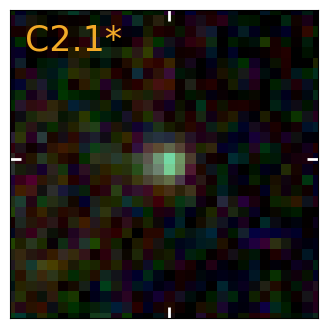}
 \includegraphics[width=0.26\linewidth]{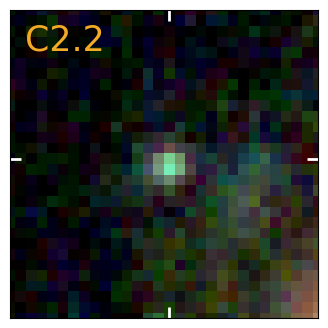}
  \includegraphics[width=0.26\linewidth]{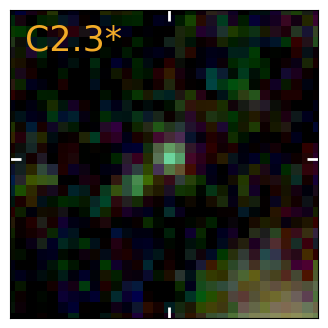}\\
\includegraphics[width=0.26\linewidth]{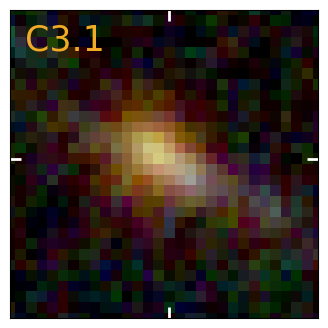}
\includegraphics[width=0.26\linewidth]{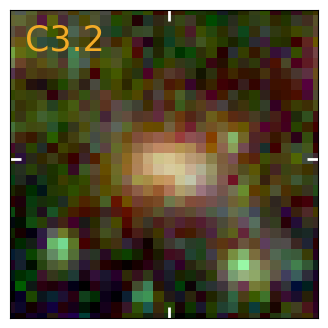}
\includegraphics[width=0.26\linewidth]{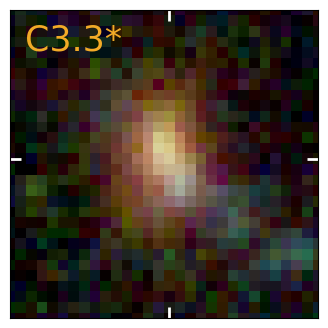}\\
\includegraphics[width=0.26\linewidth]{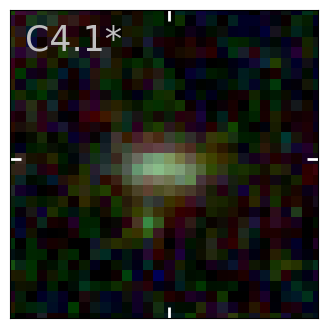}
\includegraphics[width=0.26\linewidth]{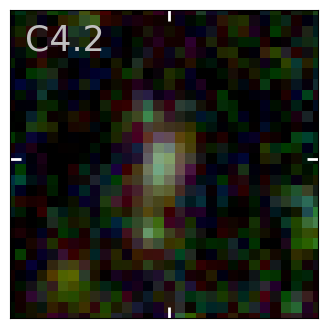}
\includegraphics[width=0.26\linewidth]{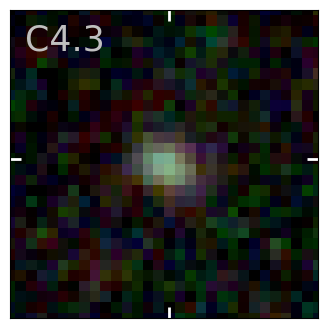}\\
\includegraphics[width=0.26\linewidth]{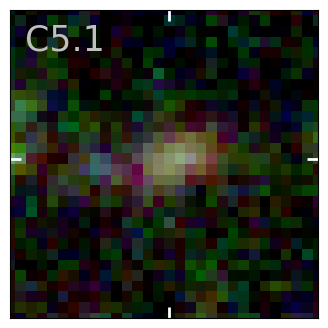}
\includegraphics[width=0.26\linewidth]{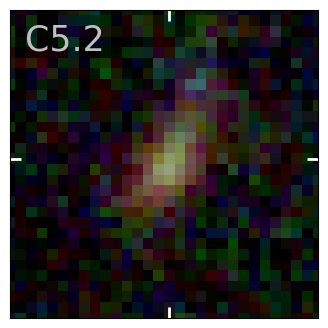}
\includegraphics[width=0.26\linewidth]{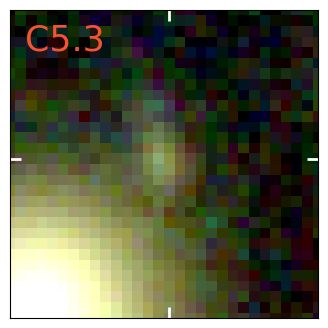}\\
\includegraphics[width=0.26\linewidth]{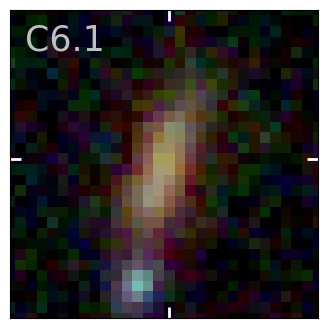}
\includegraphics[width=0.26\linewidth]{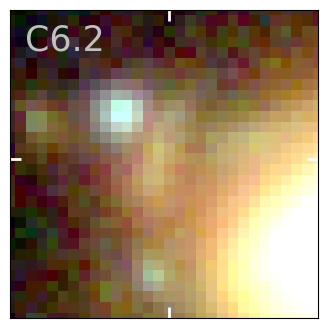}\\
\includegraphics[width=0.26\linewidth]{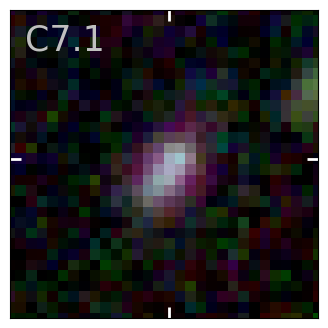}
\includegraphics[width=0.26\linewidth]{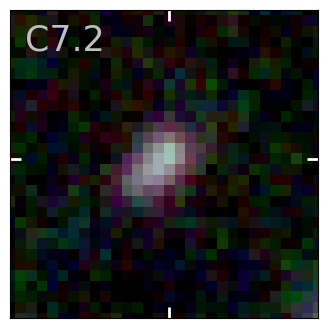}
\includegraphics[width=0.26\linewidth]{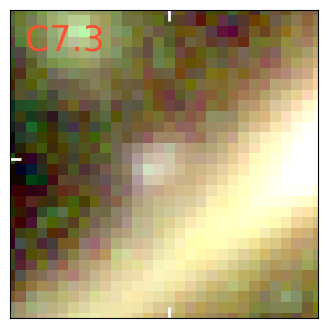}

    \caption{JWST/NIRCam and HST/WCS  (F277W, F356W, F410M, F444W in red, F115W, F150W, F200W in green and F814W, F606W, F435W, F090W in blue) cutouts spanning  $1.2''$ showing multiple images of new CANUCS systems, not included in the \citetalias{Pearls23} catalogue. Multiple images, marked with asterisk ($*$) are spectroscopically confirmed and are also shown in Fig. \ref{fig:nirspec_canucs0} with their grism spectra. The colour of the labels indicates the category (gold, silver, or bronze, depending on their reliability).    }
    \label{fig:newcandidatescutouts}
\end{figure}

\begin{figure}
    \ContinuedFloat
    
\includegraphics[width=0.26\linewidth]{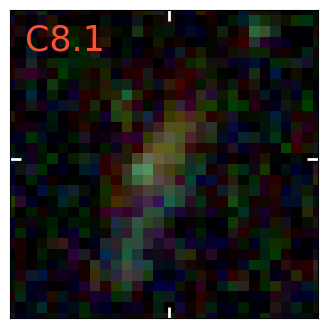}
\includegraphics[width=0.26\linewidth]{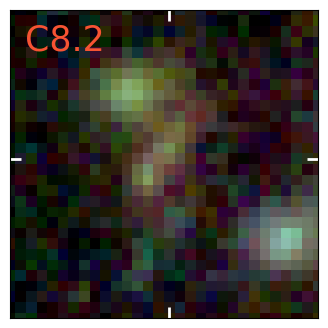}\\
\includegraphics[width=0.26\linewidth]{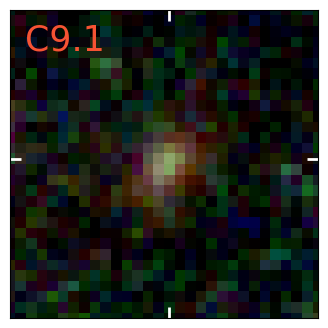}
\includegraphics[width=0.26\linewidth]
{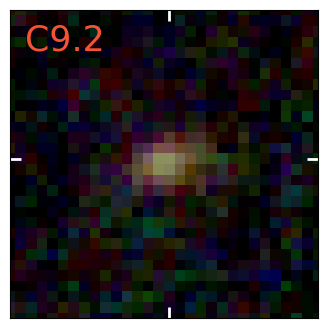}
\includegraphics[width=0.26\linewidth]{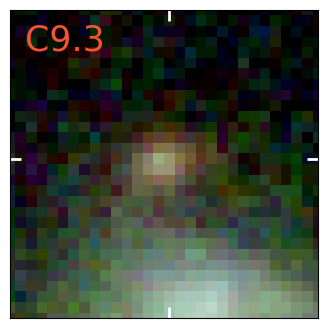}\\
\includegraphics[width=0.26\linewidth]{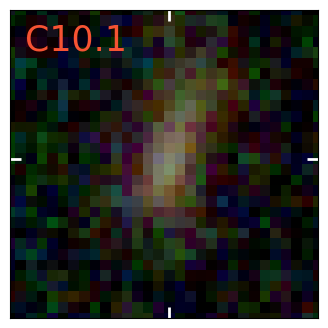}
\includegraphics[width=0.26\linewidth]{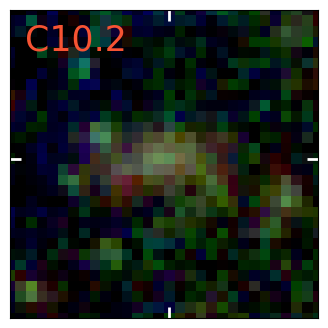}
\includegraphics[width=0.26\linewidth]{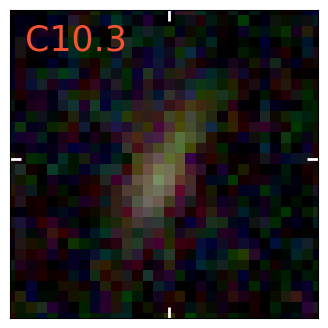}\\\includegraphics[width=0.26\linewidth]{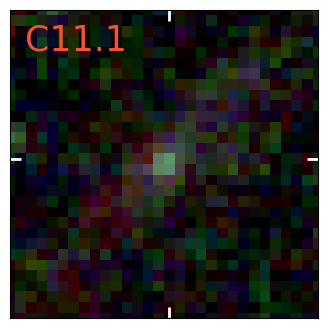}
\includegraphics[width=0.26\linewidth]{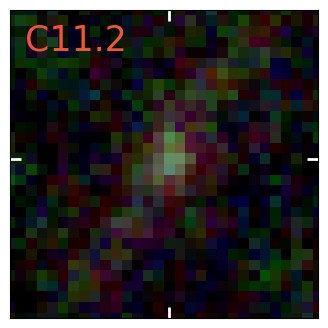}\\
\includegraphics[width=0.26\linewidth]{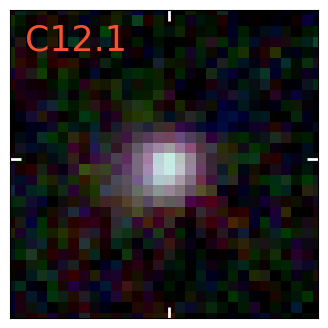}
\includegraphics[width=0.26\linewidth]{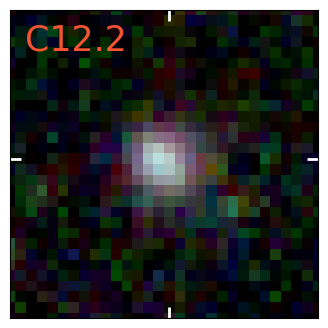}\\
\includegraphics[width=0.26\linewidth]{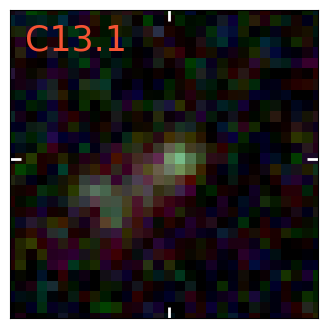}
\includegraphics[width=0.26\linewidth]{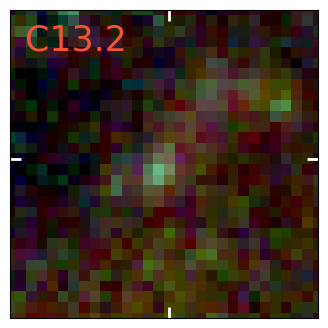}\\
\includegraphics[width=0.26\linewidth]{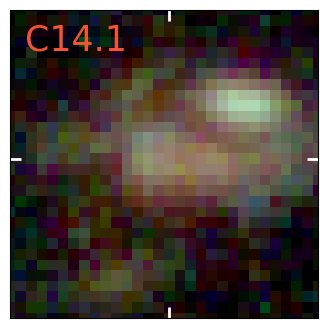}
\includegraphics[width=0.26\linewidth]{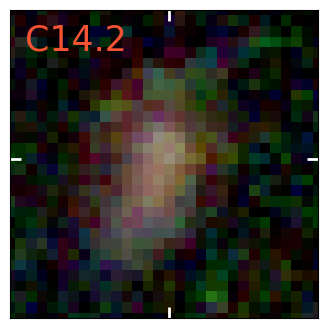}
\includegraphics[width=0.26\linewidth]{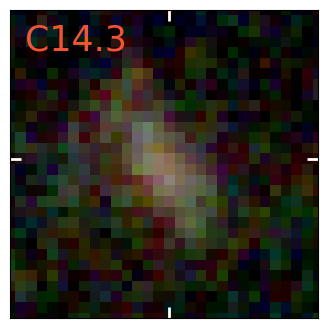}\\
\includegraphics[width=0.26\linewidth]{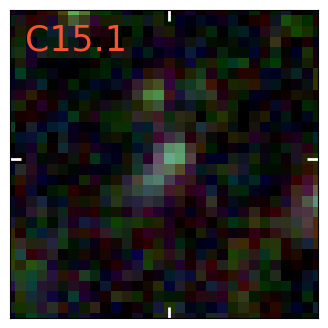}
\includegraphics[width=0.26\linewidth]{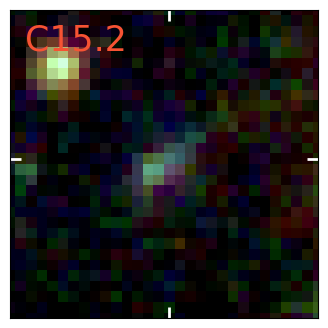}
    \caption{contiuned}
    \label{fig:enter-label}
\end{figure}

\onecolumn

\section{NIRSpec spectra}
\label{Appendix:NIRSpec spectra}

\begin{figure}
\centering
\includegraphics[width=\linewidth]{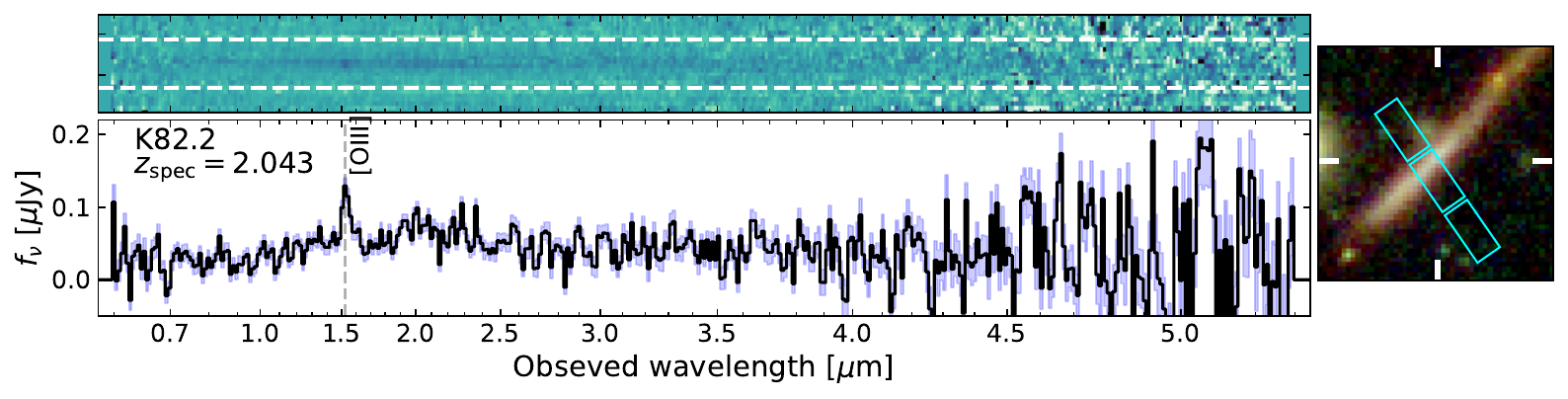}\\ 
\includegraphics[width=\linewidth]{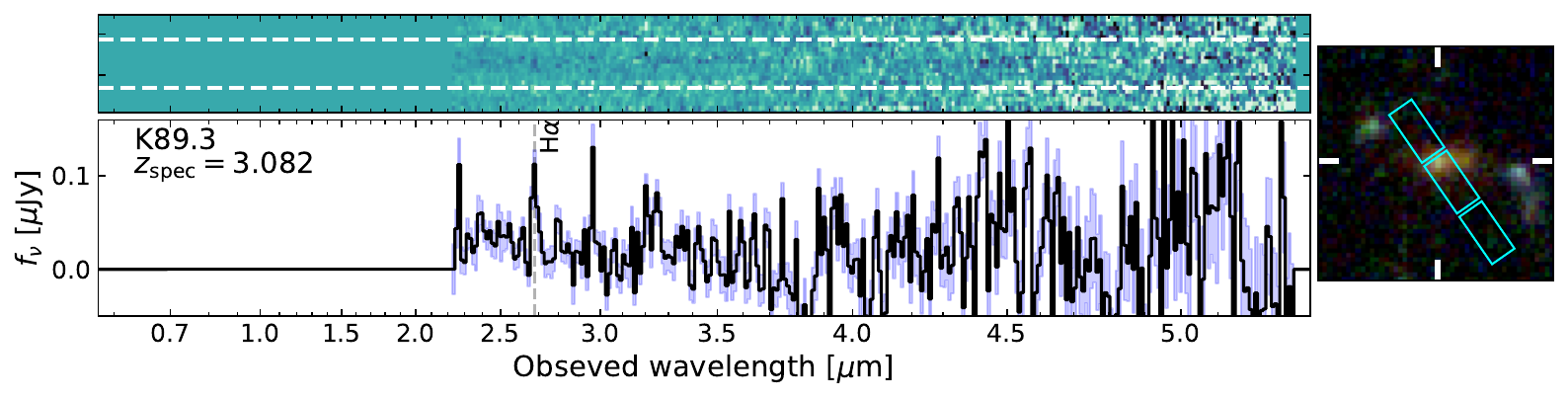}\\ 
\includegraphics[width=\linewidth]{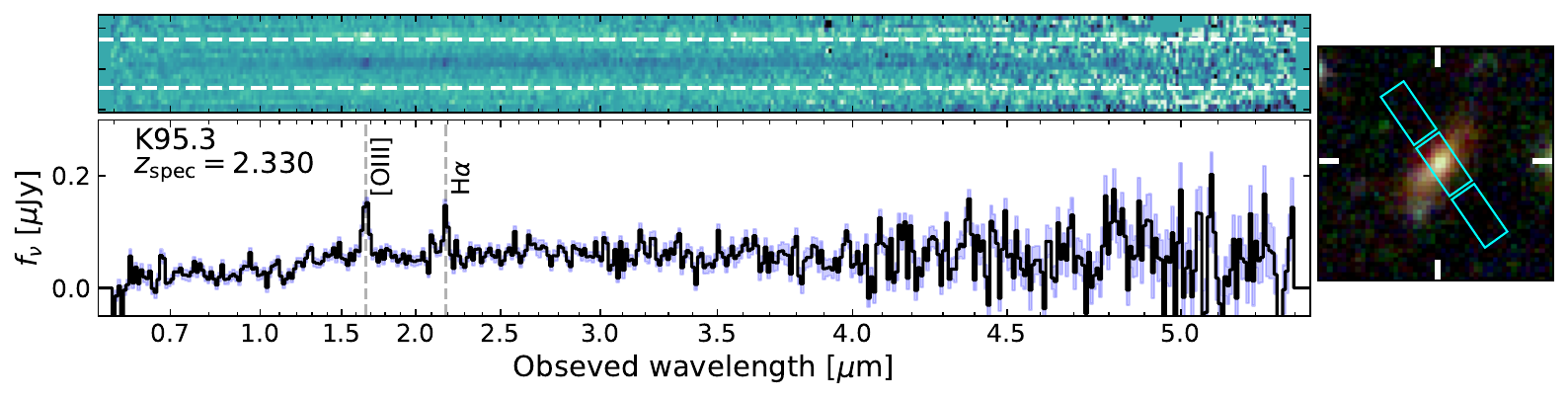}\\ 
\includegraphics[width=\linewidth]{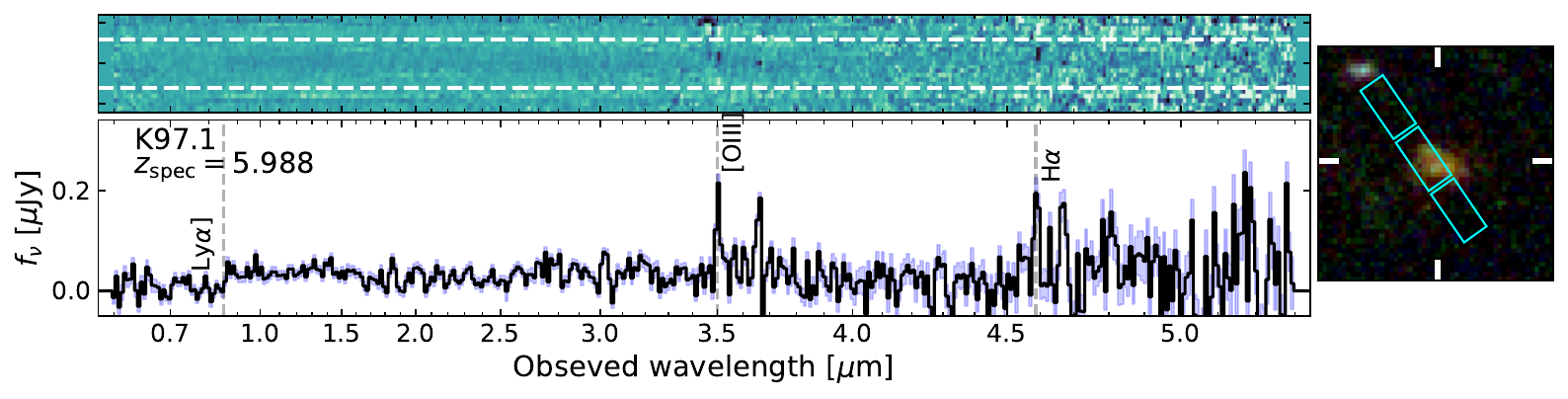}\\ 
\includegraphics[width=\linewidth]
{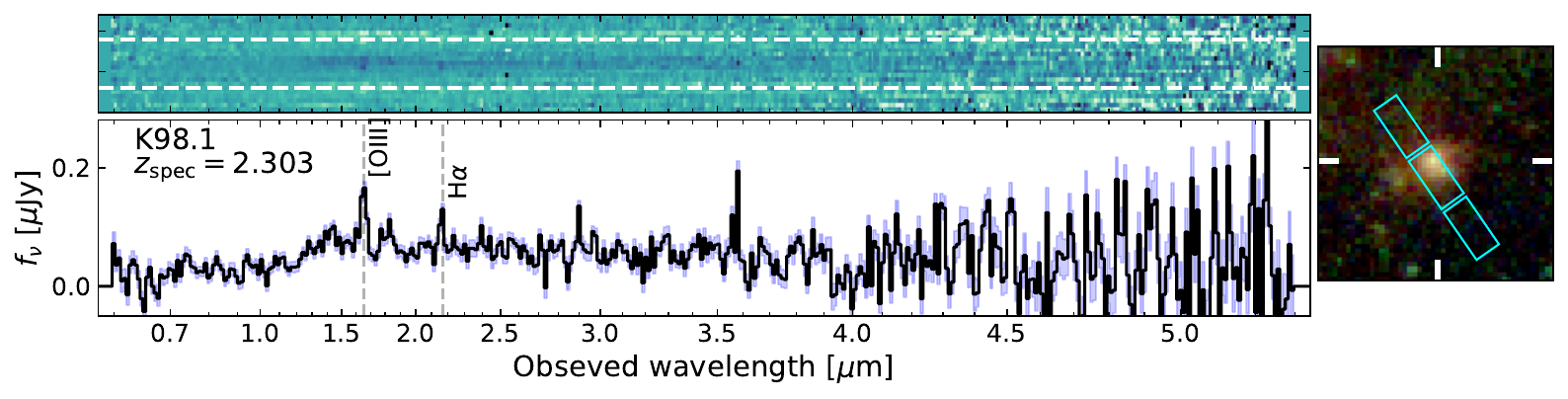}\\ 
\includegraphics[width=\linewidth]{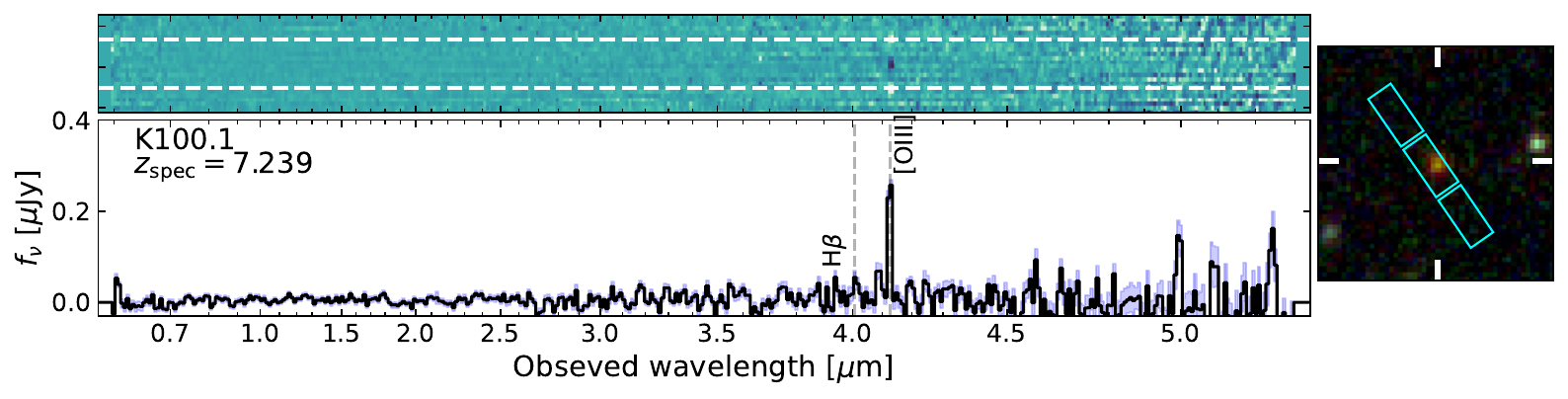}\\ 
\includegraphics[width=\linewidth]{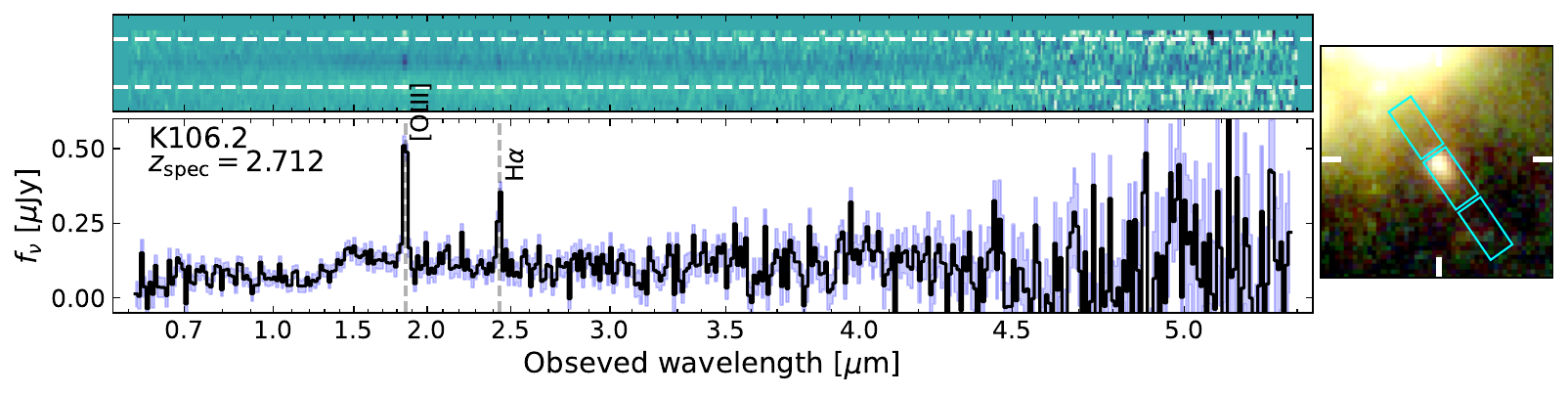}\\ 
\includegraphics[width=\linewidth]{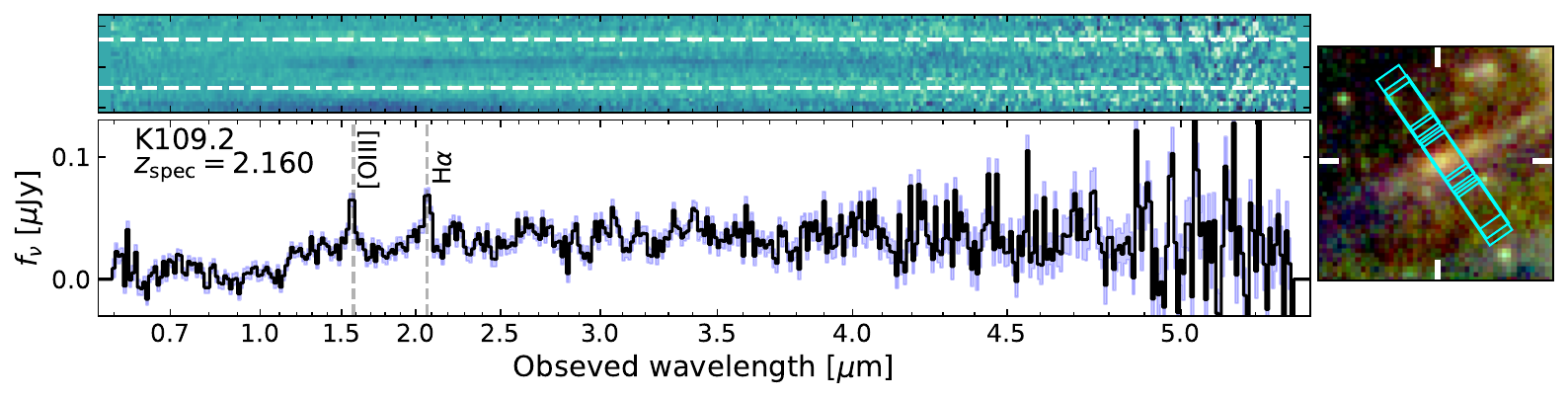}\\ 
\includegraphics[width=\linewidth]{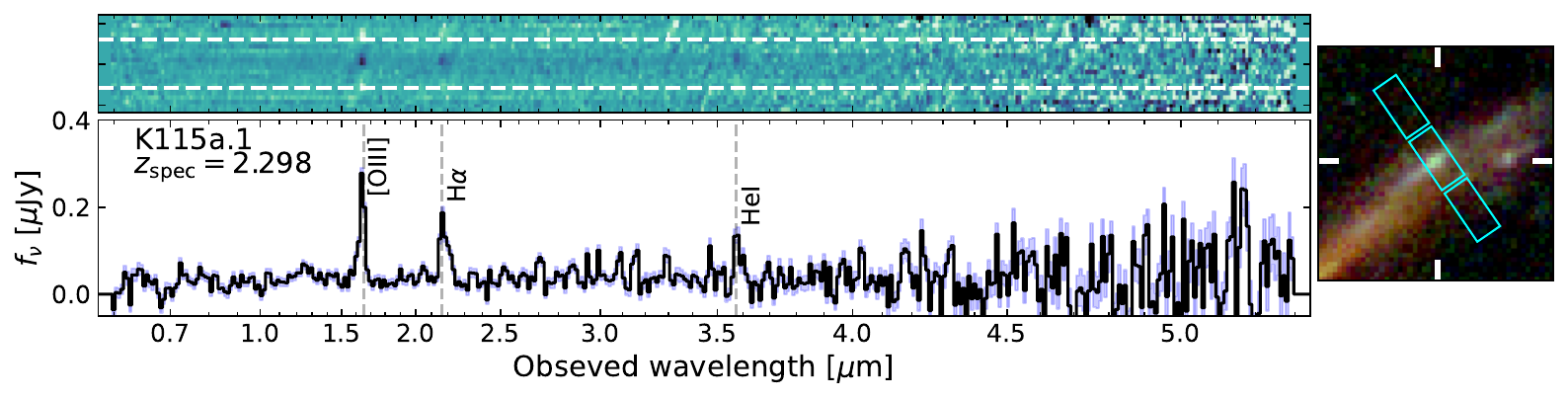}\\
\caption{2D and 1D NIRSpec prism spectra of multiple images with previously unknown system redshift. Vertical dashed lines indicate some of the prominent emission lines. The cutouts show the RGB composition of JWST/NIRCam and HST/WFC3 images (F277W, F356W, F410M, F444W in red, F115W, F150W, F200W in green and F814W, F606W, F435W and F090W in blue) with the size of $2''$. Cyan rectangles indicate the MSA slit positions.}
\label{fig:nirspec_canucs_new}
\end{figure}

\begin{figure}
\centering
\includegraphics[width=\linewidth]{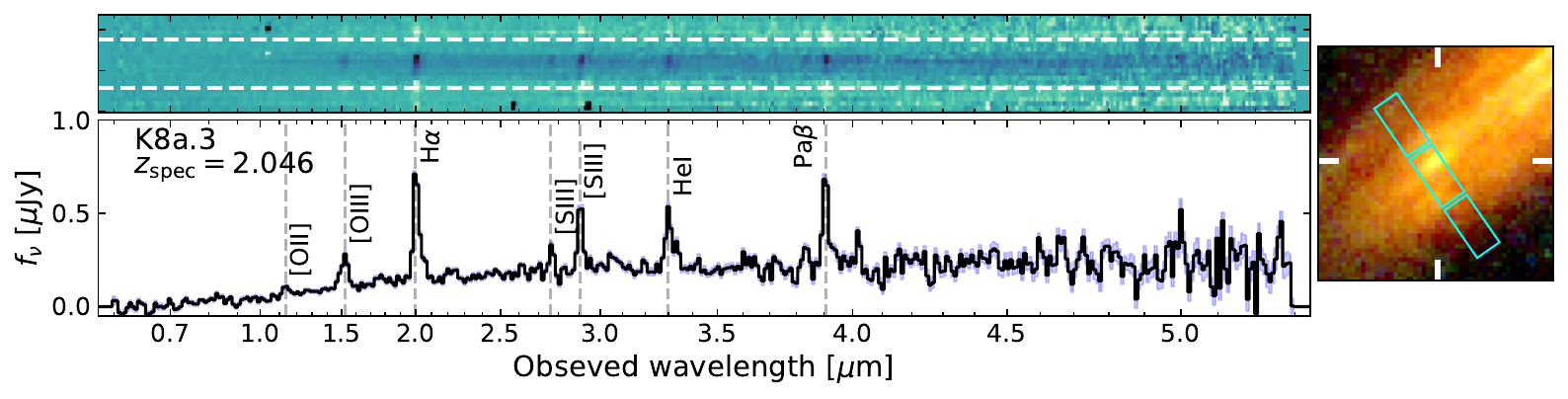}\\ 
\includegraphics[width=\linewidth]{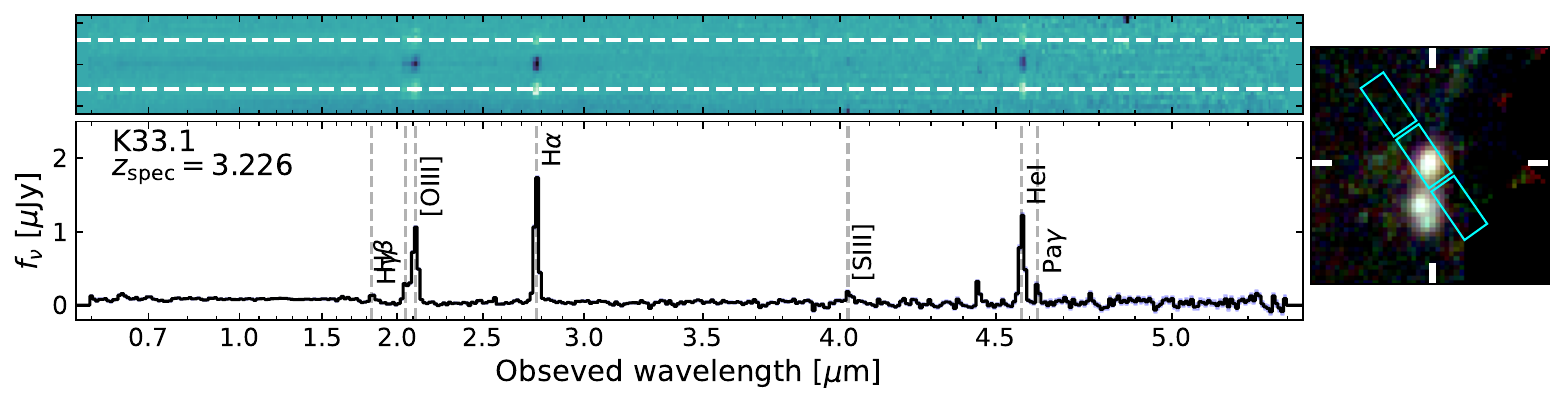}\\ 
\includegraphics[width=\linewidth]{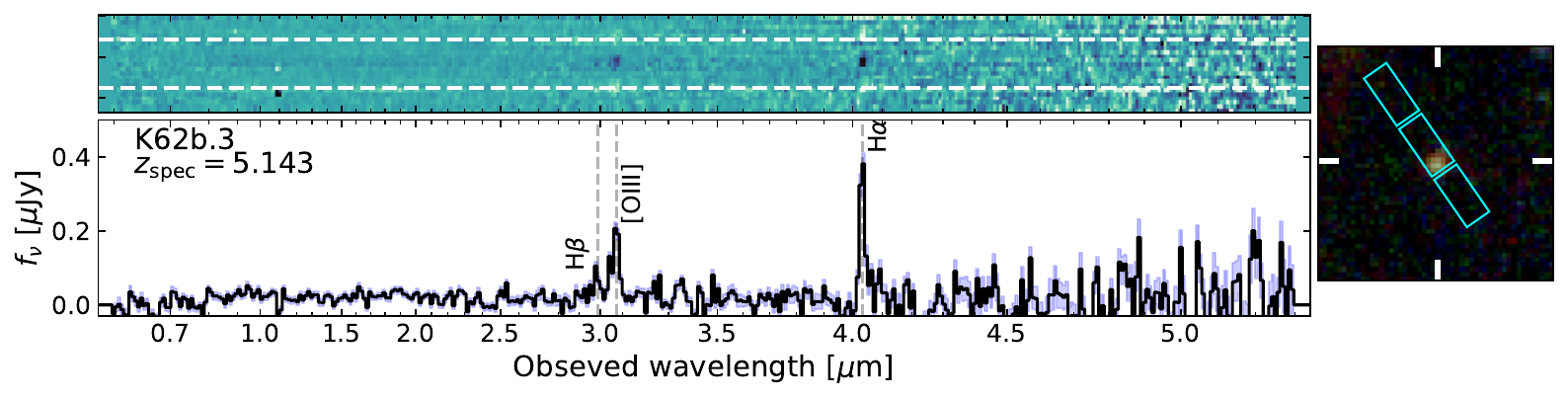}\\

\caption{2D and 1D NIRSpec prism spectra, not included in Fig. \ref{fig:nirspec_canucs_new} but also used in this work, either for updating the system redshift (K8a.3), confirming a new multiple image candidate (K62b.3) or for showing the emission line, detected in the grism spectra of the counter images (K33.1).  }
\label{fig:nirspec_canucs_old}
\end{figure}

In Figs. \ref{fig:nirspec_canucs_new} and \ref{fig:nirspec_canucs_old} we show the NIRSpec prism spectra used for obtaining our updated multiple image catalogue. The reduction and redshift extraction is outlined in Sect. \ref{sec:nirspec}. In the following, we describe two cases where \texttt{msaexp} fitting with wide redshift prior did not produce satisfying results. In the spectrum of K82.2 we find a noticeable emission line consistent with \OIII\ emission in K82.1 grism spectrum (Fig. \ref{fig:nirspec_canucs1}, different part of the same arc), thus, we used a narrow redshift prior between 1.8 and 2.4, obtaining a spectroscopic redshift of 2.053. The spectrum of K89.3 shows an \Halpha\ emission line, consistent with redshift obtained from \OII\ and \OIII\ emission lines in the grism spectrum of the same image, shown in Fig. \ref{fig:nirspec_canucs1} (the spectral range containing \OII\ and \OIII\ lines is missing in the NIRSpec spectrum). As the full spectral fit could not produce reliable redshift, we fitted the \Halpha\ emission line with a Gaussian profile and obtained the redshift of $3.082\pm0.008$. 


\twocolumn

\end{document}